\pdfoutput=1
\documentclass[twocolumn,showkeys,floatfix]{aastex631}
\usepackage{fix-cm}
\usepackage{epstopdf}
\usepackage{xcolor}
\usepackage{xspace}
\usepackage{graphicx}
\usepackage{amssymb, amsmath}
\usepackage{float}
\usepackage{verbatim}
\usepackage{textcomp}
\usepackage{latexsym}
\usepackage{multirow}
\usepackage[integrals]{wasysym}
\usepackage{academicons}
\usepackage{gensymb}
\usepackage{longtable}
\usepackage{ifthenx}
\usepackage{comment}
\usepackage{silence}
\usepackage{placeins}
\usepackage{soul}
\usepackage{longfigure}
\usepackage{hyperref}
\usepackage{cleveref}

\usepackage{supertabular}

\shorttitle{Long term variability of PKS~2155$-$304}
\shortauthors{Thimmappa et al.}

\accepted{2026 November 16}
\revised{2026 November 12} 
\received{2026 August 03} 

\graphicspath{{./}{figures/}}

\begin{document}
%\linenumbers

\title{Long-Term Spectral and Timing Variability of PKS 2155$-$304 from {\it Chandra} and {\it NuSTAR} Observations}
\correspondingauthor{R.~Thimmappa}
\email{rameshan.thimmappa@villanova.edu}

\author[0000-0001-5122-8425]{Rameshan Thimmappa}
\affiliation{Department of Physics, Villanova University, Villanova, PA 19085, USA}

%\author[0000-0002-8247-786X]{Joey Neilsen}
%\affiliation{Department of Physics, Villanova University, Villanova, PA 19085, USA}

\author[0000-0002-9019-9441]{ S. B. Gudennavar}
\affiliation{Department of Physics and Electronics, CHRIST University, Bengaluru 560029, Karnataka, India}

\begin{abstract}
\noindent Blazars exhibit strong variability across the electromagnetic spectrum, over a wide range of timescales from minutes to decades. We present a comprehensive study of the spectral and timing variability of the TeV blazar PKS~2155$-$304 using {\it Chandra} and {\it NuSTAR} observations spanning 25 years (1999$-$2024). Timing and flux distribution analyses reveal a wide range of variability, with fractional variability amplitudes of up to $F_{\rm var}\approx 24\%$ in the hard X-ray band and $F_{\rm var}\lesssim 7.5\%$ in the soft X-ray band, along with variable power-spectral indices across individual observations. The minimum variability timescale, $\tau_{\rm var}$ = 1.15$\pm$0.26~ks, shows an upper limit on the X-ray emitting region size of $R \lesssim 7.7 \times 10^{14}$~cm, consistent with synchrotron emission from a compact region in the relativistic jet. We fitted the X-ray spectra using power-law (PL), broken power-law (BPL), log-parabola (LP), and bremsstrahlung plus power-law (BR+PL) models, the latter used only as a phenomenological description of the spectral curvature. The BPL model is statistically preferred (F-test and AIC) for most {\it Chandra} spectra. An F-test confirms this preference over PL for all {\it Chandra} spectra, whereas most {\it NuSTAR} spectra are consistent with a single power law. Hardness ratio analysis shows that many observations exhibit limited spectral evolution with flux, while some epochs display significant hardness variations correlated with brightness, consistent with a harder-when-brighter trend in the Chandra band. 
Joint fitting of the simultaneous {\it Chandra} and {\it NuSTAR} observations obtained during the 2013 flare shows that the broadband 0.4$-$79~keV spectral curvature is confined to the soft X-ray band, with the spectrum steepening above $\sim$2~keV and the hard X-ray spectrum following a steep power law ($\Gamma \simeq 2.7$). Overall, our results reveal energy-dependent X-ray variability in PKS~2155$-$304 over more than two decades, and show that broadband X-ray coverage is needed to locate where the curvature of the synchrotron tail occurs in high-synchrotron-peaked blazars.

\end{abstract}


\section{Introduction}
\label{sec:intro}
\noindent Blazars, a subclass of active galactic nuclei (AGNs), are highly luminous sources whose relativistic jets are aligned nearly along our line of sight. This geometric alignment results in strong Doppler boosting of non-thermal emission, making the AGNs among the most luminous and variable sources \citep{Urry95}. These sources show strong spectral and flux variability across the electromagnetic spectrum on timescales ranging from minutes to decades. This makes them ideal laboratories for studying particle acceleration, radiative cooling, and emission processes in relativistic jets. The broadband spectral energy distribution (SED) of blazars is characterized by two different components: a low-energy hump produced by synchrotron radiation from relativistic electrons in the jet, and a high-energy hump. The origin of the high-energy hump was widely debated between two classes of models, leptonic and hadronic. In leptonic scenarios, the high-energy emission is explained through inverse-Compton scattering of either synchrotron photons within the jet (synchrotron self-Compton, SSC) or photons from external radiation fields, whereas hadronic scenarios attribute it to proton-synchrotron emission and/or photo-meson-induced cascades \citep[e.g.,][]{Mannheim93, Mucke01, Bottcher13}. Variability studies, in particular the rapid and correlated X-ray and TeV flux variations observed in high-synchrotron-peaked blazars, have supported leptonic models, which are now widely used to describe the SEDs of these sources \citep[e.g.,][]{Bottcher07, Ghisellini09}. Hadronic models are, however, still discussed, mainly in connection with the possible association of blazars with high-energy neutrinos \citep{IceCube18}. Detailed spectral and temporal investigations are therefore crucial for constraining particle acceleration and cooling mechanisms and understanding the physical conditions that govern jet emission. 
\\[6pt]
PKS~2155$-$304 ($z = 0.116$; \citealt{Falomo93}) is one of the brightest and most extensively studied high-synchrotron-peaked (HSP) blazars. It is widely observed from radio up to very-high-energy (VHE; $E > 100$ GeV) $\gamma$-rays. In particular, it is notable for its extreme variability and minute-scale TeV flares \citep{Aharonian07}. Also, long-term multiwavelength monitoring campaigns have characterized the baseline variability of PKS~2155$-$304 over multiple decades \citep{HESS2014, Wierzcholska15}. Recent observations of {\it Swift} and {\it Fermi} from 2005 to 2024 confirm that flux distributions in the optical, X-ray, and $\gamma$-ray bands are all consistent with log-normality, with no stable temporal lags found between bands \citep{Wierzcholska26}. Furthermore, multi-epoch SED modeling from 2008 to 2023 reveals that the X-ray fractional variability amplitude reaches $\sim$0.75, exceeding that in the optical/UV and $\gamma$-ray bands. These identify distinct flux states including X-ray flares, $\gamma$-ray flares, and multiwavelength outbursts \citep{Harutyunyan26}.
\\[6pt]
In the X-ray band, PKS~2155$-$304 offers the high-energy tail of the synchrotron component, which also gives direct constraints on the acceleration and cooling of relativistic particles in the jet. Several studies show that its X-ray spectra are often curved, favoring models like the log-parabola to describe the gradual steepening at higher energies \citep{Massaro04,Tramacere09}. A recent study on PKS~2155$-$304 with {\it Imaging X-ray Polarimetry Explorer (IXPE)} reveals a significant X-ray polarization variability, displaying one of the highest polarization degrees among all {\it IXPE} detected blazars ($\sim$22\% in the 2$-$8 keV band), which provides new constraints on the jet magnetic field geometry and particle acceleration mechanisms \citep{Hu24, Kouch24}.
\\[6pt]
The capabilities of {\it Chandra} and {\it NuSTAR} have significantly advanced X-ray studies of blazars, enabling detailed probes of the particle acceleration and emission mechanisms. With its sub-arcsecond angular resolution and sensitivity over the 0.1$-$10.0 keV band, {\it Chandra} provides precise measurements of soft X-ray emission, while {\it NuSTAR} extends this coverage into the hard X-ray regime over 3.0$-$79.0 keV with unprecedented sensitivity \citep{Harrison13}. Together, these facilities allow us to study the broadband spectral and timing variability over more than two decades, bridging the synchrotron peak and the inverse-Compton component. The first joint NuSTAR and XMM-Newton study of PKS~2155$-$304 during a low-flux state revealed a log-parabola spectrum with a hard power-law tail extending to $\sim$60 keV, challenging the simple one-zone synchrotron self-Compton model \citep{Madejski16}.
\\[6pt]
The X-ray variability of PKS~2155$-$304 has been extensively studied with several observatories, including {\it BeppoSAX}, {\it Chandra}, {\it NuSTAR}, {\it XMM-Newton}, {\it Swift}-XRT, {\it AstroSat}, {\it Suzaku}, and {\it RXTE} \citep[e.g.,][]{Zhang99, Edelson01, Kapanadze14, Bhattacharyya20, Zhang21, Bhatta25}. These studies have revealed pronounced flux variability on both short and long timescales, accompanied by spectral changes that often follow a harder-when-brighter trend \citep{Kapanadze14}. Such behavior is generally associated with particle acceleration and radiative cooling processes in relativistic jets, possibly driven by shocks, turbulence, or magnetic reconnection \citep{Giannios09, Hu25}. In some cases, instabilities in the accretion flow may also influence the jet emission, indicating a possible disk$-$jet connection \citep{Chatterjee09}. Detailed X-ray spectral and timing studies of PKS~2155$-$304 are therefore crucial for constraining the physical origin of its variability, identifying the dominant jet-emission processes, and probing the coupling between the accretion flow and relativistic outflow.
\\[6pt]
In this work, we present a comprehensive investigation of the long-term X-ray timing and spectral variability of PKS~2155$-$304 using 16 {\it Chandra} and 11 {\it NuSTAR} observations. These datasets enable us to examine (i) the evolution of the X-ray spectrum across the soft and hard energy bands, (ii) the relationship between flux and spectral shape, and (iii) the presence of spectral curvature and its variability. While several of these diagnostics have been applied to subsets of this data previously (e.g., \citealt{Pandey17, Bhatta18} for the 2012–2013 NuSTAR epochs; \citealt{Madejski16} for a joint hard X-ray spectral fit with XMM-Newton), the joint broadband  {\it Chandra}+{\it NuSTAR} spectral fitting presented here extends curvature constraints to 79 keV using this specific instrument pairing, and the light curve is extended to the 2023–2024 {\it NuSTAR} epochs, both presented here for the first time. We retain the earlier {\it NuSTAR} epochs (2012–2013) and the full 25-year Chandra baseline in our analysis, rather than adopting the previously published values directly, in order to provide a single, uniformly reprocessed and self-consistent dataset using a common calibration (CALDB 20230705), extraction pipeline, and variability/spectral methodology across the full time span, and to enable direct comparison with the newly analyzed 2023–2024 epochs within the same framework.
\\[6pt]
This paper is organized as follows. In Section~\ref{sec:data_reduction}, we describe the {\it Chandra} and {\it NuSTAR} observations and data reduction procedures. In Section~\ref{sec:results}, we present the spectral and timing analyses and the corresponding results. In Section~\ref{sec:discussion}, we discuss the implications of our findings for blazar emission models, and summarize the main conclusions in Section~\ref{sec:conclusion}. 
\begin{table*}[!htbp]
\centering
\caption{{\it Chandra} and {\it NuSTAR} observational data of PKS~2155$-$304.}
\label{tab:grouped_columns}
\begin{tabular}{ccccccccc}
\hline
\hline
\multicolumn{4}{c}{{\it Chandra}} & \multicolumn{4}{c}{{\it NuSTAR}} \\
\hline
ObsID & Date & MJD & Exposure & ObsID & Date & MJD & Exposure \\
      & (yyyy-mm-dd) & & (ks) & & (yyyy-mm-dd) & & (ks) \\
\hline
337   & 1999-10-20 & 51471 & 38.66 & 10002010001 & 2012-07-08 & 56116 & 33.833 \\
1705  & 2000-05-31 & 51695 & 25.50 & 60002022002 & 2013-04-23 & 56405 & 45.050 \\
1014  & 2001-03-28 & 51996 & 26.69 & 60002022004 & 2013-07-16 & 56489 & 13.857 \\
3167  & 2001-11-30 & 52243 & 29.65 & 60002022006 & 2013-08-02 & 56506 & 10.971 \\
3708  & 2002-11-29 & 52607 & 26.62 & 60002022008 & 2013-08-08 & 56512 & 13.494 \\
3706  & 2002-11-29 & 52607 & 27.71 & 60002022010 & 2013-08-14 & 56518 & 10.533 \\
5173  & 2004-11-23 & 53332 & 26.71 & 60002022012 & 2013-08-26 & 56530 & 11.356 \\
6926  & 2006-07-06 & 53922 & 13.82 & 60002022014 & 2013-09-04 & 56539 & 12.282 \\
7291  & 2006-07-07 & 53923 & 13.81 & 60002022016 & 2013-09-28 & 56563 & 11.524 \\
8380  & 2007-07-14 & 54295 & 15.29 & 11001611002 & 2023-12-07 & 60285 & 42.190 \\
8436  & 2007-07-14 & 54295 & 15.29 & 11001611004 & 2024-11-06 & 60620 & 51.968 \\
9705  & 2008-07-02 & 54649 & 14.93 \\
9712  & 2008-07-03 & 54650 & 14.80 \\
14265 & 2012-04-28 & 56045 & 26.50 \\
15475 & 2013-04-24 & 56406 & 28.39 \\
16423 & 2014-04-25 & 56772 & 28.14 \\
\hline
\end{tabular}
\label{table1_observation_details}
\end{table*}
\section{Observations and Data Analysis}
\label{sec:data_reduction}
\noindent We use {\it Chandra} and {\it NuSTAR} observational data to study the spectral and timing variability of the blazar PKS~2155$-$304, observed from 1999 to 2024. 
\\[6pt]
The blazar PKS~2155$-$304 was observed on-axis with the Advanced CCD Imaging Spectrometer – Spectroscopic array \citep[ACIS-S,][]{Garmire03} onboard the {\it Chandra} X-ray Observatory \citep{Weisskopf00} between 20 October 1999 and 25 April 2014 with a total exposure time of 372.5~ks. The details of ObsID, date, MJD, and exposure time are given in Table~\ref{table1_observation_details}. The observational data were reprocessed using the {\ttfamily chandra\_repro} script as per CIAO v4.16 \citep{Fruscione06} analysis threads\footnote{\url{http://cxc.harvard.edu/ciao/threads/}}, using Chandra Calibration Database (CALDB)\,v4.12.0. Pixel randomization and readout streaks were removed from the data during processing, and astrometric correction was performed. The {\ttfamily wcs\_match} and {\ttfamily wcs\_update} tools were used to match all the observation IDs (ObsIDs) with reference ObsID 337. All event and aspect solution files were subsequently updated. 
For our spectral and timing analysis, we extracted spectra and photon counts for the source and background regions from individual event files using the {\ttfamily specextract} script in the energy range 0.4$-$8.0~keV. For each ObsID, the source spectrum was selected from a circular region with a radius of 10~px ($\simeq 5^{\prime\prime}$; pixel scale $0^{\prime\prime}.492,{\rm px}^{-1}$), and the background spectrum was selected from a concentric annulus region of 12$-$20~px, respectively. The centroid of each source region was selected using SAOImage DS9\footnote{\url{https://ds9.si.edu}} \citep{Joye03}. 
The spectra were fitted using the \textsc{Sherpa}\footnote{\url{https://cxc.cfa.harvard.edu/sherpa}} package \citep{Freeman01}. We calculated the background-subtracted flux in three energy bands: the soft band (0.4$-$2.0~keV), the hard band (2.0$-$8.0~keV), and the full band (0.4$-$8.0~keV).
The count rates for all exposures place us in the regime where calibration uncertainties dominate over statistical uncertainties \citep{Drake06}. Methods to account for calibration uncertainties in {\it Chandra} data analysis have been discussed by \citet{Lee11} and \citet{Xu14}. 
The moderate source count rate of $\simeq$ 0.16$-$0.61~{counts~s$^{-1}$}, for the source located at the center of the S3 chip, indicates a low probability of detector pile-up \citep{Davis01}. We have further verified this during the spectral analysis, as discussed in Section \ref{sec:spectral_fitting}.
\\[6pt]
We analyzed {\it NuSTAR} data from July 2012 to November 2024 for the blazar PKS~2155$-$304, with a total exposure time of $\sim$257 ks. The details of ObsID, date, MJD, and exposure time are given in Table~\ref{table1_observation_details}. The observational data cover the hard X-ray band (3.0$-$79.0 keV). We used the two co-aligned telescopes onboard {\it NuSTAR}, each focusing onto independent focal plane modules (FPMA and FPMB). The data reduction was performed using the NuSTAR Data Analysis Software ({\ttfamily NuSTARDAS}) package with the recent calibration files ({\ttfamily CALDB 20230705}). The {\ttfamily nupipeline} script was used to produce calibrated and cleaned level 2 event files. The source spectra and flux were extracted using the {\ttfamily nuproducts} script from a circular region of $30''$ radius around the source location, and the background was extracted from nearby source-free circular regions with a radius of $60''$. For the light curve, time was binned at 500\,s, and spectra were binned with 20 counts per channel. Spectral fitting was performed using the \textsc{Sherpa} package with {\ttfamily chi2gehrels} statistics and uncertainties at the $1\sigma$ confidence level. We calculated the background-subtracted flux in three energy bands: soft band (3.0$-$10.0~keV), hard band (10.0$-$79.0~keV), and full band (3.0$-$79.0~keV). 
\\[6pt]
Throughout this paper, quantities that are positive by definition (variability amplitude, $F_{\rm var}$, $\tau_{\rm var}$, break energy, temperature and column density) are reported as upper limits when they are measured with a signal-to-noise ratio below 2; in that case we quote the best-fit value plus twice its $1\sigma$ uncertainty. Spectral parameters that are not constrained by the fit (i.e., the parameter runs to a hard limit and no finite confidence interval can be computed) are not reported and are flagged in the tables. Where the fit is poorly constrained and the covariance errors are unreliable, we instead quote asymmetric $1\sigma$ confidence intervals computed with the Sherpa \texttt{conf} command.
\\[6pt]
ObsIDs 14265, 15475 and 16423 were observed in subarray mode. For ObsID 16423, the source region periodically dithered partly off the active chip, producing artificial dips in the light curve on the $\sim$1~ks dither period. We therefore corrected its light curve for the fraction of the extraction region on the active chip, computed with the CIAO tool \texttt{dither\_region}, and excluded intervals in which less than half of the region was on the chip.

\begin{figure*}[!htbp]
	\centering 
    \includegraphics[width=0.90\textwidth]{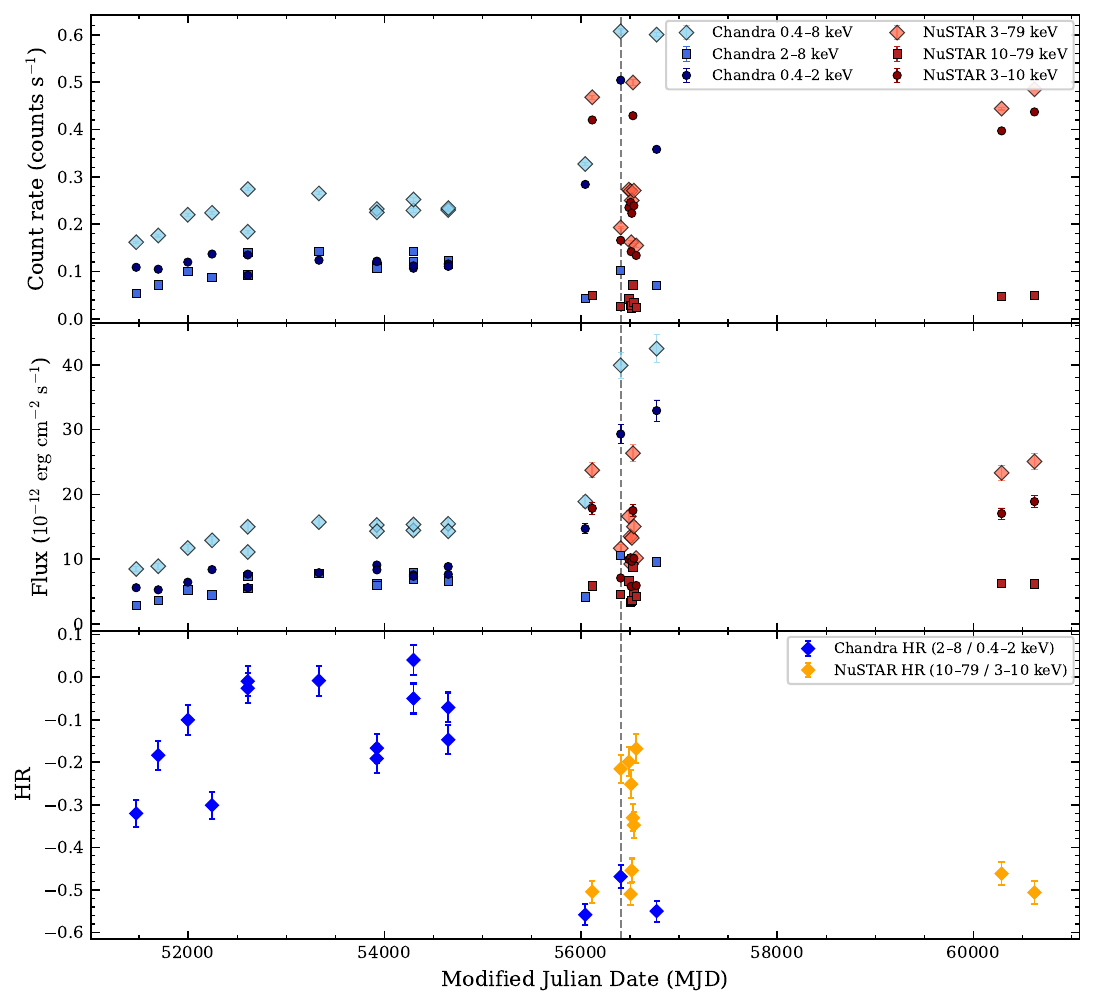}
	\caption{HR, flux and light curve of PKS~2155$-$304 from {\it Chandra} and {\it NuSTAR}. The gray colored line indicates the same time observations of {\it NuSTAR} ObsID 60002022002 (56405.99 days) and {\it Chandra} ObsID 15475 (56406.17 days).}
   \label{fig:Lightcurve_Chandra_NuSTAR}
\end{figure*}
\section{Results}
\label{sec:results}
\noindent PKS~2155$-$304 is known to exhibit rapid X-ray variability across a broad range of timescales. We analyzed 16 {\it Chandra} and 11 {\it NuSTAR} observations spanning 1999$-$2024 to investigate its timing and spectral variability. The results are presented in the following sections.
\subsection{The Light Curves and Hardness Ratio}
\label{sec:lightcurve_results}
\noindent The background-subtracted light curves from 16 \textit{Chandra} and 11 \textit{NuSTAR} observations are presented in Figure~\ref{fig:Lightcurve_Chandra_NuSTAR}, which shows the hardness ratio (HR), flux, and count rate as a function of Modified Julian Date (MJD). The source exhibits significant flux variability over the 25 year observational timescale. The HR evolution indicates both long-term spectral changes and significant variations between individual epochs. Owing to its high-energy coverage, \textit{NuSTAR} systematically samples harder spectral states than \textit{Chandra}.
\subsection{X-ray Flux Variability}
\label{sec:variability}
\begin{figure*}[!ht]
\centering
\includegraphics[width=0.85\textwidth]{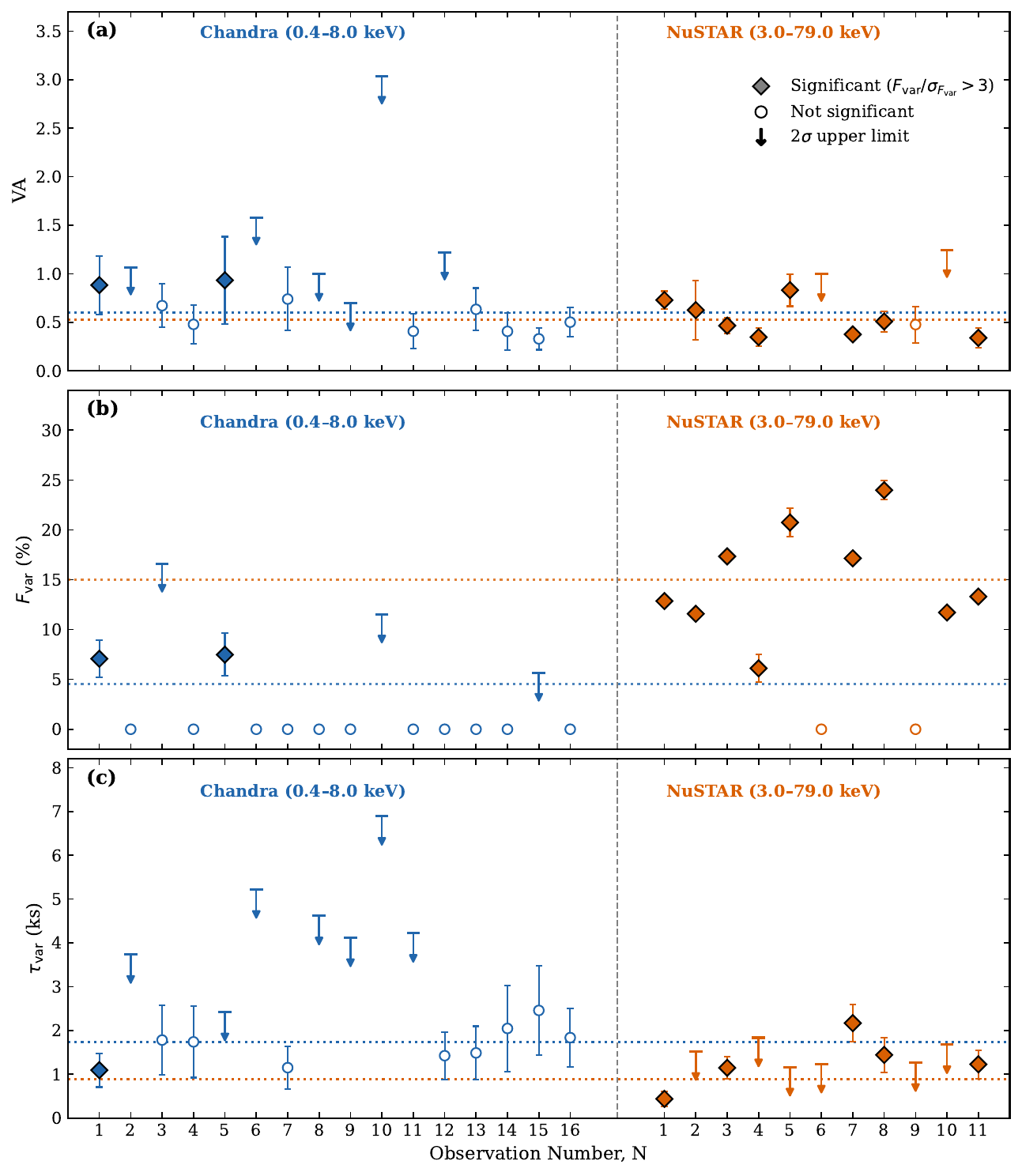}
\caption{Intraday variability properties ((a) VA, (b) $F_{\rm var}$ (\%) and (c) $\tau_{\rm var}$ (ks)) of PKS~2155$-$304 from {\it Chandra} (blue, left) and {\it NuSTAR} (orange, right) observations using 500\,s binned light curves, plotted against observation number (see Tables~\ref{tab:chandra_var} and~\ref{tab:nustar_var} for the corresponding ObsID). Filled diamonds indicate significant variability \textbf{($F_{\rm var}/\sigma_{F_{\rm var}} > 3$)}; open circles indicate non-significant detections; downward arrows indicate $2\sigma$ upper limits; observations with negative excess variance are plotted at $F_{\rm var} = 0$}. The dotted horizontal lines mark the mean value for each instrument in each panel. The dashed vertical line separates the two instruments.
\label{fig:variability_comparison}
\end{figure*}

\noindent In order to quantify the observed X-ray flux variability, we computed variability amplitude ($VA$), the fractional root-mean-square variability amplitude ($F_{\rm var}$) and the minimum variability timescale ($\tau_{\rm var}$) for each observation using light curves binned at 500~s. These quantities were estimated as per the following methods.
\\[6pt]
Variability amplitude ($VA$) gives information about peak-to-peak flux variations and is defined as:
\begin{equation}
    VA = \frac{F_{\rm max}-F_{\rm min}}{F_{\rm min}},
    \label{eqn:VA}
\end{equation}
\noindent where, $F_{\rm max}$ and $F_{\rm min}$ are the maximum and minimum flux, respectively. The uncertainty of $VA$ is estimated using
\begin{equation}
    \sigma_{VA} = (VA+1)\cdot\sqrt{\left(\frac{\sigma_{F_{\rm max}}}
    {F_{\rm max}}\right)^2+\left(\frac{\sigma_{F_{\rm min}}}
    {F_{\rm min}}\right)^2},
\end{equation}
\noindent where, $\sigma_{F_{\rm max}}$ and $\sigma_{F_{\rm min}}$ are the uncertainties of $F_{\rm max}$ and $F_{\rm min}$, respectively. 
\\[6pt]
Since $VA$ quantifies only the peak-to-peak flux variation, it may not adequately represent the overall variability behavior of the source. Therefore, the fractional root-mean-square variability amplitude ($F_{\rm var}$) is often preferred, as it provides a more comprehensive measure of the observed intrinsic flux variability. To quantify the flux variability across different X-ray bands, we calculated $F_{\rm var}$, which measures the intrinsic variability of the source after accounting for measurement uncertainties. It is defined as:
\begin{equation}
    F_{\rm var} = \sqrt{\frac{S^{2} - \bar{\sigma}^{2}_{\rm err}}{\bar{x}^2}},
    \label{equation:fractional_variability}
\end{equation}
\noindent where, $\bar{x}$ is the arithmetic mean of the measured count rates $x_i$, $S^2$ is the sample variance, and $\bar{\sigma}^{2}_{\rm err}$ is the mean squared measurement error. The uncertainty of the $F_{\rm var}$ is calculated using the equation:
\begin{equation}
   \sigma_{F_{\rm var}} =\sqrt{\left(\frac{1}{\sqrt{2N}} \,
   \frac{\bar{\sigma}_{\rm err}^{2}}{\bar{x}^{2} F_{\rm var}}
   \right)^{2}+\left(\sqrt{\frac{\bar{\sigma}_{\rm err}^{2}}{N}} \,
   \frac{1}{\bar{x}}\right)^{2}}
   \end{equation}
\\[6pt]
Since blazars exhibit variability over a wide range of timescales, we additionally estimated the minimum variability timescale ($\tau_{\rm var}$, in seconds) for each observation. The minimum variability timescale is defined as:
\begin{equation}
    \tau_{\rm var} = \left|\frac{dt}{d\ln F}\right|,
\end{equation}
\noindent where, $dt$ is the time interval between flux ($F$) measurements. The uncertainty of $\tau_{\rm var}$ is given by
\begin{equation}
    \sigma_{\tau_{\rm var}} \approx \sqrt{\frac{F_1^2\Delta F_2^2 + 
    F_2^2\Delta F_1^2}{F_1^2 F_2^2 
    (\ln[F_1/F_2])^4}}\cdot\Delta t,
\end{equation}
\noindent where, $\Delta F_1$ and $\Delta F_2$ are the flux uncertainties used to estimate the minimum variability timescale for the fluxes $F_1$ and $F_2$, respectively.
\\[6pt]
The results of the intraday variability analysis are presented in Tables~\ref{tab:chandra_var} and \ref{tab:nustar_var}, respectively, and are illustrated in Figure~\ref{fig:variability_comparison}. The {\it NuSTAR} observations exhibit systematically larger fractional variability amplitudes 
($F_{\rm var}\approx 6-24\%$) than the {\it Chandra} observations ($F_{\rm var}\lesssim 7.5\%$), indicating stronger variability in the hard X-ray band. The minimum variability timescales measured with a signal-to-noise ratio $\geq 2$ span 0.44$-$2.17~ks for {\it NuSTAR} and 1.1$-$2.5~ks for {\it Chandra}; for the remaining observations only upper limits are obtained (Tables~\ref{tab:chandra_var} and \ref{tab:nustar_var}), suggesting that the hard X-ray emission varies on shorter timescales and likely originates from a more compact emitting region. Since a fixed threshold on $\chi^2/{\rm dof}$ does not correspond to a fixed significance level when the number of bins varies from $\sim$30 to $\sim$130 between observations, we quantify the significance of the intraday variability through the ratio $F_{\rm var}/\sigma_{F_{\rm var}}$, and consider the variability significant when $F_{\rm var}/\sigma_{F_{\rm var}} > 3$ (bold values in Tables~\ref{tab:chandra_var} and \ref{tab:nustar_var}). With this criterion, significant intraday variability is detected in 9 of the 11 {\it NuSTAR} observations, compared to only 2 of the 16 {\it Chandra} observations (ObsIDs 337 and 3708), reinforcing the stronger and more frequent variability at higher X-ray energies. Nine of the eleven {\it NuSTAR} epochs analyzed here (2012--2013) were previously studied by \citet{Pandey17, Bhatta18}; as noted in Section~\ref{sec:intro}, we reprocess and re-analyze them alongside the new 2023--2024 epochs to provide a single, uniformly calibrated baseline across the full 25-year span.
% ============================================================
% TABLE — Chandra Variability (500\,s binning)
% ============================================================
\begin{table*}[!htbp]
\centering
\caption{Intraday variability properties of PKS~2155$-$304 from \textit{Chandra} observations using 500~s binned light curves in the energy range 0.4$-$8.0 keV. $N$: number of valid time bins. $F_{\rm var}$ values in bold are significant at $F_{\rm var}/\sigma_{F_{\rm var}} > 3$. Upper limits ($2\sigma$) are given for quantities with S/N $< 2$. ``---'' indicates a negative excess variance, i.e., no intrinsic variability above the measurement noise level.}
\label{tab:chandra_var}
\begin{tabular}{clccccc}
\hline\hline
Observation & ObsID & $N$ & Mean Count Rate & VA &
$F_{\rm var}$ & $\tau_{\rm var}$ \\
  Number &  & & (counts s$^{-1}$) & & (\%) & (ks) \\
\hline
1 & 337 & 79 & 0.170 $\pm$ 0.002 & 0.88 $\pm$ 0.30 & 7.1 $\pm$ 1.8 & 1.09 $\pm$ 0.39 \\
2 & 1705 & 53 & 0.184 $\pm$ 0.003 & $<$1.1 & --- & $<$3.7 \\
3 & 1014 & 55 & 0.228 $\pm$ 0.003 & 0.67 $\pm$ 0.23 & $<$17 & 1.78 $\pm$ 0.80 \\
4 & 3167 & 61 & 0.234 $\pm$ 0.003 & 0.48 $\pm$ 0.20 & --- & 1.74 $\pm$ 0.82 \\
5 & 3708 & 55 & 0.189 $\pm$ 0.003 & 0.93 $\pm$ 0.45 & 7.5 $\pm$ 2.1 & $<$2.4 \\
6 & 3706 & 58 & 0.279 $\pm$ 0.004 & $<$1.6 & --- & $<$5.2 \\
7 & 5173 & 56 & 0.271 $\pm$ 0.004 & 0.74 $\pm$ 0.33 & --- & 1.15 $\pm$ 0.49 \\
8 & 6926 & 29 & 0.238 $\pm$ 0.004 & $<$1.0 & --- & $<$4.6 \\
9 & 7291 & 29 & 0.232 $\pm$ 0.004 & $<$0.70 & --- & $<$4.1 \\
10 & 8380 & 32 & 0.232 $\pm$ 0.004 & $<$3.0 & $<$12 & $<$6.9 \\
11 & 8436 & 32 & 0.257 $\pm$ 0.004 & 0.41 $\pm$ 0.18 & --- & $<$4.2 \\
12 & 9705 & 32 & 0.237 $\pm$ 0.004 & $<$1.2 & --- & 1.42 $\pm$ 0.54 \\
13 & 9712 & 31 & 0.241 $\pm$ 0.004 & 0.63 $\pm$ 0.22 & --- & 1.49 $\pm$ 0.61 \\
14 & 14265 & 57 & 0.329 $\pm$ 0.004 & 0.41 $\pm$ 0.19 & --- & 2.04 $\pm$ 0.98 \\
15 & 15475 & 61 & 0.612 $\pm$ 0.005 & 0.33 $\pm$ 0.11 & $<$5.7 & 2.5 $\pm$ 1.0 \\
16 & 16423 & 60 & 0.600 $\pm$ 0.005 & 0.50 $\pm$ 0.15 & --- & 1.83 $\pm$ 0.67 \\

\hline
\end{tabular}
\end{table*}
% ============================================================
% TABLE — NuSTAR Variability (500\,s binning)
% ============================================================
\begin{table*}[!htbp]
\centering
\caption{
Intraday variability properties of PKS~2155$-$304 from \textit{NuSTAR} (FPMA+FPMB combined) observations using 500~s binned light curves in the energy range 3.0$-$79.0 keV. Notation as in Table~\ref{tab:chandra_var}.}
\label{tab:nustar_var}
\begin{tabular}{clccccc}
\hline\hline
Observation & ObsID & $N$ & Mean Count Rate & VA &
$F_{\rm var}$ & $\tau_{\rm var}$ \\
Number &  & & (counts s$^{-1}$) & & (\%) & (ks) \\
\hline
1 & 10002010001 & 91 & 1.303 $\pm$ 0.021 & 0.729 $\pm$ 0.091 & 12.86 $\pm$ 0.20 & 0.44 $\pm$ 0.17 \\
2 & 60002022002 & 117 & 0.563 $\pm$ 0.010 & 0.62 $\pm$ 0.31 & 11.58 $\pm$ 0.31 & $<$1.5 \\
3 & 60002022004 & 34 & 0.802 $\pm$ 0.027 & 0.465 $\pm$ 0.084 & 17.34 $\pm$ 0.51 & 1.15 $\pm$ 0.26 \\
4 & 60002022006 & 31 & 0.738 $\pm$ 0.019 & 0.346 $\pm$ 0.095 & 6.1 $\pm$ 1.4 & $<$1.8 \\
5 & 60002022008 & 45 & 0.474 $\pm$ 0.023 & 0.83 $\pm$ 0.17 & 20.7 $\pm$ 1.4 & $<$1.2 \\
6 & 60002022010 & 32 & 0.751 $\pm$ 0.028 & $<$1.0 & --- & $<$1.2 \\
7 & 60002022012 & 31 & 1.354 $\pm$ 0.048 & 0.375 $\pm$ 0.044 & 17.15 $\pm$ 0.53 & 2.17 $\pm$ 0.43 \\
8 & 60002022014 & 32 & 0.788 $\pm$ 0.038 & 0.51 $\pm$ 0.11 & 23.98 $\pm$ 0.96 & 1.44 $\pm$ 0.40 \\
9 & 60002022016 & 34 & 0.454 $\pm$ 0.015 & 0.48 $\pm$ 0.19 & --- & $<$1.3 \\
10 & 11001611002 & 110 & 1.283 $\pm$ 0.019 & $<$1.2 & 11.71 $\pm$ 0.19 & $<$1.7 \\
11 & 11001611004 & 134 & 1.368 $\pm$ 0.019 & 0.34 $\pm$ 0.10 & 13.31 $\pm$ 0.16 & 1.23 $\pm$ 0.33 \\
\hline
\end{tabular}
\end{table*}
\subsection{Long-term Flux Distribution}
\label{sec:flux_dist}
\begin{figure*}[!htbp]
    \centering
    \includegraphics[width=0.98\textwidth]{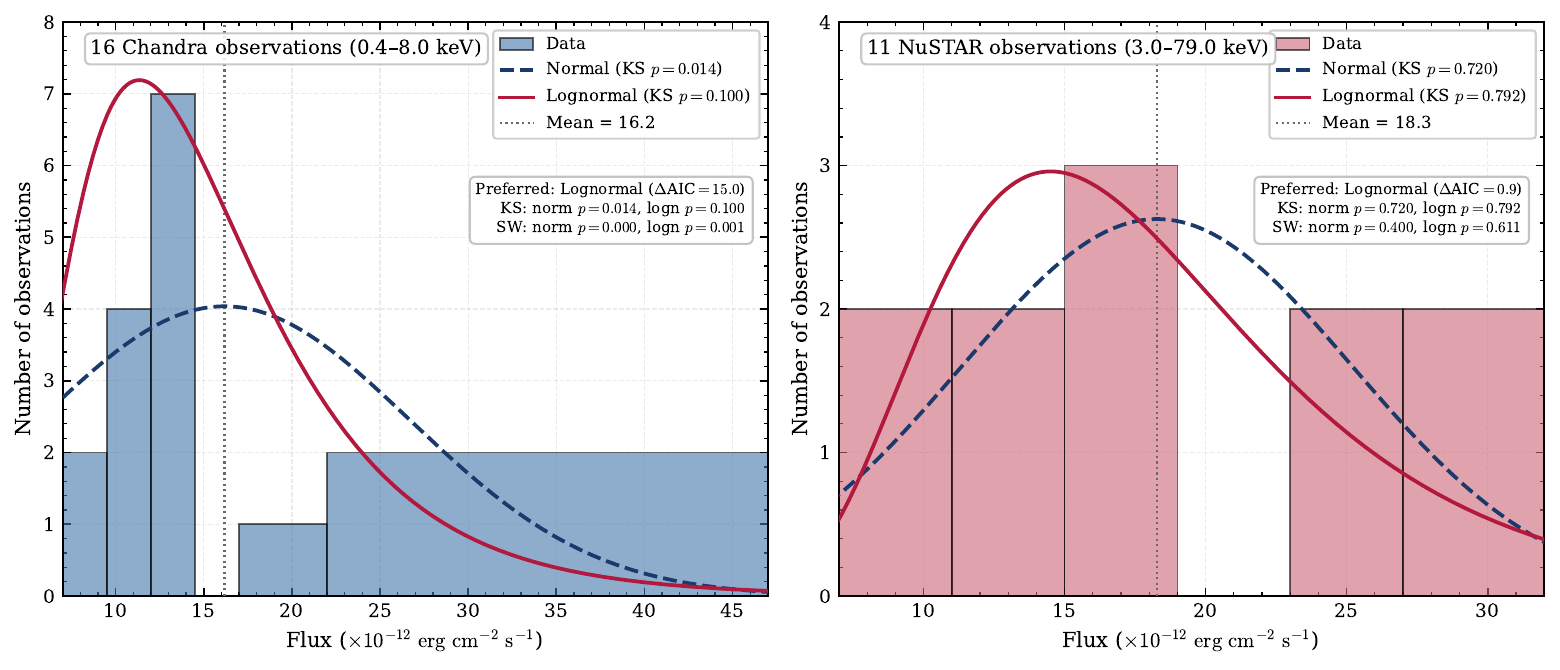}
    \caption{Long-term X-ray flux distributions of PKS~2155$-$304 from {\it Chandra} (Left Panel) and {\it NuSTAR} (Right Panel) using the epoch-averaged fluxes. The dashed blue and solid red curves show the best-fit normal and lognormal PDFs fitted by MLE. The dotted vertical line marks the mean flux. The annotation box in each panel lists the KS and SW test $p$-values and the preferred model based on AIC.}
    \label{fig:flux_dist}
\end{figure*}

\noindent To investigate the long-term X-ray variability, we analyzed the distribution of epoch-averaged fluxes derived from 16 {\it Chandra} and 11 {\it NuSTAR} observations over 25-year timescales. The flux distributions were modeled using normal and lognormal probability density functions (PDFs), with the parameters estimated through maximum likelihood estimation (MLE). The quality of the fits was evaluated using the Kolmogorov$-$Smirnov (KS) test, the Shapiro$-$Wilk (SW) test, and the Akaike Information Criterion (AIC). The resulting distributions and best-fitting models are presented in Figure~\ref{fig:flux_dist}.
\\[6pt]
The \textit{Chandra} flux distribution is strongly right-skewed, with 12 of the 16 observations clustered within $(8$-$15) \times 10^{-12}$~erg~cm$^{-2}$~s$^{-1}$ and two high-flux epochs (ObsID 15475 and 16423) reaching $\sim(40$-$45) \times 10^{-12}$~erg~cm$^{-2}$~s$^{-1}$, indicating occasional bright flaring activity against a predominantly quiescent baseline. Statistical model comparison strongly favors a lognormal PDF over a normal PDF ($\Delta\mathrm{AIC}=15.0$). Consistently, the KS test rejects normality ($p=0.014$) while remaining compatible with a lognormal distribution ($p=0.100$). The substantial difference between the mean flux ($16.2 \times 10^{-12}$~erg~cm$^{-2}$~s$^{-1}$) and the median flux ($12.4 \times 10^{-12}$~erg~cm$^{-2}$~s$^{-1}$) further supports the presence of a heavy-tailed and asymmetric distribution. However, the SW test rejects lognormality ($p=0.001$), most likely because the two extreme flaring observations produce an excess high-flux tail that is not adequately captured by a simple lognormal model.
\\[6pt]
For the \textit{NuSTAR} observations, both the normal and lognormal PDFs provide statistically acceptable representations of the flux distribution, with KS-test probabilities of $p = 0.720$ and $p = 0.792$, respectively. The small AIC difference ($\Delta\mathrm{AIC} = 0.9$) indicates no significant preference for either model. The SW test is also consistent with a lognormal distribution ($p = 0.611$), while the Anderson-Darling test does not reject lognormality at any of the standard significance levels. These results suggest that the \textit{NuSTAR} flux distribution is relatively well behaved and lacks the pronounced high-flux tail seen in the \textit{Chandra} data, likely reflecting the more uniform flux sampling of the \textit{NuSTAR} observations.
\\[6pt]
Lognormal flux distributions are commonly observed in blazars and AGN across the optical, X-ray, and $\gamma$-ray bands, and are widely regarded as signatures of multiplicative, non-linear variability processes \citep{Uttley05, Bhatta21}. In such a framework, fluctuations generated on different timescales are coupled and propagate through the jet in a multiplicative manner, naturally giving rise to a lognormal flux distribution. The preference for a lognormal flux distribution in the \textit{Chandra} data, together with the positive $\mathrm{RMS}-$ flux correlation observed in the \textit{NuSTAR} band (Section~\ref{sec:rms_flux}), suggests that the long-term X-ray variability of PKS~2155$-$304 is governed by multiplicative variability mechanisms acting over a broad range of timescales and spanning both the soft and hard X-ray regimes over more than two decades of observations.
\subsection{Power Spectral Density Analysis}
\label{sec:psd}
\begin{figure*}[!htbp]
    \centering
    \includegraphics[width=\textwidth]{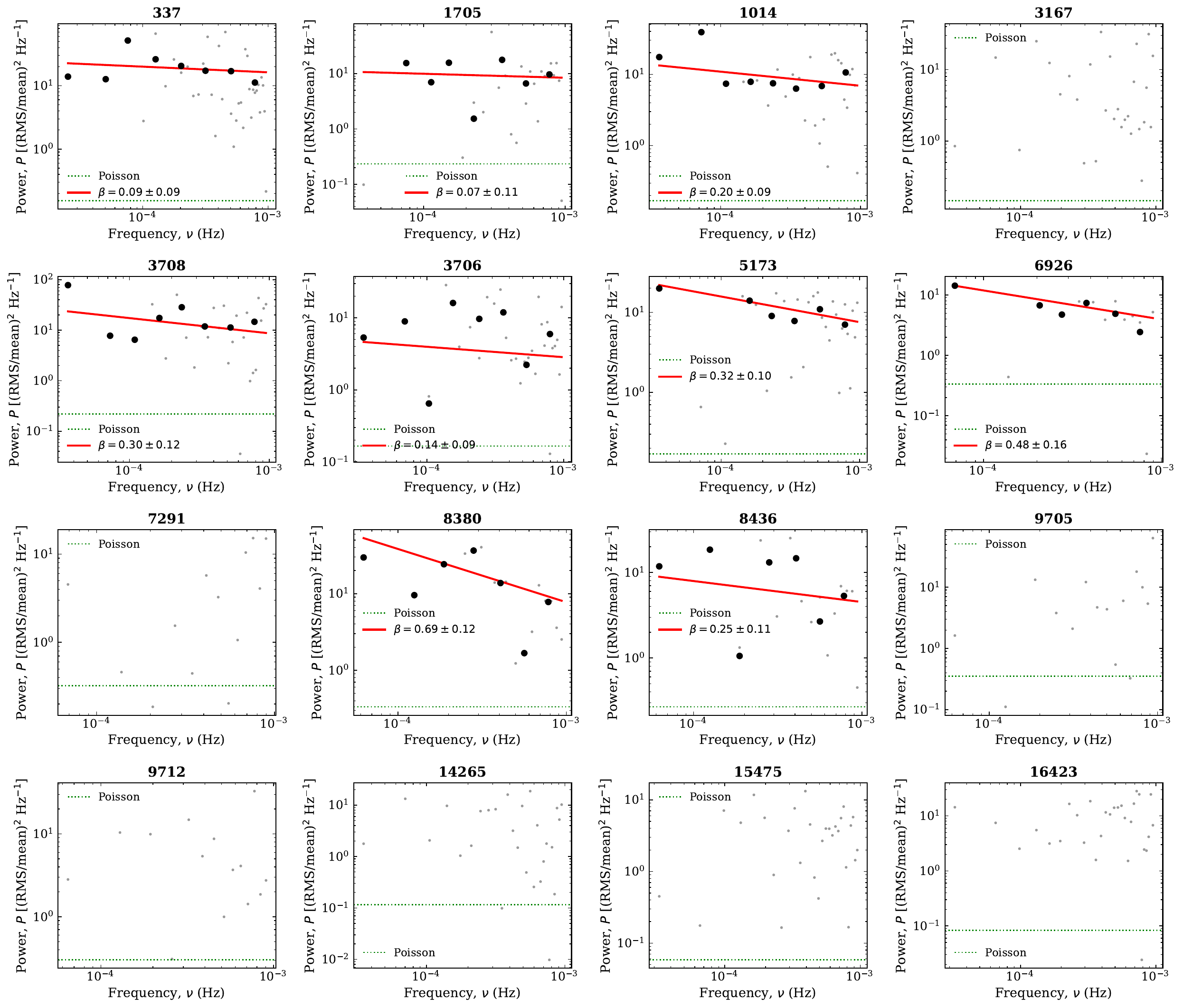}
    \caption{Intraday PSD of PKS~2155--304 from {\it Chandra} observations using 500~s binned light curves. Gray points show the raw periodogram; black circles show the geometrically rebinned PSD; the red line shows the best-fit power-law $P(\nu) \propto \nu^{-\beta}$ with $\beta$ annotated; the green dotted line marks the Poisson noise level ($P_{noise}$). Panels without a fit were classified as noise-dominated ($\langle P(\nu)\rangle$ $<$ 3$P_{noise}$)}
    \label{fig:psd_chandra}
\end{figure*}

\begin{figure*}[!htbp]
    \centering
    \includegraphics[width=\textwidth]{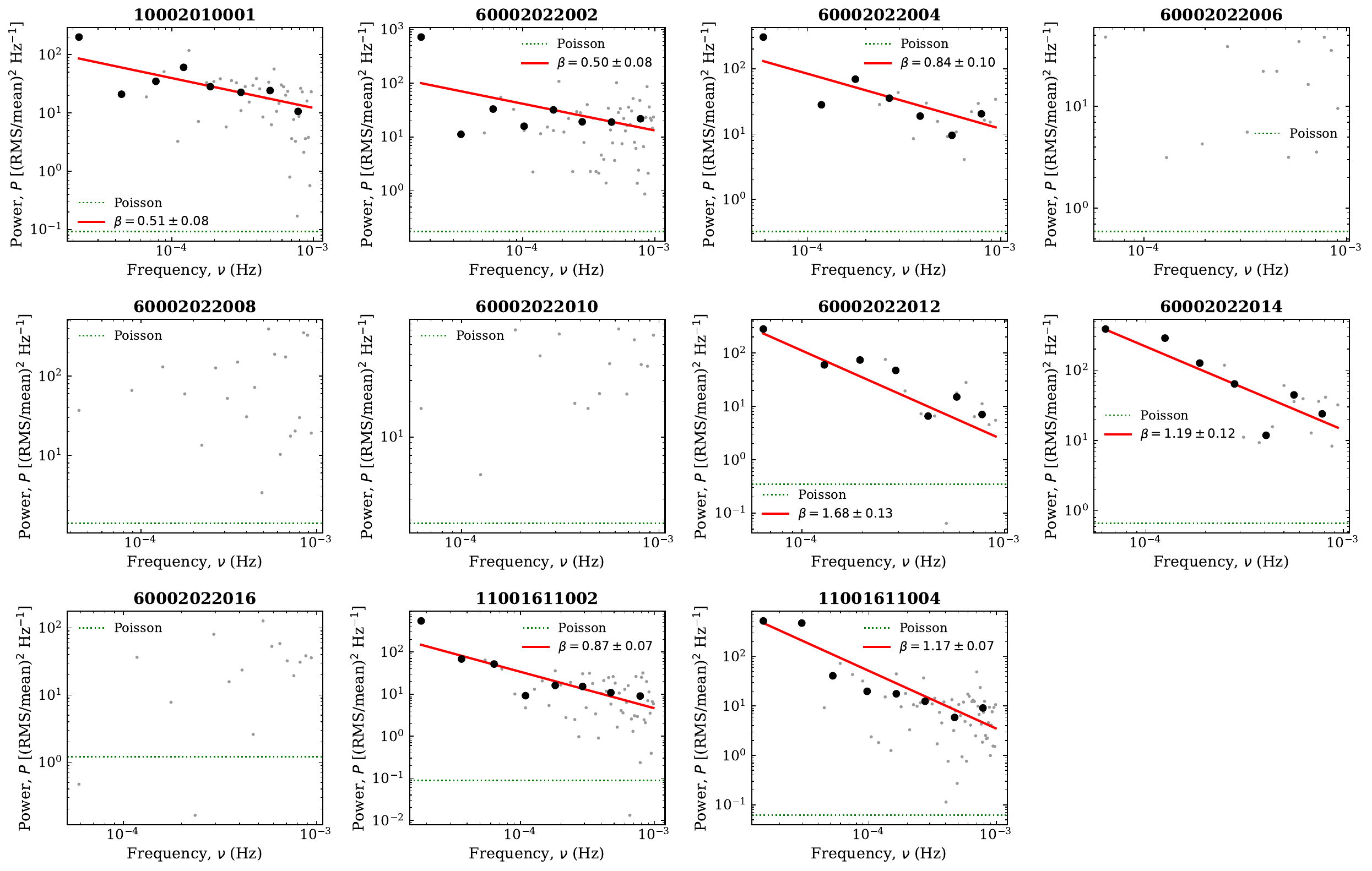}
    \caption{
    Intraday PSD of PKS~2155--304 from {\it NuSTAR} observations using 500~s binned light curves. Gray points show the raw periodogram; black circles show the geometrically rebinned PSD; the red line shows the best-fit power-law $P(\nu) \propto \nu^{-\beta}$ with $\beta$ annotated; the green dotted line marks the Poisson noise level ($P_{noise}$). Panels without a fit were classified as noise-dominated ($\langle P(\nu)\rangle$ $<$ 3$P_{noise}$).}
    \label{fig:psd_nustar}
\end{figure*}

\begin{figure}[!htbp]
    \centering
    \includegraphics[width=\columnwidth]{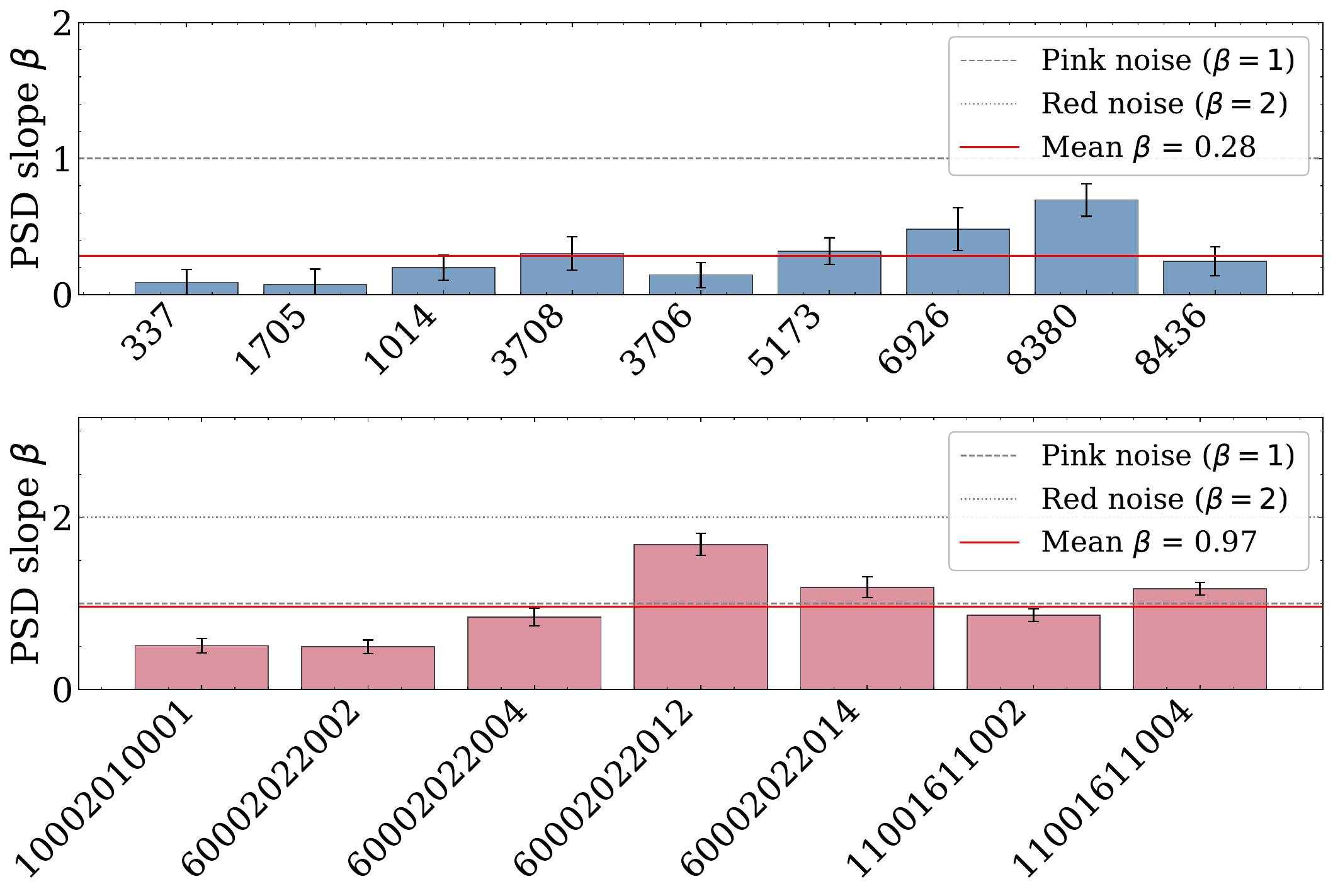}    
    \caption{Summary of PSD slopes $\beta$ for PKS~2155--304 from {\it Chandra} (top) and {\it NuSTAR} (bottom). Only observations with reliable fits ($\langle P(\nu)\rangle$ $<$ 3$P_{noise}$) are shown. The dashed and dotted horizontal lines mark the pink-noise ($\beta=1$) and red-noise ($\beta=2$) levels, respectively. The red horizontal line shows the mean $\beta$ for each instrument.}
    \label{fig:psd_summary}
\end{figure}

\noindent To characterize the temporal X-ray variability, we computed the power spectral density (PSD) for each observation using a 500~s binning timescale. We follow the periodogram approach of \citet{Vaughan03}, computing the one-sided discrete Fourier transform (DFT) with $\mathrm{RMS}^2$ normalization:
\begin{equation}
P(\nu_j) = \frac{2\,\Delta t}{N\,\bar{x}^2}
            \left|\sum_{k=1,j =1}^{N,~N/2} x_k\, e^{2\pi i \nu_j t_k}\right|^2,
\end{equation}
\noindent where, $\Delta t = 500$~s is the bin size, $N$ is the number of valid time bins, $\bar{x}$ is the mean count rate, and $\nu_j = j/(N\Delta t)$ are the discrete Fourier frequencies for $j = 1, 2, \ldots, \lfloor N/2 \rfloor$, spanning from the lowest frequency $1/(N\Delta t)$ to the Nyquist frequency $1/(2\Delta t)$. The Poisson noise level was estimated analytically as 
\begin{equation}
P_{\rm noise} = \frac{2\bar{\sigma}^2_{\rm err}\,\Delta t}{N\bar{x}^2}
\end{equation}
\noindent where, $\bar{\sigma}^2_{\rm err}$ is the mean squared measurement error, and subtracted before fitting. Observations in which the mean power did not exceed three times the Poisson noise level were classified as noise-dominated and excluded from further PSD analysis; this applied to 7 of 16 {\it Chandra} and 4 of 11 {\it NuSTAR} observations. These noise-dominated observations correspond to the majority of epochs showing a negative excess variance (for which $F_{\rm var}$ is undefined) in Tables~\ref{tab:chandra_var} and~\ref{tab:nustar_var} (8 of 11 cases); the remaining three (ObsIDs 15475, 60002022006, and 60002022008) show a positive excess variance despite a noise-dominated periodogram, indicating that the variability in these epochs, where present, is not well captured by a power-law PSD across the sampled frequency range. The remaining PSDs were geometrically rebinned into 8 frequency bins and fitted with a power-law model $P(\nu) = A\,\nu^{-\beta}$ in log-log space. The resulting PSD plots and slope summary are shown in Figures~\ref{fig:psd_chandra}, \ref{fig:psd_nustar}, and~\ref{fig:psd_summary}.
\\[6pt]
For the 9 {\it Chandra} observations with reliable PSD measurements, the PSD slopes span the range $\beta$ = 0.07$-$0.69, with a mean of $\langle\beta\rangle = 0.28 \pm 0.18$; three of them (ObsIDs 337, 1705, and 3706) are consistent with $\beta = 0$ within $2\sigma$. These shallow slopes, predominantly in the white-to-pink noise regime ($\beta < 1$), reflect the low count rates of the {\it Chandra} observations ($\sim$0.16$-$0.61 counts~s$^{-1}$). The intrinsic variability power remains close to the Poisson noise floor on intraday timescales. The steepest \textit{Chandra} slope, $\beta = 0.69 \pm 0.12$ (ObsID 8380), is found in an observation with no significant intraday flux variability ($F_{\rm var} < 12\%$; Table~\ref{tab:chandra_var}) but with the clearest intraday spectral evolution in our sample (Section~\ref{sec:HR_maps}).
\\[6pt]
For the 7 {\it NuSTAR} observations with reliable PSD measurements, the slopes span $\beta$ = 0.50$-$1.68, with a mean of $\langle\beta\rangle = 0.97 \pm 0.38$. This range encompasses both the flicker-noise ($\beta \sim 1$) and shallow red-noise ($\beta \sim 1$--$2$) regimes. The steepest slope, $\beta = 1.68 \pm 0.13$, is measured in ObsID~60002022012, which also exhibits significant intraday fractional variability ($F_\mathrm{var} = 17.15\%$, Table~\ref{tab:nustar_var}). The highest fractional variability among the {\it NuSTAR} observations is found in ObsID~60002022014 ($F_\mathrm{var} = 23.98\%$), which similarly shows a steep PSD slope ($\beta = 1.19 \pm 0.12$), reinforcing the association between strong flux variability and red-noise-dominated power spectra in this band.
\\[6pt]
The {\it NuSTAR} PSD slopes are systematically steeper than those of {\it Chandra} ($\langle\beta\rangle$ = 0.97 versus 0.28), consistent with stronger and more correlated variability in the hard X-ray band. Our results, while limited by the shorter {\it Chandra} exposures and lower count rates, extend this picture to the hard X-ray band, where the flicker-noise character ($\beta \sim 1$) of the {\it NuSTAR} PSD is consistent with stochastic particle acceleration processes operating within the jet on intraday timescales \citep{Marscher14, Bhatta18}.

\begin{table*}[!htbp]
\centering
\caption{Best-fitting parameters of PKS~2155$-$304 from {\it Chandra} with different models. 
The model with the lowest AIC for each observation is shown in bold. F-test and AIC model comparison results are reported in Section~\ref{sec:spectral_fitting}. Curvature parameters consistent with zero within $2\sigma$ are given as limits on $|\beta|$; upper limits ($2\sigma$) are given for $E_{\rm b}$ and $kT$ measured with S/N $<2$. $^{\rm a}$Parameter not constrained (pegged at a hard limit, or no finite confidence interval could be computed).}
\begin{tabular}{clc c c c}
\hline \hline
ObsID & Model & $\alpha | \Gamma_1$ & $E_{b}$ (keV)$|\beta$ & $\Gamma_2 | kT$ (keV) & $\chi^2_{\nu}$ \\
\hline
337   & PL & 2.370 $\pm$ 0.025 & {} & {} & 0.77 \\
      & BPL & 2.418 $\pm$ 0.027 & 5.39 $\pm$ 0.18 & $-$2.23 $\pm$ 0.71 & \textbf{0.55} \\
      & LP & 2.447 $\pm$ 0.037 & $-$0.231 $\pm$ 0.076 & {} & 0.74 \\
      & BR+PL & 1.11 $\pm$ 0.24 & {} & 0.746 $\pm$ 0.064 & 0.62 \\
1705  & PL & 2.170 $\pm$ 0.030 & {} & {} & 0.86 \\
      & \textbf{BPL} & 2.231 $\pm$ 0.033 & 5.41 $\pm$ 0.19 & $-$2.50 $\pm$ 0.71 & \textbf{0.62} \\
      & LP & 2.287 $\pm$ 0.051 & $-$0.281 $\pm$ 0.094 & {} & 0.83 \\
      & BR+PL & $-$0.07 $\pm$ 0.41 & {} & 1.19 $\pm$ 0.11 & 0.72 \\
1014  & PL & 2.064 $\pm$ 0.027 & {} & {} & 1.03 \\
      & \textbf{BPL} & 2.145 $\pm$ 0.030 & 5.38 $\pm$ 0.14 & $-$2.91 $\pm$ 0.53 & \textbf{0.65} \\
      & LP & 2.296 $\pm$ 0.047 & $-$0.504 $\pm$ 0.082 & {} & 0.93 \\
      & BR+PL & $-$1.23 $\pm$ 0.36 & {} & 1.52 $\pm$ 0.10 & 0.78 \\
3167  & PL & 2.353 $\pm$ 0.026 & {} & {} & 1.09 \\
      & \textbf{BPL} & 2.436 $\pm$ 0.029 & 5.13 $\pm$ 0.14 & $-$2.20 $\pm$ 0.49 & \textbf{0.72} \\
      & LP & 2.588 $\pm$ 0.041 & $-$0.575 $\pm$ 0.076 & {} & 0.93 \\
      & BR+PL & $2.18^{+0.08}_{-4.5}$ & {} & $0.24^{+0.56}_{-0.03}$ & 1.02 \\
3708  & PL & 1.916 $\pm$ 0.031 & {} & {} & 0.98 \\
      & \textbf{BPL} & 2.054 $\pm$ 0.037 & 4.81 $\pm$ 0.18 & $-$0.99 $\pm$ 0.39 & \textbf{0.69} \\
      & LP & 2.356 $\pm$ 0.055 & $-$0.822 $\pm$ 0.088 & {} & 0.73 \\
      & BR+PL & 0.68 $\pm$ 0.22 & {} & 0.87 $\pm$ 0.12 & 0.71 \\
3706  & PL & 1.927 $\pm$ 0.023 & {} & {} & 0.95 \\
      & \textbf{BPL} & 1.963 $\pm$ 0.024 & 6.04 $\pm$ 0.18 & $-$2.73 $\pm$ 0.89 & \textbf{0.82} \\
      & LP & 2.079 $\pm$ 0.047 & $-$0.272 $\pm$ 0.075 & {} & 0.92 \\
      & BR+PL & 1.931 $\pm$ 0.024 & {} & \nodata$^{\rm a}$ & 0.96 \\
5173  & PL & 1.877 $\pm$ 0.024 & {} & {} & 0.93 \\
      & \textbf{BPL} & 1.956 $\pm$ 0.028 & 5.17 $\pm$ 0.21 & $-$0.56 $\pm$ 0.43 & \textbf{0.77} \\
      & LP & 2.238 $\pm$ 0.050 & $-$0.606 $\pm$ 0.076 & {} & 0.78 \\
      & BR+PL & $1.87^{+0.01}_{-1.3}$ & {} & $0.12^{+0.88}_{-0.04}$ & 0.87 \\
6926  & PL & 2.158 $\pm$ 0.041 & {} & {} & 1.03 \\
      & BPL & 2.379 $\pm$ 0.053 & 4.24 $\pm$ 0.20 & $-$0.58 $\pm$ 0.37 & 0.69 \\
      & LP & 2.690 $\pm$ 0.069 & $-$1.06 $\pm$ 0.11 & {} & 0.72 \\
      & \textbf{BR+PL} & 0.86 $\pm$ 0.22 & {} & 0.642 $\pm$ 0.081 & \textbf{0.69} \\
7291  & PL & 2.119 $\pm$ 0.041 & {} & {} & 0.93 \\
      & \textbf{BPL} & 2.318 $\pm$ 0.055 & 4.05 $\pm$ 0.25 & 0.12 $\pm$ 0.35 & \textbf{0.70} \\
      & LP & 2.561 $\pm$ 0.072 & $-$0.86 $\pm$ 0.12 & {} & 0.73 \\
      & BR+PL & 0.80 $\pm$ 0.28 & {} & 0.77 $\pm$ 0.11 & 0.70 \\
8380  & PL & 1.945 $\pm$ 0.039 & {} & {} & 1.03 \\
      & \textbf{BPL} & 2.129 $\pm$ 0.048 & 4.68 $\pm$ 0.19 & $-$1.18 $\pm$ 0.41 & \textbf{0.70} \\
      & LP & 2.491 $\pm$ 0.070 & $-$0.99 $\pm$ 0.11 & {} & 0.76 \\
      & BR+PL & 0.27 $\pm$ 0.29 & {} & 0.97 $\pm$ 0.13 & 0.73 \\
8436  & PL & 1.794 $\pm$ 0.036 & {} & {} & 0.83 \\
      & \textbf{BPL} & 1.885 $\pm$ 0.040 & 5.56 $\pm$ 0.20 & $-$2.39 $\pm$ 0.71 & \textbf{0.63} \\
      & LP & 2.275 $\pm$ 0.073 & $-$0.78 $\pm$ 0.11 & {} & 0.68 \\
      & BR+PL & 1.23 $\pm$ 0.17 & {} & 0.62 $\pm$ 0.15 & 0.71 \\
\hline
\end{tabular}
\label{spectral_fit_chandra1}
\end{table*}
\begin{table*}[!ht]
\centering
{\textbf{Table 4} (continued)
\vspace{0.1 cm} } \\
\begin{tabular}{clc c c c}
\hline \hline
ObsID & Model & $\alpha | \Gamma_1$ & $E_{b}$ (keV)$|\beta$ & $\Gamma_2 | kT$ (keV) & $\chi^2_{\nu}$ \\
\hline
9705  & PL & 2.059 $\pm$ 0.039 & {} & {} & 0.91 \\
      & BPL & 2.196 $\pm$ 0.045 & 5.10 $\pm$ 0.18 & $-$2.09 $\pm$ 0.56 & 0.61 \\
      & \textbf{LP} & 2.640 $\pm$ 0.069 & $-$1.06 $\pm$ 0.11 & {} & \textbf{0.60} \\
      & BR+PL & 1.12 $\pm$ 0.18 & {} & 0.562 $\pm$ 0.083 & 0.63 \\
9712  & PL & 1.969 $\pm$ 0.038 & {} & {} & 0.95 \\
      & \textbf{BPL} & 2.060 $\pm$ 0.042 & 5.53 $\pm$ 0.18 & \nodata$^{\rm a}$ & \textbf{0.69} \\
      & LP & 2.392 $\pm$ 0.074 & $-$0.75 $\pm$ 0.11 & {} & 0.81 \\
      & BR+PL & $-$0.48 $\pm$ 0.45 & {} & 1.46 $\pm$ 0.19 & 0.79 \\
14265 & PL & 2.655 $\pm$ 0.022 & {} & {} & 1.46 \\
      & BPL & 2.443 $\pm$ 0.038 & 2.06 $\pm$ 0.20 & 3.25 $\pm$ 0.13 & 1.13 \\
      & \textbf{LP} & 2.509 $\pm$ 0.031 & 0.514 $\pm$ 0.072 & {} & \textbf{1.12} \\
      & BR+PL & 2.74 $\pm$ 0.10 & {} & 1.37 $\pm$ 0.21 & 1.15 \\
15475 & PL & 2.529 $\pm$ 0.015 & {} & {} & 1.31 \\
      & \textbf{BPL} & 2.323 $\pm$ 0.028 & 2.08 $\pm$ 0.15 & 3.030 $\pm$ 0.079 & \textbf{0.94} \\
      & LP & 2.389 $\pm$ 0.023 & 0.399 $\pm$ 0.048 & {} & 1.04 \\
      & BR+PL & 2.614 $\pm$ 0.084 & {} & 1.67 $\pm$ 0.21 & 1.01 \\
16423 & PL & 2.679 $\pm$ 0.019 & {} & {} & 1.00 \\
      & \textbf{BPL} & 2.491 $\pm$ 0.034 & 2.06 $\pm$ 0.19 & 3.137 $\pm$ 0.098 & \textbf{0.75} \\
      & LP & 2.540 $\pm$ 0.029 & 0.402 $\pm$ 0.061 & {} & 0.79 \\
      & BR+PL & 2.728 $\pm$ 0.078 & {} & 1.40 $\pm$ 0.22 & 0.79 \\
\hline
\end{tabular}
\label{spectral_fit_chandra2}
\end{table*}

\begin{deluxetable*}{llcccc}[!htbp]
\tiny
\tablecaption{{\it NuSTAR} spectral fit results of PKS~2155$-$304 with different models. The {\it NuSTAR} background is subtracted and $N_{\rm H,Gal}$ is fixed at the HI4PI value. The preferred model for each observation (F-test probability $p<0.01$ relative to PL, then lowest AIC; PL where no curved model is significant) is shown in bold. F-test and AIC model comparison results are reported in Section~\ref{sec:spectral_fitting}. \label{tab:spectral_fit_NuSTAR}}
\tablehead{
\colhead{Observation ID} & \colhead{Model} & \colhead{$\alpha\,|\,\Gamma_1$} & \colhead{$E_{\rm b}$ (keV)$\,|\,\beta$} & \colhead{$\Gamma_2\,|\,kT$ (keV)} & \colhead{$\chi^2_\nu$}}
\startdata
10002010001 & PL    & $3.008 \pm 0.022$ & -- & -- & 0.79 \\
            & \textbf{BPL} & $2.868^{+0.058}_{-0.072}$ & $5.85^{+0.69}_{-0.74}$ & $3.120^{+0.053}_{-0.050}$ & 0.77 \\
            & LP    & $3.087^{+0.048}_{-0.045}$ & $0.21 \pm 0.10$ & -- & 0.78 \\
            & BR+PL & $3.008 \pm 0.022$ & -- & \nodata$^{\rm a}$ & 0.79 \\
60002022002 & \textbf{PL} & $2.728 \pm 0.031$ & -- & -- & 0.64 \\
            & BPL   & \nodata$^{\rm b}$ & \nodata$^{\rm b}$ & $2.732^{+0.034}_{-0.033}$ & 0.64 \\
            & LP    & $2.655^{+0.051}_{-0.047}$ & $|\beta| < 0.47$ & -- & 0.63 \\
            & BR+PL & $2.48^{+0.17}_{-0.20}$ & -- & $1.96^{+0.43}_{-0.72}$ & 0.63 \\
60002022004 & \textbf{PL} & $2.607 \pm 0.046$ & -- & -- & 0.73 \\
            & BPL   & \nodata$^{\rm b}$ & \nodata$^{\rm b}$ & $2.607 \pm 0.046$ & 0.74 \\
            & LP    & $2.604^{+0.074}_{-0.067}$ & $|\beta| < 0.38$ & -- & 0.74 \\
            & BR+PL & $2.53^{+0.29}_{-0.32}$ & -- & \nodata$^{\rm a}$ & 0.74 \\
60002022006 & \textbf{PL} & $3.049 \pm 0.056$ & -- & -- & 0.64 \\
            & BPL   & \nodata$^{\rm c}$ & \nodata$^{\rm c}$ & \nodata$^{\rm c}$ & 0.64 \\
            & LP    & $3.10^{+0.11}_{-0.10}$ & $|\beta| < 0.63$ & -- & 0.64 \\
            & BR+PL & $3.21^{+0.57}_{-0.25}$ & -- & \nodata$^{\rm a}$ & 0.64 \\
60002022008 & \textbf{PL} & $2.870 \pm 0.068$ & -- & -- & 0.67 \\
            & BPL   & \nodata$^{\rm c}$ & \nodata$^{\rm c}$ & \nodata$^{\rm c}$ & 0.66 \\
            & LP    & $2.73^{+0.11}_{-0.09}$ & $|\beta| < 0.93$ & -- & 0.65 \\
            & BR+PL & $2.79^{+0.08}_{-0.58}$ & -- & $<3.9$ & 0.66 \\
60002022010 & \textbf{PL} & $2.992 \pm 0.059$ & -- & -- & 0.59 \\
            & BPL   & \nodata$^{\rm c}$ & \nodata$^{\rm c}$ & \nodata$^{\rm c}$ & 0.60 \\
            & LP    & $3.00^{+0.11}_{-0.10}$ & $|\beta| < 0.52$ & -- & 0.60 \\
            & BR+PL & $2.96^{+0.70}_{-0.30}$ & -- & \nodata$^{\rm a}$ & 0.60 \\
60002022012 & PL    & $2.660 \pm 0.035$ & -- & -- & 0.71 \\
            & \textbf{BPL} & $2.555 \pm 0.050$ & $11.3^{+1.7}_{-1.5}$ & $3.18^{+0.27}_{-0.20}$ & 0.67 \\
            & LP    & $2.767^{+0.059}_{-0.056}$ & $0.40 \pm 0.15$ & -- & 0.68 \\
            & BR+PL & $2.81^{+0.50}_{-0.19}$ & -- & $8.0^{+2.2}_{-2.3}$ & 0.67 \\
60002022014 & \textbf{PL} & $2.768 \pm 0.049$ & -- & -- & 0.51 \\
            & BPL   & \nodata$^{\rm c}$ & \nodata$^{\rm c}$ & \nodata$^{\rm c}$ & 0.51 \\
            & LP    & $2.811^{+0.087}_{-0.079}$ & $|\beta| < 0.56$ & -- & 0.51 \\
            & BR+PL & $2.78^{+0.38}_{-0.19}$ & -- & \nodata$^{\rm a}$ & 0.52 \\
60002022016 & \textbf{PL} & $2.658 \pm 0.072$ & -- & -- & 0.72 \\
            & BPL   & $4.0^{+2.2}_{-1.0}$ & $3.72^{+0.66}_{-0.42}$ & $2.597 \pm 0.082$ & 0.70 \\
            & LP    & $2.59^{+0.11}_{-0.09}$ & $|\beta| < 0.75$ & -- & 0.72 \\
            & BR+PL & $2.594^{+0.087}_{-0.090}$ & -- & $<1.2$ & 0.71 \\
11001611002 & PL    & $2.947 \pm 0.020$ & -- & -- & 0.74 \\
            & BPL   & $2.72^{+0.13}_{-0.20}$ & $4.6^{+1.5}_{-0.7}$ & $3.008^{+0.034}_{-0.031}$ & 0.72 \\
            & LP    & $3.006^{+0.041}_{-0.039}$ & $|\beta| < 0.35$ & -- & 0.73 \\
            & \textbf{BR+PL} & $2.63^{+0.14}_{-0.16}$ & -- & $2.61^{+0.21}_{-0.17}$ & 0.71 \\
11001611004 & PL    & $3.017 \pm 0.017$ & -- & -- & 0.76 \\
            & BPL   & $2.901 \pm 0.058$ & $5.6^{+1.4}_{-0.6}$ & $3.094^{+0.049}_{-0.035}$ & 0.74 \\
            & \textbf{LP} & $3.091 \pm 0.034$ & $0.20 \pm 0.08$ & -- & 0.74 \\
            & BR+PL & $2.951^{+0.069}_{-0.082}$ & -- & $3.1^{+1.0}_{-0.5}$ & 0.74 \\
\enddata
\tablecomments{Uncertainties are $1\sigma$; upper limits are $2\sigma$. $^{\rm a}$Temperature not constrained; the bremsstrahlung component is not required by the data. $^{\rm b}$Break not constrained (it falls at or below the lower edge of the {\it NuSTAR} band); the BPL reduces to a PL. $^{\rm c}$Break not constrained within the band.}
\end{deluxetable*}

\subsection{{Spectral Analysis}}
\label{sec:spectral_fitting}
\begin{figure}[!htbp]
    \centering 
    \includegraphics[width=0.48\textwidth]{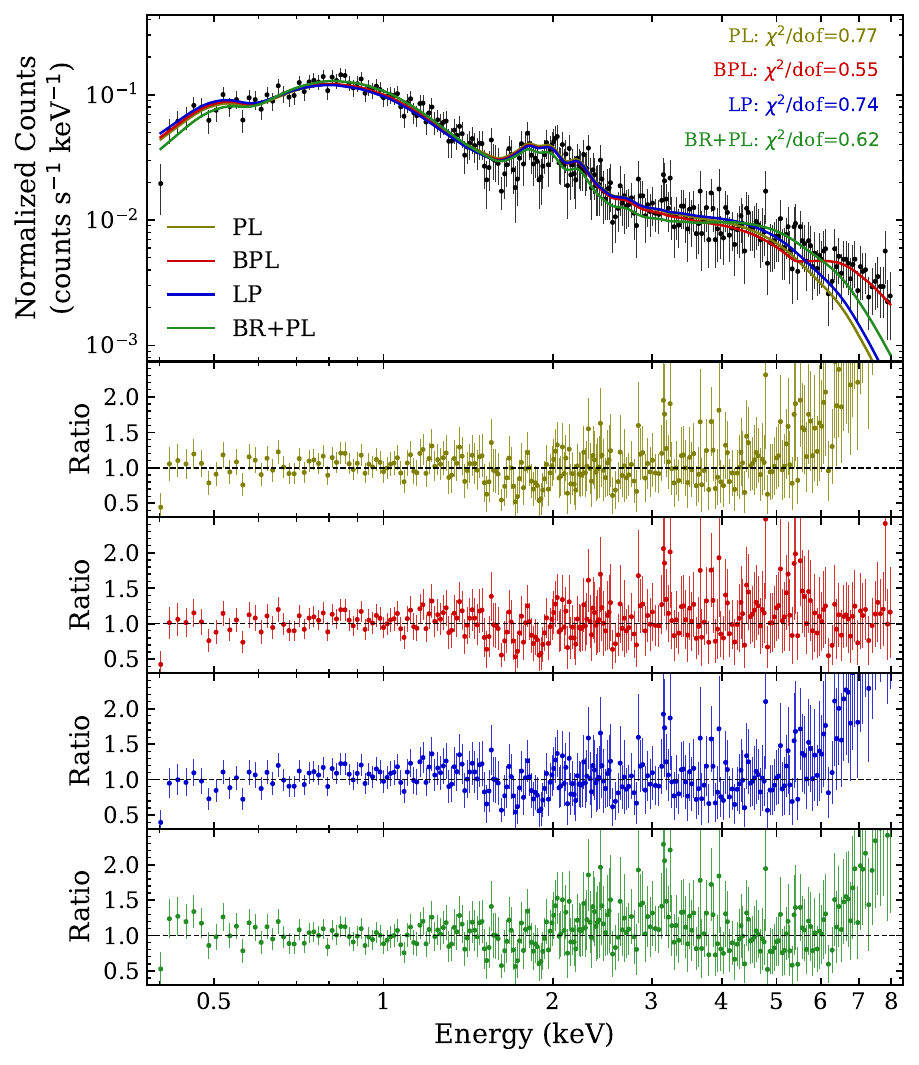}
    \caption{Spectral fitting of PKS 2155--304 in the 0.4$-$8.0~keV range using PL, BPL, LP, and BR+PL models for ObsID 337. Ratio (= data/model) is shown in the sub-panels for each model fitting in olive, red, blue, and green colors.}
    \label{fig:spec_example}
\end{figure}

We performed spectral analysis of PKS~2155$-$304 in the range of 0.4$-$8.0~keV for {\it Chandra} and the 3.0$-$79.0~keV band for {\it NuSTAR}. The spectra were extracted using the {\ttfamily specextract} script for {\it Chandra} and {\ttfamily nuproducts} for {\it NuSTAR}, applying the source and background regions as described in Section~\ref{sec:data_reduction}. To characterize the X-ray spectral shape, we fitted the {\it Chandra} and {\it NuSTAR} spectra of each ObsID with four models using \textsc{Sherpa}\footnote{\url{https://cxc.cfa.harvard.edu/sherpa/}} \citep{Freeman01}: absorbed power law (PL), broken-power law (BPL), log-parabola (LP), and bremsstrahlung plus power law (BR+PL).
One additional model, a cutoff power law (CPL), was used only for the joint broadband fitting presented in Section~\ref{sec:joint_fit}.
\\[6pt]
The background-subtracted source spectra from {\it Chandra} and {\it NuSTAR} were fitted in the 0.4$-$8.0~keV and 3.0$-$79.0~keV bands, respectively, using the PL, BPL, LP, and BR+PL models. In all fits, the Galactic hydrogen column density was fixed at $N_{\rm H,Gal} = 1.28 \times 10^{20}$~cm$^{-2}$ \citep{HI4PI16}. We note that analogous PL fits with the Galactic absorption set free (for {\it Chandra} band) returned similar results, only with slightly decreased values of $N_{\rm H,\,Gal}$ and the photon index, $\Gamma$. Furthermore, including the {\ttfamily jdpileup} model in the fitting procedure does not affect the best-fit values of the model parameters, as the fraction of piled-up events that result in a good grade turns out to be very low. 
\\[6pt]
Among the four models, we first applied the simple PL model; the curved models improve the fit significantly (F-test), and then explored more complex models, like BPL, LP, and BR+PL, as presented in Tables~\ref{spectral_fit_chandra1} and \ref{tab:spectral_fit_NuSTAR}. The LP model was motivated by the spectral curvature observed at lower energies, which has the lowest AIC for two {\it Chandra} observations (ObsIDs 9705 and 14265). Overall, all three complex models substantially improve upon the PL fits. To test whether the improvement from PL to each curved model reflects genuine spectral curvature rather than the addition of free parameters, we performed an F-test for each nested model pair (PL vs. BPL, PL vs. LP, PL vs. BR+PL), using the raw $\chi^2$ and degrees of freedom from each fit. For \textit{Chandra}, BPL is preferred over PL at high significance ($p < 10^{-8}$) in all 16 observations. LP and BR+PL are similarly favored in most cases, with the exception of BR+PL for ObsID 3706, where the bremsstrahlung normalization converges to zero and BR+PL reduces to PL (p=1.00). For ObsIDs 3167 and 5173, BR+PL improves the fit (p $<10^{-4}$), but the bremsstrahlung temperature is not constrained. For {\it NuSTAR}, a curved model is statistically required (F-test $p < 0.01$) in only 4 of the 11 observations: ObsIDs 10002010001 (BPL, $p = 3\times10^{-3}$), 60002022012 (BPL, LP, and BR+PL, $p \leq 10^{-3}$), 11001611002 (BPL and BR+PL, $p \leq 3\times10^{-3}$), and 11001611004 (all three curved models, $p \leq 3\times10^{-3}$). In all four, the spectrum steepens with energy ($\Gamma_2 > \Gamma_1$, $\beta > 0$). In the remaining seven observations, the break energy is not constrained, the curvature parameter is consistent with zero, and the bremsstrahlung component is not required, so the spectra are adequately described by a single PL with $\Gamma \simeq 2.6-3.0$ (Table~\ref{tab:spectral_fit_NuSTAR}).
\\[6pt]
Since BPL, LP, and BR+PL are not nested with respect to one another, we used the Akaike Information Criterion (AIC) to select among them for each observation where more than one model is statistically justified over PL. For \textit{Chandra}, BPL has the lowest AIC in 13 of 16 observations,
BR+PL in 1 (ObsID 6926), and LP in 2 (ObsIDs 9705 and 14265), consistent with the lowest-$\chi^2_\nu$ model identified in Table~\ref{spectral_fit_chandra1}. For the four {\it NuSTAR} observations with significant curvature, BPL has the lowest AIC in two (ObsIDs 10002010001 and 60002022012), BR+PL in one (ObsID 11001611002), and LP in one (ObsID 11001611004); the differences in AIC between the curved models are, however, small ($\Delta{\rm AIC} \lesssim 5$), and the data do not clearly favor any one of them. We stress that BPL, LP, and BR+PL are all used here as phenomenological descriptions of the spectral curvature. In particular, the bremsstrahlung component of the BR+PL model is not interpreted as genuine thermal emission (Section~\ref{sec:spectral_analysis}). The best-fit parameters are listed in Tables~\ref{spectral_fit_chandra1} and~\ref{tab:spectral_fit_NuSTAR}. An example model of PL, BPL, LP, and BR+PL is shown in Figure~\ref{fig:spec_example} for ObsID 337. In addition, for \textit{Chandra} data, we present HR vs $\Gamma$ of PL in Figure~\ref{fig:PL_Gamma_vs_HR_Chandra} left; the correlation plots between the LP parameters of $\alpha$, $\beta$, $E_p$ and flux are shown in Figure~\ref{fig:PL_Gamma_vs_HR_Chandra} right. Similarly, the BR+PL parameters of $kT$ vs. flux and $kT$ vs. $\Gamma$ are shown in Figure~\ref{fig:Flux_vs_kT_Gamma_vs_kT} (left and right, respectively).
\begin{figure*}[!htbp]
    \centering
    \includegraphics[width=\textwidth]{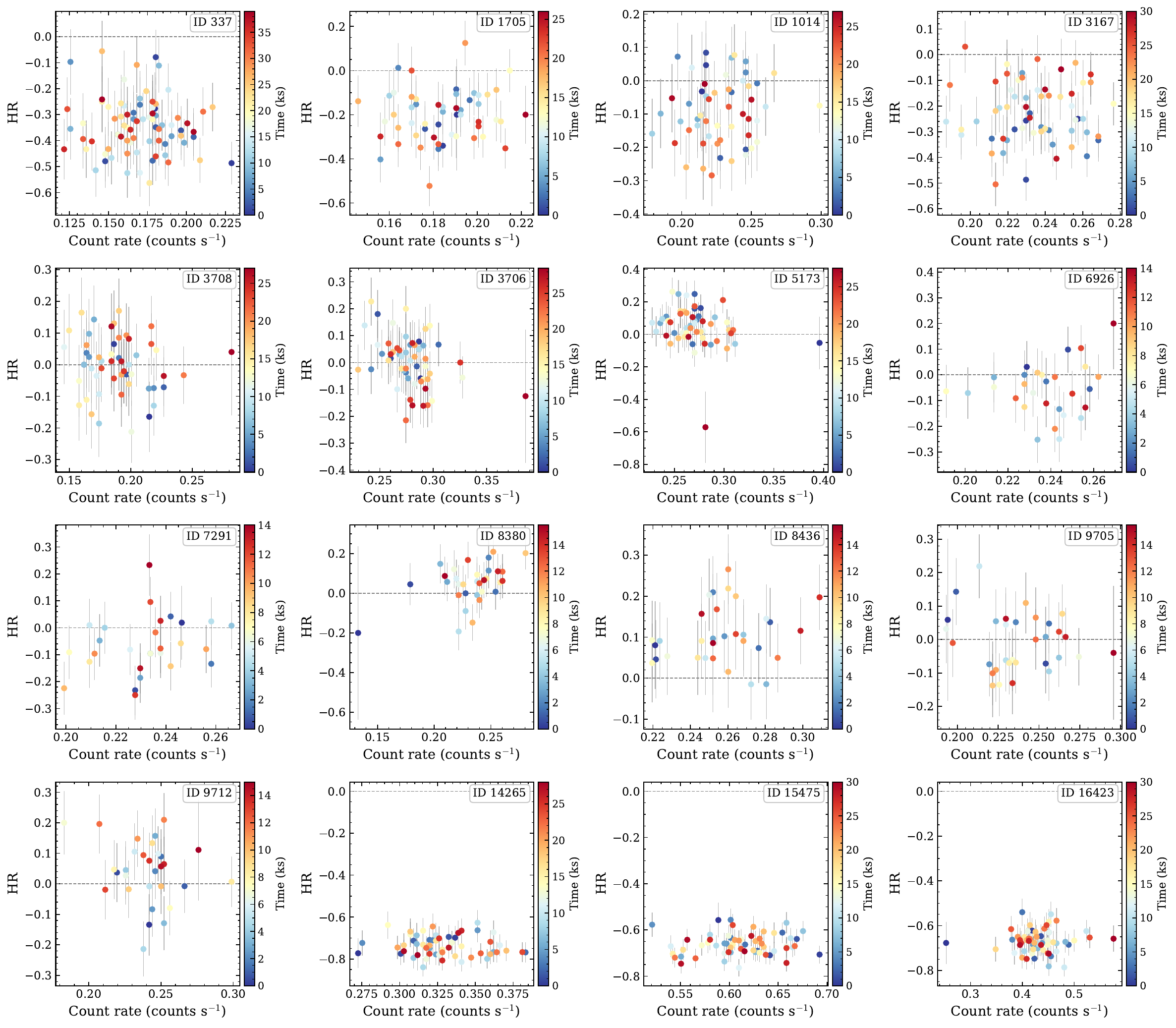}
    \caption{Intraday HR plotted against the total count rate (0.4$-$8.0~keV) for all 16 {\it Chandra} observations of PKS~2155--304. Each point represents a 500~s bin; colors indicate the elapsed time within the observation from blue (start) to red (end). The dashed horizontal line marks HR\,$=$\,0. Soft (0.4$-$2.0~keV) and hard (2.0$-$8.0~keV) band count rates were extracted separately using \texttt{dmextract} with PI channel filters.
    \label{fig:hr_per_obsid}}
\end{figure*}

\subsection{Hardness Ratio Analysis}
\label{sec:HR_maps}
\noindent
The hardness ratio (HR) analysis is widely used for different classes of astrophysical sources \citep[e.g.,][]{Balucinska05, Siemiginowska07, Nandra15, Haggard19, Thimmappa22}, and is particularly useful when the available source counts are low. To investigate spectral evolution on both long-term and intraday timescales, we performed an HR analysis. For the long-term study  (Figure~\ref{fig:Lightcurve_Chandra_NuSTAR}), we defined ${HR} = (H - S)/(H + S)$, where, for {\it Chandra}, $H$ is the flux in the 2.0$-$8.0~keV hard energy band, while $S$ corresponds to the flux in the 0.4$-$2.0~keV soft energy band. For {\it NuSTAR}, $H$ corresponds to the flux in the 10.0$-$79.0~keV hard energy band, while $S$ corresponds to the flux in the 3.0$-$10.0~keV energy band. For {\it Chandra}, HR shows a positive correlation with flux, consistent with the harder-when-brighter trend also seen in the photon index (Section \ref{sec:flux_conversion}).

For {\it NuSTAR}, we find no significant correlation between HR and flux (Spearman $\rho = -0.47$, $p = 0.14$ for all 11 epochs; $\rho = -0.37$, $p = 0.33$ for the nine 2012--2013 epochs alone). The weak tendency, to which the two high-flux 2023--2024 epochs with soft HR values contribute, is toward softer-when-brighter, consistent with the $\Gamma$--flux relation (Section~\ref{sec:flux_conversion}). Given the small {\it NuSTAR} sample ($N=11$), we do not consider this tendency significant, and discuss the implications in Section~\ref{sec:spectral_analysis}.
\\[6pt]
For the intraday analysis, we used the same definition of $HR$ (the $\pm1$ bound), which allows direct comparison among observations with different flux levels. To examine intraday spectral evolution within individual observations, we extracted soft band (0.4$-$2.0~keV) and hard band (2.0$-$8.0~keV) light curves for all 16 {\it Chandra} observations and computed the bin-by-bin $HR$ at a 500~s time resolution. The resulting HR evolution is shown in Figure~\ref{fig:hr_per_obsid},  where each data point is color-coded according to the elapsed time within the corresponding observation.
\\[6pt]
Most observations exhibit HR values in the range of $-0.6 \lesssim \mathrm{HR} \lesssim 0.2$, consistent with soft-dominated X-ray emission ($\Gamma \sim 2.1$-$2.7$; Table~\ref{spectral_fit_chandra1}). In the majority of cases, HR remains likely constant over the observed count-rate range, indicating predominantly \textit{achromatic} intraday variability; that is, the spectral shape does not change significantly despite changes in flux. This behavior is particularly evident in ObsIDs 3706, 7291, 8436, 9705, and 9712, for which the observed HR scatter is consistent with measurement uncertainties alone. ObsIDs 14265, 15475, and 16423 exhibit the most negative HR values ($\mathrm{HR} \sim -0.7$ to $-0.4$), consistent with their high-flux states and relatively steep photon indices ($\Gamma \sim 2.5$--$2.7$; Table~\ref{spectral_fit_chandra1}). These observations correspond to the brightest {\it Chandra} epochs in our sample (fluxes of $20$--$45 \times 10^{-12}$~erg~cm$^{-2}$~s$^{-1}$), suggesting that the X-ray spectrum becomes softer during the high-flux state. The most notable spectral evolution is observed in ObsID~8380, where the HR increases from ${\sim}-0.6$ at early times, represented by the blue points in Figure \ref{fig:hr_per_obsid}, to nearly zero at later times, indicating spectral hardening during the observation. This observation also shows the steepest measured {\it Chandra} PSD slope ($\beta = 0.69 \pm 0.12$; Section~\ref{sec:psd}), consistent with efficient particle acceleration extending the synchrotron emission to higher X-ray energies during this epoch.
\\[6pt]
Overall, significant intraday spectral variability, characterized by a clear HR$-$flux trend, is detected in only 1$-$2 out of the 16 {\it Chandra} observations, while the majority exhibit predominantly achromatic flux variations. This result is consistent with the findings of \citet{Bhatta25} for PKS~2155--304 using {\it XMM-Newton} observations, where clear HR variability was reported in only 2 out of 22 observations. This supports the interpretation that the intraday X-ray variability of PKS~2155--304 is driven mainly by changes in the normalization of the emission component, rather than by substantial changes in the spectral shape.

\begin{figure}[!htbp]
	\centering 
    {\includegraphics[width=\columnwidth]{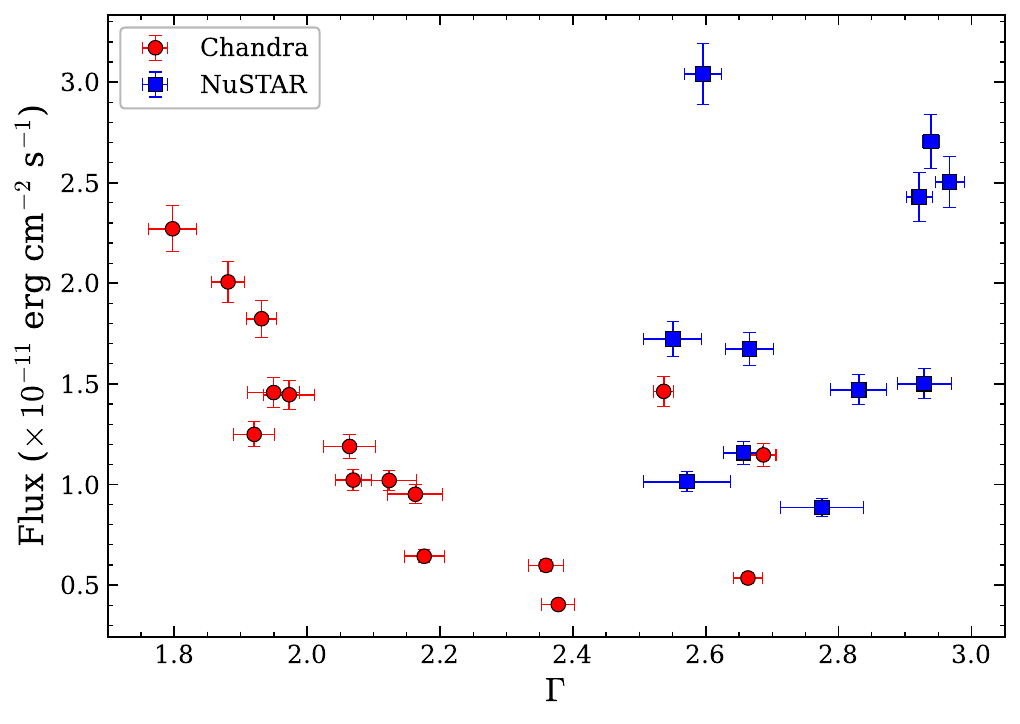}
   	\caption{Flux versus photon index ($\Gamma$) for {\it Chandra} (red) and {\it NuSTAR} (blue) using the PL model for PKS~2155$-$304.}
   \label{fig:PL_Flux_vs_Gamma_Chandra+NuSTAR}}
\end{figure}

\subsection{Flux versus \texorpdfstring{$\Gamma$}{Gamma} of PKS~2155\texorpdfstring{$-$}{-}304}
\label{sec:flux_conversion}
\noindent To compare the flux measured in the 0.4$-$8.0~keV band with that in the 3.0$-$79.0~keV band, we converted the measured fluxes assuming a simple power-law spectrum. For a photon index $\Gamma$ obtained from spectral fitting, the energy flux in a given energy band is expressed as
\begin{equation}
F(E) = \int_{E_1}^{E_2} K\, E^{-\Gamma}\, dE,
\end{equation}
\noindent where, $K$ is the normalization. For $\Gamma = 2$, the integral takes a logarithmic form, while for other values of $\Gamma$ it can be evaluated analytically. The conversion factor was computed as the ratio of the integrated fluxes in the 3.0--79.0~keV to  0.4--8.0~keV bands. The corresponding flux uncertainties were scaled by the same factor, thereby enabling a consistent comparison between instruments and energy ranges.
\\[6pt]
The flux versus $\Gamma$ plot for {\it Chandra} and {\it NuSTAR} is shown in Figure~\ref{fig:PL_Flux_vs_Gamma_Chandra+NuSTAR}. For {\it Chandra}, the flux decreases as $\Gamma$ increases from $\sim$1.8 to $\sim$2.6, indicating a harder-when-brighter trend in which higher flux states are associated with harder spectra. In contrast, for {\it NuSTAR}, the flux increases as $\Gamma$ rises from $\sim$2.6 to $\sim$3.0, with a weak, non-significant tendency toward a softer-when-brighter trend and suggesting a softer-when-brighter trend in the hard X-ray band. This apparent reversal may reflect the fact that the two instruments probe different spectral regimes. In the {\it Chandra} band, the X-ray emission is the rising part of the synchrotron component, where enhanced particle acceleration produces harder spectra at higher flux levels. The {\it NuSTAR} band, on the other hand, probes emission closer to or beyond the synchrotron peak, where increased particle injection may raise the flux while simultaneously softening the spectrum, possibly due to a shift in the peak energy. 

\subsection{Joint Spectral Fitting with {\it Chandra} and {\it NuSTAR}}
\label{sec:joint_fit}
\begin{figure*}[!htbp]
	\centering 
    \includegraphics[width=0.69\columnwidth]{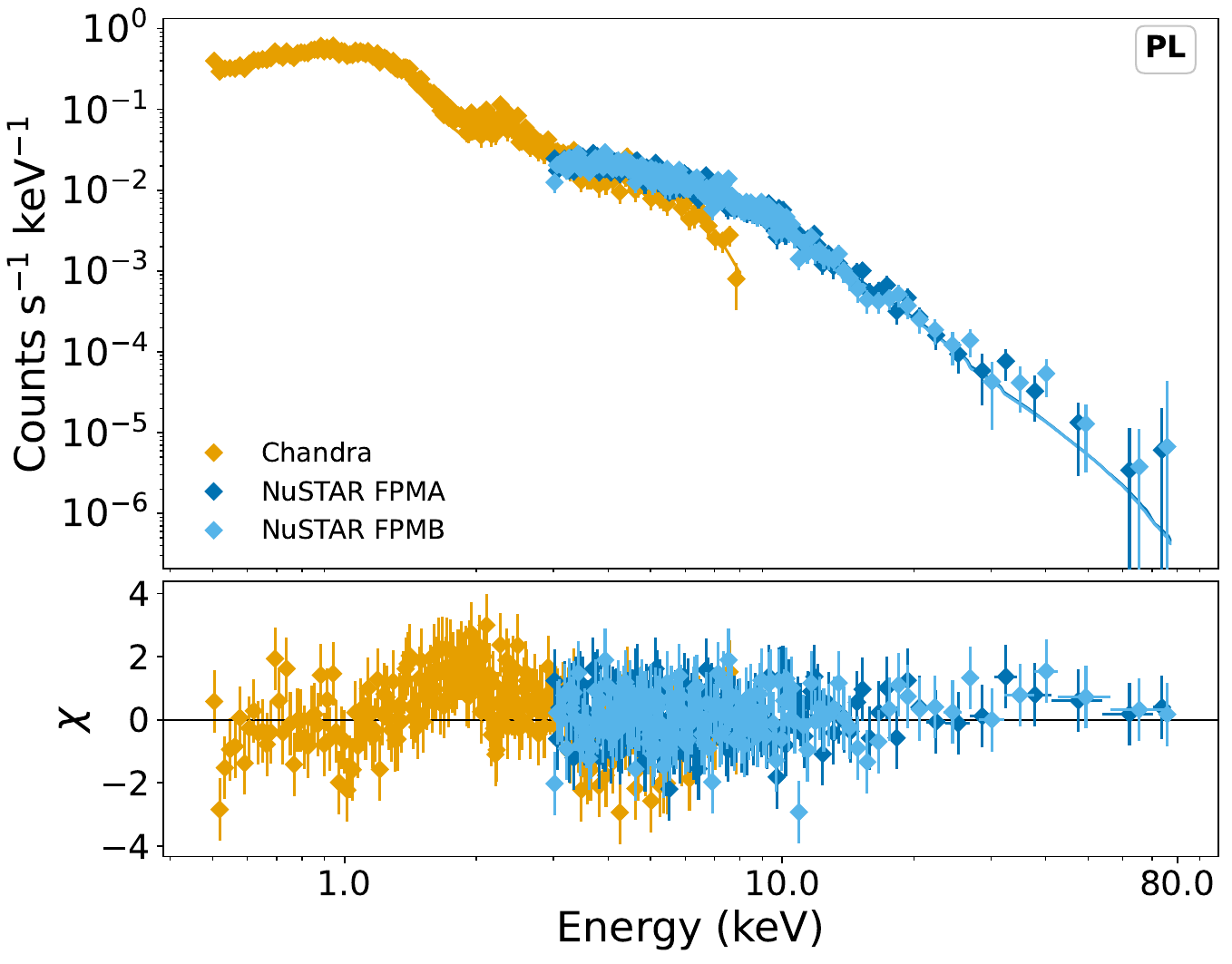}
    \includegraphics[width=0.69\columnwidth]{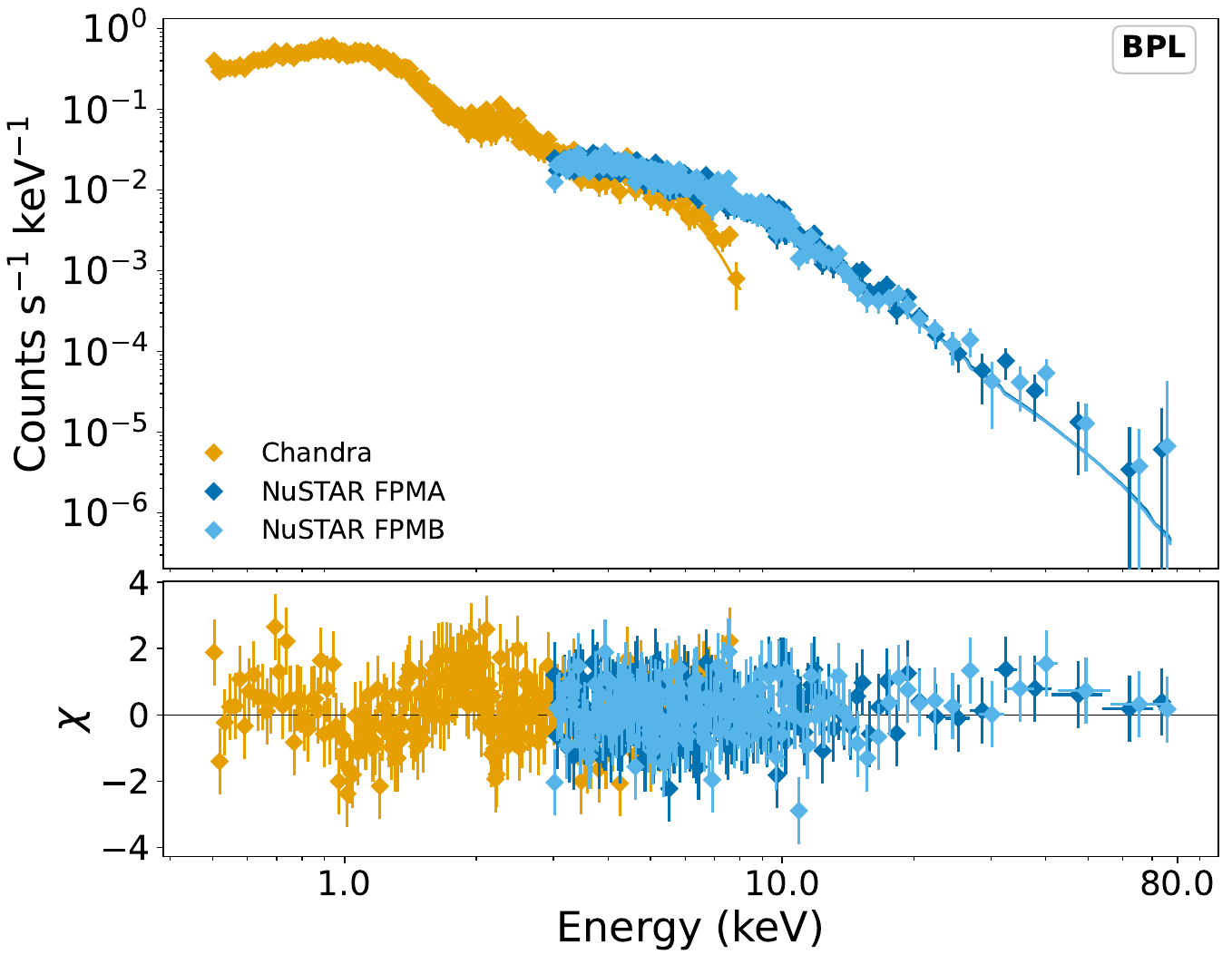}
    \includegraphics[width=0.69\columnwidth]{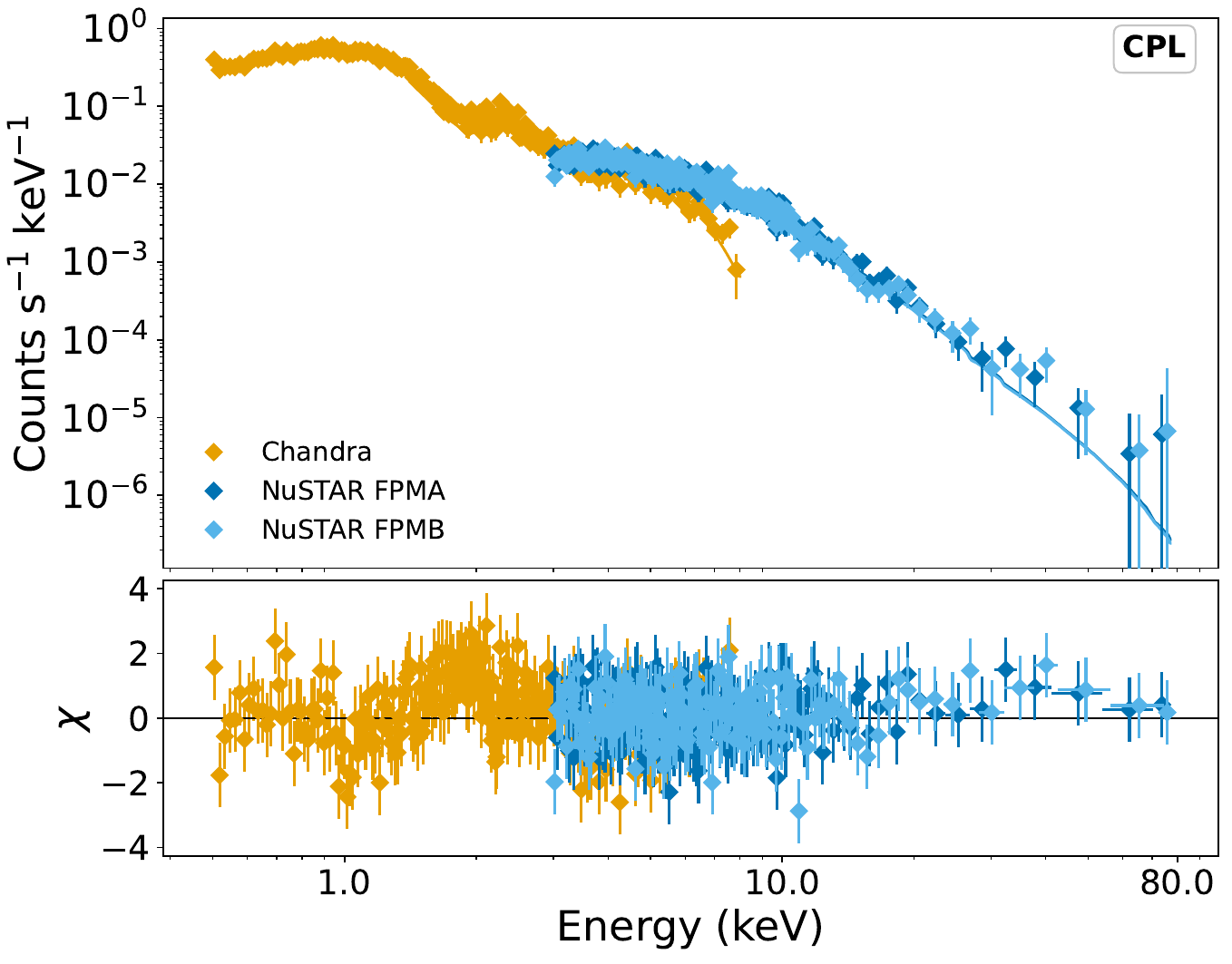}
    \includegraphics[width=0.69\columnwidth]{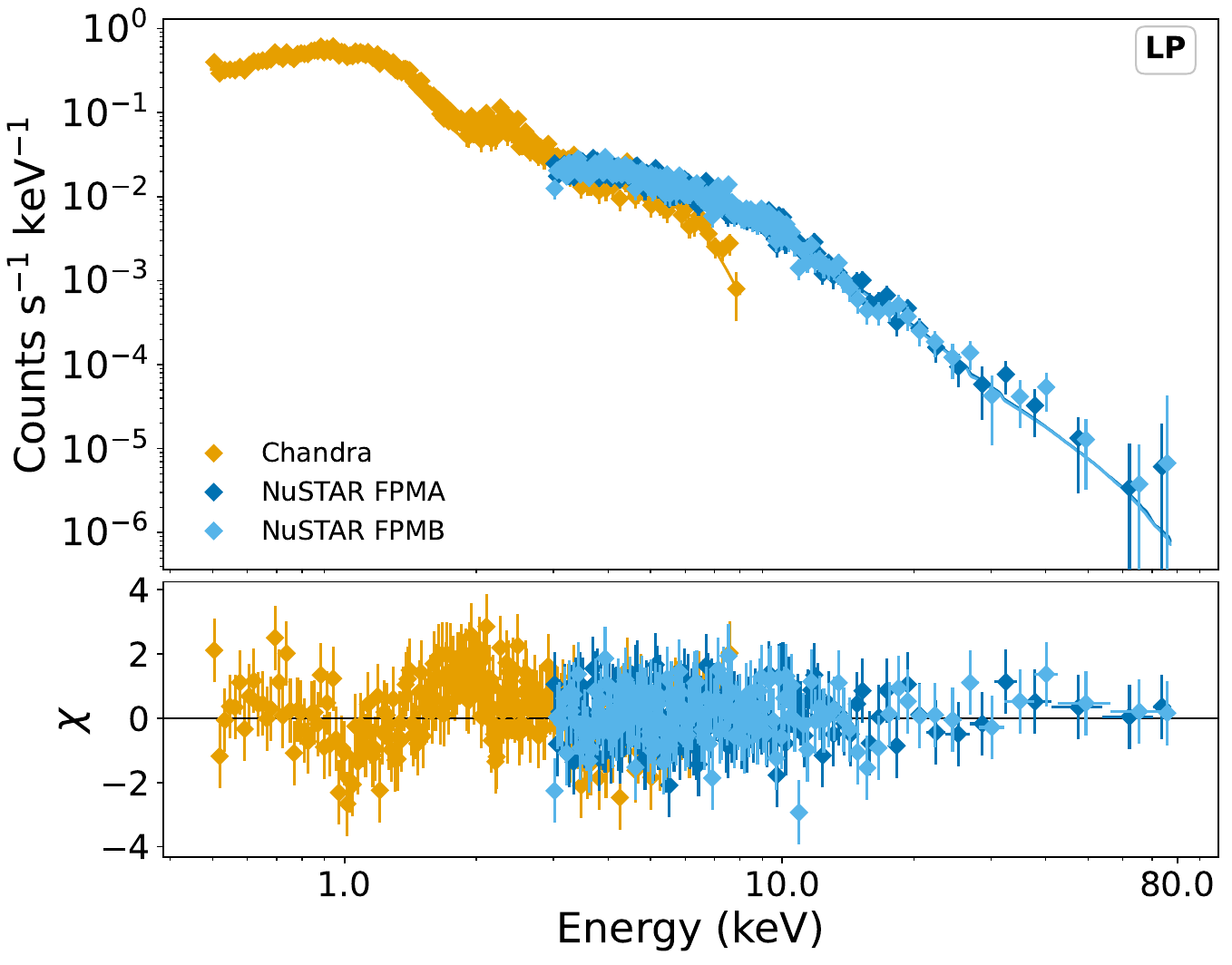}
    \includegraphics[width=0.69\columnwidth]{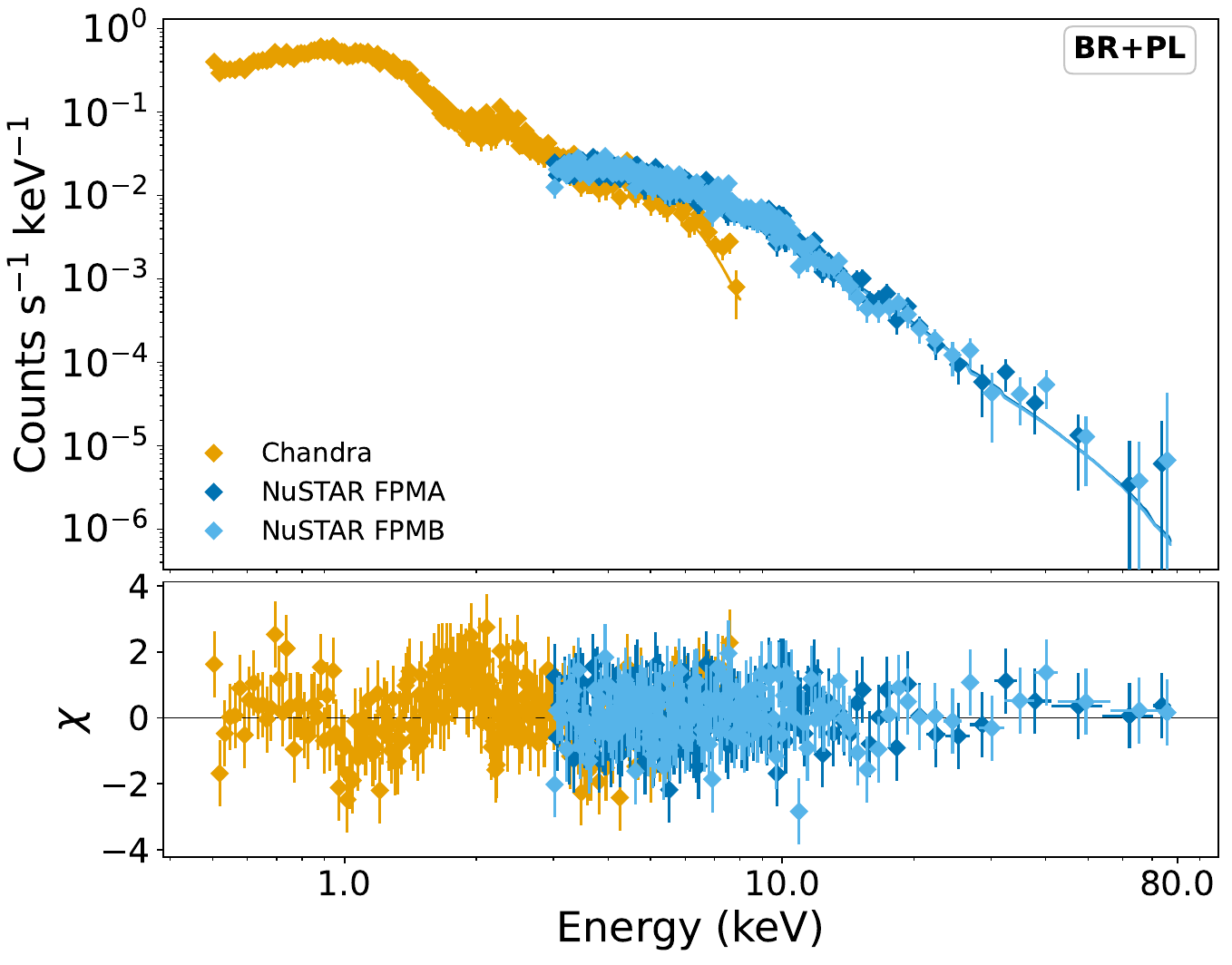}   
	\caption{Joint spectral fits of PKS~2155$-$304 from simultaneous {\it Chandra} (ObsID 15475; 0.4$-$8.0~keV) and {\it NuSTAR} (ObsID 60002022002; FPMA and FPMB, 3.0$-$79.0~keV) observations. Each panel shows the spectrum and the corresponding residuals ($\chi$ = (data $-$ model)/error) for the five models tested: PL, BPL, CPL, LP, and BR+PL (left to right, top to bottom). The Galactic column density is fixed at $N_{\rm H,Gal} = 1.28\times10^{20}$~cm$^{-2}$ and the {\it NuSTAR} background is subtracted (Section~\ref{sec:joint_fit}).}
   \label{fig:j}
\end{figure*}

\noindent In Section~\ref{sec:spectral_fitting}, we applied the PL, BPL, LP, and BR+PL models separately to the {\it Chandra} and {\it NuSTAR} spectra. To investigate the broadband spectrum, we further performed a joint fit of the simultaneous {\it Chandra} (ObsID 15475; 0.4$-$8.0~keV) and {\it NuSTAR} (ObsID 60002022002; FPMA and FPMB, 3.0$-$79.0~keV) spectra with five models: PL, BPL, CPL, LP, and BR+PL. The Galactic column density was fixed at $N_{\rm H,Gal} = 1.28 \times 10^{20}$~cm$^{-2}$ \citep{HI4PI16}, the {\it NuSTAR} background was subtracted, and the spectral parameters of FPMA and FPMB were tied, with a free cross-normalization constant for FPMB ($C_{\rm FPMB} = 1.02 \pm 0.03$). The {\it Chandra} and {\it NuSTAR} spectra were fitted with independent spectral parameters. The results are summarized in Table~\ref{tab:jointfits} and shown in Figure~\ref{fig:j}.
\\[6pt]
A simple PL yields $\Gamma = 2.56 \pm 0.02$ for {\it Chandra} and $\Gamma = 2.73 \pm 0.03$ for {\it NuSTAR}, with $\chi^2_\nu = 0.96$ (494/517), but leaves systematic residuals in the {\it Chandra} band. All curved models improve the fit significantly (F-test probabilities $p < 10^{-12}$ relative to PL). The BPL model gives $\chi^2_\nu = 0.81$ ($\Delta\chi^2 = 78$ for four additional parameters), with a break at $E_{\rm b} = 2.25^{+0.32}_{-0.35}$~keV in the {\it Chandra} spectrum, where the photon index steepens from $\Gamma_1 = 2.37^{+0.04}_{-0.09}$ to $\Gamma_2 = 3.02^{+0.20}_{-0.15}$. Excluding the 1.8$-$2.4~keV range, which contains the instrumental Si K edge of the ACIS detector and the Ir M edges of the \textit{Chandra} mirrors, does not change the break ($E_{\rm b} = 2.37 \pm 0.39$~keV, $\Gamma_2 = 3.02 \pm 0.36$), and BPL remains strongly preferred over PL ($\Delta\chi^2 = 51$ for four additional parameters, F-test $p < 10^{-12}$). The CPL and LP models describe the same curvature through a cutoff at $E_{\rm cut} = 5.3^{+0.9}_{-0.7}$~keV and a curvature parameter $\beta = 0.36^{+0.06}_{-0.05}$, respectively. The BR+PL model also provides a good fit ($\chi^2_\nu = 0.83$), but, as for the individual spectra, we regard it only as a phenomenological description of the curvature (Section~\ref{sec:spectral_analysis}).
\\[6pt]
In contrast, the {\it NuSTAR} spectrum shows no significant curvature: the BPL break is not constrained and the BPL reduces to a PL, the CPL cutoff energy runs to the upper limit of the allowed range, and the LP curvature parameter is consistent with zero ($|\beta| < 0.46$). The {\it NuSTAR} spectrum is therefore well described by a single power law with $\Gamma \simeq 2.7$, comparable to the {\it Chandra} photon index above the break. The broadband 0.4$-$79~keV spectrum is thus described by a curved soft X-ray spectrum that steepens above $\sim$2~keV, followed by a steep power law extending to the hard X-ray band. An extended log-parabola parameterized by the peak energy is equivalent to the LP model and is not considered further, since the peak energy is undefined for the {\it NuSTAR} spectrum, whose curvature is consistent with zero.

\begin{deluxetable*}{lccccccc}[!htbp]
%\rotate
\tiny
%\footnotesize
\tablecaption{Best-fitting parameters of PKS~2155$-$304 from joint \textit{Chandra}  + \textit{NuSTAR} fits observed simultaneously. The Galactic column density is fixed at $N_{\rm H,Gal} = 1.28\times10^{20}$~cm$^{-2}$.} \label{tab:jointfits}
\tablehead{
 & \multicolumn{3}{c}{\textit{Chandra}} & \multicolumn{3}{c}{\textit{NuSTAR}} & \\
\cline{2-7}
\colhead{Model} &
\colhead{$\alpha \,|\, \Gamma_{1}$} &
\colhead{$\beta \,|\, \Gamma_{2}$} &
\colhead{$E_{\rm b}\,|\,E_{\rm cut}\,|\,kT$} &
\colhead{$\alpha \,|\, \Gamma_{1}$} &
\colhead{$\beta \,|\, \Gamma_{2}$} &
\colhead{$E_{\rm b}\,|\,E_{\rm cut}\,|\,kT$} &
\colhead{$\chi^2_\nu$ ($\chi^2$/dof)} \\
\colhead{} & \colhead{} & \colhead{} & \colhead{(keV)} &
\colhead{} & \colhead{} & \colhead{(keV)} & \colhead{}}
\startdata
PL    & $2.56 \pm 0.02$ & -- & -- & $2.73 \pm 0.03$ & -- & -- & 0.96 (494/517) \\
BPL   & $2.37^{+0.04}_{-0.09}$ & $3.02^{+0.20}_{-0.15}$ & $2.25^{+0.32}_{-0.35}$ & \nodata$^{\rm a}$ & $2.76_{-0.09}^{\,\rm b}$ & \nodata$^{\rm a}$ & 0.81 (416/513) \\
CPL   & $2.20 \pm 0.05$ & -- & $5.3^{+0.9}_{-0.7}$ & $2.71 \pm 0.03$ & -- & \nodata$^{\rm c}$ & 0.85 (440/515) \\
LP    & $2.40 \pm 0.03$ & $0.36^{+0.06}_{-0.05}$ & -- & $2.66 \pm 0.05$ & $|\beta| < 0.46$ & -- & 0.86 (443/515) \\
BR+PL & $2.70^{+0.19}_{-0.10}$ & -- & $2.14 \pm 0.29$ & $2.48^{+0.17}_{-0.21}$ & -- & $2.19^{+0.48}_{-0.81}$ & 0.83 (427/513) \\
\enddata
\tablecomments{Uncertainties are $1\sigma$; upper limits are $2\sigma$. $^{\rm a}$Break not constrained (it falls at or below the lower edge of the {\it NuSTAR} band); the {\it NuSTAR} BPL reduces to a PL. $^{\rm b}$Upper confidence bound not constrained. $^{\rm c}$Cutoff energy not constrained (pegged at the upper limit of 500~keV).}
\end{deluxetable*}

\section{Discussion} 
\label{sec:discussion}
\noindent We performed a detailed investigation of the temporal and spectral behavior of PKS~2155$-$304 using {\it Chandra} and {\it NuSTAR} observations spanning 25 years. In the following subsections, we discuss the implications of our results for the physical mechanisms driving the observed X-ray variability, the nature of the spectral components, and the constraints they place on the relativistic jet.
\subsection{X-ray Flux Variability}
\label{sec:X-ray_flux_variability}
\noindent The short variability timescales measured (Section~\ref{sec:variability}) provide direct constraints on the physical properties of the X-ray emitting region. Assuming that the hard X-ray emission from PKS~2155$-$304 is dominated by synchrotron radiation, the relevant physical parameters can be estimated using synchrotron cooling arguments. The synchrotron cooling timescale ($t_{\rm cool}(\gamma$), s) in the observer's frame \citep{Rybicki79}, for an electron with energy $E = \gamma m_e c^2$, is given by
\begin{equation}
    t_{\rm cool}(\gamma) \simeq 7.74 \times 10^{8} 
    \frac{(1+z)}{\delta}\, B^{-2}\, \gamma^{-1},
\end{equation}
where, $\delta$ is the Doppler factor, $B$ is the magnetic field strength (in G), $z$ is the cosmological redshift and $\gamma$ is the Lorentz factor. The characteristic synchrotron frequency ($\nu$, Hz) is given by \citep{Tavecchio98}
\begin{equation}
    \nu \simeq 4.2 \times 10^{6} \frac{\delta}{1+z}\, 
    B\gamma^{2}.
\end{equation}

\noindent We find X-ray variability timescales (measured with S/N $\geq 2$) in the range of 0.44$-$2.17~ks for {\it NuSTAR} band and 1.1$-$2.5~ks for {\it Chandra} band. Among the {\it NuSTAR} measurements, the shortest variability timescale measured with S/N $> 3$ is $\tau_{\rm var} = 1.15 \pm 0.26$~ks, measured from ObsID 60002022004; the shorter value of $0.44 \pm 0.17$~ks found for ObsID 10002010001 is significant at only $2.6\sigma$. We use $\tau_{\rm var} = 1.15$~ks for the physical parameter estimates discussed below.
\\[6pt]
By requiring the observed minimum variability timescale to be greater than or equal to the synchrotron cooling timescale, we obtain the following constraint for PKS~2155$-$304: 
($\tau_{\rm var} = 1.15$ ks, $z = 0.116$):
\begin{equation}
    B \geq 0.60\; \delta^{-1/3}\, \nu_{19}^{-1/3} \; {\rm G},
\end{equation}
where, $\nu_{19}$ is the synchrotron frequency in units of $10^{19}$~Hz. For a typical Doppler factor of $\delta = 25$, this gives $B \gtrsim 0.20\,\nu_{19}^{-1/3}$~G, consistent with values inferred from SED modeling. Correspondingly, the Lorentz factor of the electrons is constrained to
\begin{equation}
    \gamma \leq 7.2 \times 10^{5}\, \nu_{19}^{2/3}.
\end{equation}
\noindent The size of the X-ray emitting region ($R$, cm) is estimated from the variability timescale as
\begin{equation}
    R \leq \frac{c\,\tau_{\rm var}\,\delta}{1+z} 
    \approx 7.7 \times 10^{14}~{\rm cm},
    \label{eq:emitting_region}
\end{equation}
Here, we used $\tau_{\rm var} = 1.15$~ks and a Doppler factor of $\delta = 25$. 
In the soft X-ray band, the shortest timescale is $1.09 \pm 0.39$~ks ({\it Chandra} ObsID 337, S/N $\approx 2.8$), which would give a similar size estimate, $R \lesssim 7.3\times10^{14}$~cm. These estimates suggest that the observed X-ray variability of PKS~2155$-$304 is consistent with synchrotron emission from a compact region within the relativistic jet, where rapid particle acceleration and radiative cooling occur.
\\[6pt]
We can compare our intraday variability results of PKS~2155$-$304 with {\it XMM-Newton} observations obtained between 2000 and 2014 \citep{Bhatta25}. The {\it XMM-Newton} results show that the $F_{\rm var}$ ranges from $1.32\% \pm 0.35\%$ to $21.06\% \pm 0.14\%$, with a mean value of $\sim$6.99\%, and minimum variability timescales in the range of 0.36--5.03~ks. Our {\it Chandra} results, with $F_{\rm var} \lesssim 7.5$\% and $\tau_{\rm var} = 1.1$-$2.5$~ks, are broadly consistent with these {\it XMM-Newton} measurements, despite differences in energy band and source flux state. In contrast, the {\it NuSTAR} observations show systematically higher fractional variability, with $F_{\rm var} = 6$-$24$\%, and variability timescales extending to shorter values, $\tau_{\rm var} = 0.44$-$2.17$~ks. This suggests that the hard X-ray-emitting region is more compact and varies more rapidly than the soft X-ray-emitting region. Such energy-dependent variability behavior is consistent with a synchrotron interpretation, in which higher-energy electrons cool more rapidly and occupy a more compact region within the jet \citep{Bhatta18, Zhang21}.

\subsection{$RMS-Flux$ Relationship}
\label{sec:rms_flux}
\begin{figure*}[!ht]
    \centering
    \includegraphics[width=\textwidth]{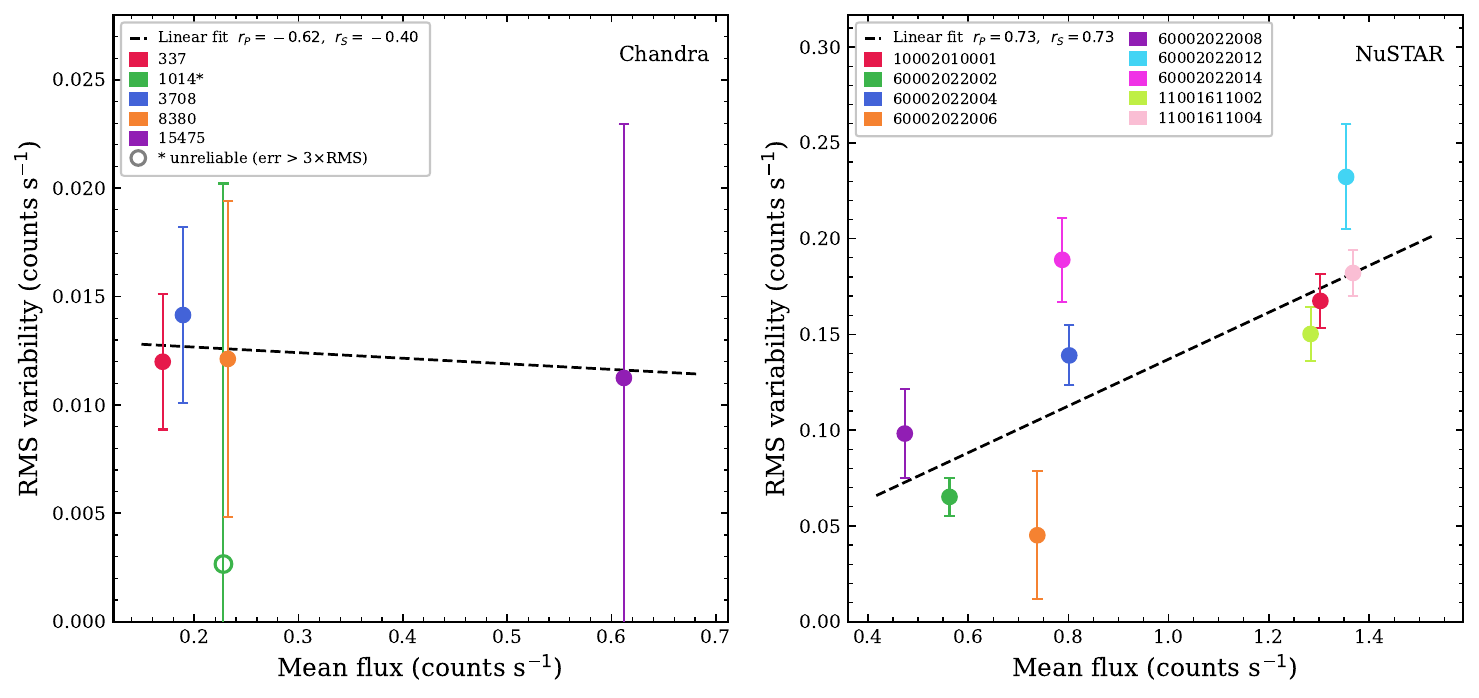}
\caption{$RMS$ variability versus mean flux for PKS~2155--304 from {\it Chandra} (\textit{left}) and {\it NuSTAR} (\textit{right}) using 500~s binned light curves. Each observation is shown in a distinct color. The dashed line shows a weighted linear fit to reliable detections (filled symbols). Open symbols indicate observations with poorly constrained rms (uncertainty $> 3\times$ $RMS$ value) and are excluded from the fit. The Pearson and Spearman correlation coefficients are indicated in the legend.}
\label{fig:rms_flux}
\end{figure*}

\noindent 
Furthermore, to characterize the X-ray variability of PKS~2155$-$304, we examined the relationship between the $RMS$ variability and the mean flux for each observation, following the approach of \citet{Uttley01, Bhattacharyya20} and \citet{Bhatta25}. The $RMS$ variability was calculated as the square root of the excess variance, $\sigma_{\rm RMS} = \sqrt{S^2 - \bar{\sigma}_{\rm err}^2}$, with observations yielding negative excess variance excluded from the analysis (see Tables~\ref{tab:chandra_var} and~\ref{tab:nustar_var}). The resulting $RMS-flux$ relationship is shown in Figure~\ref{fig:rms_flux}.
\\[6pt]
For {\it NuSTAR} observations, we find a positive correlation, significant at the $\sim$2$\sigma$ level, between the $\mathrm{RMS}$ variability and the mean flux, with Pearson and Spearman correlation coefficients of $r_P = 0.73$ ($p = 0.025$) and $r_S = 0.73$ ($p = 0.025$), respectively. This is consistent with a linear $RMS-flux$ relation. Such a relation is commonly interpreted as evidence for multiplicative, non-linear variability processes, in which fluctuations on different timescales are coupled \citep{Uttley05}. In contrast, only four {\it Chandra} observations have a measurable excess variance with a reliable rms (ObsIDs 337, 3708, 8380, and 15475), and these show no significant correlation ($r_P = -0.62$, $p = 0.38$; $r_S = -0.40$, $p = 0.60$). With so few observations, the rms--flux relation in the soft X-ray band cannot be constrained, consistent with the weak intraday variability detected in only 2 out of 16 {\it Chandra} observations (Section~\ref{sec:variability}).

\subsection{Spectral Analysis}
\label{sec:spectral_analysis}

\noindent In blazars, the X-ray emission is dominated by the relativistic jet. Any unbeamed thermal component, for example from a hot corona or from diffuse hot gas in the host galaxy, is expected to be negligible. For the Doppler factor used in Section~\ref{sec:X-ray_flux_variability} ($\delta = 25$), the jet luminosity is enhanced by a factor of $\sim\delta^{4} \approx 4\times10^{5}$ compared to isotropic emission. In HSP blazars such as PKS~2155$-$304, the X-ray band covers the high-energy part of the synchrotron component. Therefore, the X-ray spectra are non-thermal and are usually well described by a power-law model with mild curvature \citep{Massaro04, Aharonian05}. Thus, X-ray spectral studies of blazars provide a direct probe of the physical conditions within the relativistic jet and the efficiency of particle acceleration processes. The combination of the high spatial resolution of {\it Chandra} and the broad energy coverage of {\it NuSTAR} makes these observations particularly valuable for constraining the emission properties of PKS~2155$-$304.
\\[6pt]
We used PL, BPL, LP, and BR+PL models. Among these, the simple PL is formally rejected only for ObsIDs 14265 and 15475. An F-test confirms that at least one curved model is statistically preferred over PL in all 16 {\it Chandra} observations, but in only 4 of the 11 {\it NuSTAR} observations (ObsIDs 10002010001, 60002022012, 11001611002, and 11001611004); the remaining seven {\it NuSTAR} spectra are adequately described by a single PL (Section~\ref{sec:spectral_fitting}). Based on AIC, BPL provides the preferred description for most of the {\it Chandra} spectra (13 out of 16 observations; Table~\ref{spectral_fit_chandra1})%
, while the four curved {\it NuSTAR} spectra are described comparably well by BPL, LP, and BR+PL (Table~\ref{tab:spectral_fit_NuSTAR}).
Where curvature is detected in the {\it NuSTAR} band, the spectrum always steepens with energy, as expected for the high-energy tail of the synchrotron component. We use all three curved models only as phenomenological descriptions of the spectral curvature; as discussed below, the bremsstrahlung component of the BR+PL model does not represent physical thermal emission.
\\[6pt]
The LP model also provided satisfactory fits for several observations, particularly for the {\it Chandra} spectra (Table~\ref{spectral_fit_chandra1}). The spectral curvature captured by the LP model can arise naturally from stochastic particle acceleration, in which the acceleration probability depends on particle energy \citep{Massaro04, Tramacere07}. This interpretation is consistent with {\it XMM-Newton} observations of PKS~2155$-$304 \citep{Bhatta25}. In the joint fit of the simultaneous 2013 observations, however, the curvature is confined to the {\it Chandra} band, while the {\it NuSTAR} spectrum is consistent with a single power law (Section~\ref{sec:joint_fit}).
\\[6pt]
A notable outcome of our analysis is that flux and spectral variability do not always evolve in tandem. Although strong flux variability is observed throughout the monitoring period, spectral variations inferred from the hardness ratio occur less frequently; in many cases, the source exhibits strong flux changes with only weak or negligible spectral evolution. The {\it Chandra} band shows a robust harder-when-brighter trend across both the hardness ratio and photon index. In the \textit{NuSTAR} band, both the hardness ratio and the photon index show a weak, non-significant softer-when-brighter trend (HR--flux: Spearman $\rho = -0.47$, $p = 0.14$; spectral index--flux: $\rho = +0.47$, $p = 0.14$; Section~\ref{sec:HR_maps}). We do not consider the {\it NuSTAR} sample large enough to establish the sign of the flux--hardness relation with confidence, and flag this as a target for future monitoring with a larger sample.
\\[6pt]
The BR+PL model gives fits comparable to BPL and LP for several spectra. For the {\it Chandra} spectra, the temperatures span $kT \sim 0.56-1.7$~keV (Table~\ref{spectral_fit_chandra1}) and show no correlation with flux (Figure~\ref{fig:Flux_vs_kT_Gamma_vs_kT}, left). The strong anticorrelation between $kT$ and $\Gamma$ (Figure~\ref{fig:Flux_vs_kT_Gamma_vs_kT}, right) reflects the degeneracy between the two components in fitting the spectral curvature, rather than a physical relation. In three observations (ObsIDs 3167, 3706, and 5173) the temperature is unconstrained or only very poorly constrained ($kT = 0.21-0.80$~keV and $0.08-1.0$~keV at $1\sigma$ for ObsIDs 3167 and 5173, respectively, while for ObsID 3706 the bremsstrahlung normalization converges to zero). For ObsIDs 10002010001, 60002022004, 60002022006, 60002022010, and 60002022014, the bremsstrahlung model is not required and its temperature cannot be constrained (Table~\ref{tab:spectral_fit_NuSTAR}); in the observations where it is constrained, the bremsstrahlung term merely reproduces the curvature also captured by the BPL and LP models. We therefore treat BR$+$PL purely as a phenomenological parameterization of the spectral curvature, equivalent in this respect to the BPL and LP models, and draw no physical conclusions from its parameters. An alternative non-thermal interpretation is provided by \citet{Jagan21}, who showed that during certain epochs the soft X-ray spectrum of PKS~2155$-$304 observed with {\it XMM-Newton} becomes convex, hardening above $\sim$4~keV. They attributed this behavior to the emergence of sub-dominant inverse-Compton scattering of synchrotron photons at the high-energy end of the synchrotron spectrum. In this scenario, the curvature captured by the BR+PL model in the {\it Chandra} data may instead reflect the onset of a low-energy inverse-Compton scattering contribution within the 0.4$-$8.0~keV band, thereby mimicking a thermal-like soft excess without requiring genuine thermal emission. Distinguishing between these possibilities requires simultaneous broadband SED modeling extending to GeV energies, which is beyond the scope of the present work. 
\\[6pt]
A similar interpretation was proposed by \citet{Madejski16}, who reported a log-parabolic spectrum with a hard power-law tail extending to $\sim$60~keV in a low-flux {\it NuSTAR}+{\it XMM-Newton} observation of this source and interpreted the tail as the onset of an inverse-Compton component. This picture was further developed by \citet{Jagan21}, who used {\it XMM-Newton} data alone to show that, during convex-spectrum epochs, a double power-law decomposition representing synchrotron and inverse-Compton scattering of synchrotron photon components can reproduce the observed curvature without invoking thermal emission. Our BR+PL results are therefore best understood as a statistically convenient parameterization of the observed spectral curvature, which in the jet-dominated X-ray emission of PKS~2155$-$304 most likely arises from the gradual steepening of the synchrotron tail and/or from a sub-dominant inverse-Compton contribution. 
\\[6pt]
The joint spectral fitting with {\it Chandra} and {\it NuSTAR} (Section~\ref{sec:joint_fit}, Table~\ref{tab:jointfits}) shows that a single power law is statistically disfavored (F-test p $< 10^{-12}$) for the broadband 0.4$-$79~keV spectrum, but that the curvature is confined to the soft X-ray band: the {\it Chandra} spectrum steepens above a break at $E_{\rm b} \simeq 2.3$~keV, while the {\it NuSTAR} spectrum is consistent with a single power law of $\Gamma \simeq 2.7$, comparable to the {\it Chandra} index above the break. BPL ($\chi^2_\nu = 0.81$) and BR+PL ($\chi^2_\nu = 0.83$) describe the spectrum comparably well, and the LP, CPL, and BR+PL models describe the same soft X-ray curvature. We find no significant hard X-ray tail or flattening of the kind reported by \citet{Madejski16} for the low-flux state of April 2013, which corresponds to our {\it NuSTAR} ObsID 60002022002. In our joint fit with the simultaneous \textit{Chandra} ObsID 15475 (Section~\ref{sec:joint_fit}), the {\it NuSTAR} spectrum of this observation shows no significant curvature and is well described by a single power law with $\Gamma \simeq 2.7$. The hard tail reported by \citet{Madejski16} ($\Gamma = 2.2 \pm 0.4$) was identified in a joint {\it XMM-Newton}+{\it NuSTAR} fit with a log-parabola plus power-law model; such a weak component appears only at the highest energies, where the background dominates, and is difficult to constrain with our data and models.
%\\[6pt]
\section{Conclusions}
\label{sec:conclusion}
We have presented a comprehensive study of the X-ray spectral and timing variability of the TeV blazar PKS~2155$-$304 using 16 {\it Chandra} observations in the 0.4$-$8.0~keV band and 11 {\it NuSTAR} observations in the 3.0$-$79.0~keV band, spanning 25 years from 1999 to 2024. The main conclusions of this work are summarized below:
\begin{enumerate}

\item The source shows significant flux variability on intraday and long-term timescales from both instruments. On intraday timescales, using 500~s binning, the {\it NuSTAR} observations exhibit systematically higher fractional variability (6$-$24\%) than the {\it Chandra} observations ($\lesssim$7.5\%). Significant variability, defined by $F_{\rm var}/\sigma_{F_{\rm var}} > 3$, is detected in 9 out of 11 {\it NuSTAR} observations, compared with only 2 out of 16 {\it Chandra} observations. This indicates that the hard X-ray band is more strongly variable on intraday timescales.

\item The minimum variability timescales measured with S/N $\geq 2$ span 0.44$-$2.17~ks for {\it NuSTAR} and 1.1$-$2.5~ks for {\it Chandra}. The shortest NuSTAR timescale measured with S/N $> 3$, $\tau_{\rm var} = 1.15 \pm 0.26$~ks from {\it NuSTAR} (ObsID~60002022004), places an upper limit of $R \lesssim 7.7 \times 10^{14}$~cm on the size of the X-ray-emitting region, assuming a Doppler factor of $\delta = 25$ (Section~\ref{sec:X-ray_flux_variability}). This is consistent with synchrotron emission arising from a compact region within the relativistic jet.

\item The {\it Chandra} observations show a clear harder-when-brighter trend in both the hardness ratio and photon index. In the {\it NuSTAR} band, the hardness ratio and photon index show only a weak, non-significant tendency toward a softer-when-brighter trend, and we do not draw a firm conclusion about it with the current sample (Section~\ref{sec:spectral_analysis}).

\item X-ray spectral fitting using four models (PL, BPL, LP, and BR+PL) shows that the more complex models generally yield lower $\chi^2_\nu$ than a simple PL, though we caution that $\chi^2$ alone cannot statistically distinguish between these models without a formal test (Section \ref{sec:spectral_analysis}). An F-test confirms that curved models are statistically preferred over PL in all 16 {\it Chandra} observations, but in only 4 of the 11 {\it NuSTAR} observations; the remaining seven {\it NuSTAR} spectra are consistent with a single PL ($\Gamma \simeq 2.6-3.0$), and where curvature is detected in the {\it NuSTAR} band the spectrum steepens with energy (Section~\ref{sec:spectral_fitting}). AIC favors BPL for most {\it Chandra} spectra (13/16), while the four curved {\it NuSTAR} spectra are described comparably well by BPL, LP, and BR+PL.
All curved models are phenomenological; in particular, the bremsstrahlung component of BR+PL describes spectral curvature rather than thermal emission, which would be negligible compared with the Doppler-boosted jet emission. The LP model also provides satisfactory fits for several observations, consistent with spectral curvature produced by stochastic particle acceleration in the jet \citep{Massaro04, Tramacere07}.

\item Joint spectral fitting of the simultaneous {\it Chandra} and {\it NuSTAR} observations (2013 April) shows that a single power law is statistically disfavored (F-test p $< 10^{-12}$) to describe the broadband 0.4$-$79~keV spectrum, and that the curvature is confined to the soft X-ray band. The {\it Chandra} spectrum steepens above a break at $E_{\rm b} = 2.25^{+0.32}_{-0.35}$~keV (BPL, $\chi^2_\nu = 0.81$), while the {\it NuSTAR} spectrum is consistent with a single power law with $\Gamma = 2.73 \pm 0.03$, comparable to the {\it Chandra} index above the break. The combined coverage thus traces the steepening of the synchrotron tail from the soft to the hard X-ray band.

\item Our results are broadly consistent with those reported by \citet{Bhatta25} for {\it XMM-Newton} observations of PKS~2155$-$304, showing comparable fractional variability amplitudes and variability timescales in the soft X-ray band. The {\it NuSTAR} observations extend these findings to higher energies, revealing stronger and faster variability in the hard X-ray band and providing new constraints on the non-thermal processes governing blazar X-ray emission.
\end{enumerate}
The combination of {\it Chandra}'s high spatial resolution in the soft X-ray band and {\it NuSTAR}'s broad hard X-ray coverage provides a uniquely powerful dataset for investigating the multi-timescale variability and spectral evolution of PKS~2155$-$304 over more than two decades. Future studies combining simultaneous multiwavelength observations, including optical, UV, and $\gamma$-ray data, with the X-ray variability constraints presented here will be crucial for constructing self-consistent broadband SED models, distinguishing between a sub-dominant inverse-Compton contribution and the intrinsic curvature of the synchrotron tail as the origin of the observed X-ray spectral curvature, and further constraining the particle-acceleration mechanisms that drive blazar emission in PKS~2155$-$304.
\\[6pt]
This research used data from \textit{Chandra} and \textit{NuSTAR} Data Archive. This research employs a list of Chandra datasets, obtained by the Chandra X-ray Observatory, contained in~\dataset[DOI: 10.25574/cdc.714]{https://doi.org/10.25574/cdc.714}. This work was supported by NASA award 80NSSC20K0645 (R.T.). The authors thank the referee for the valuable comments on the manuscript. R.T. and S.B.G. thank Joey Neilsen for the useful suggestions and discussions on this project.
\\[6pt]
Facility: {\it Chandra} (ACIS-S) and {\it NuSTAR}.
\\[6pt]
\textit{Software:} 
Numpy \citep{van_der_Walt11}, Astropy \citep{Astropy_Collaboration13, Price-Whelan18}, Scipy \citep{Virtanen20}, Matplotlib \citep{Hunter07}, Pandas \citep{McKinney10}, CIAO v4.16 \citep{Fruscione06}, Sherpa v4.16.0 \citep{Freeman01}, DS9 \citep{Joye03}
\bibliography{ms}

@ARTICLE{Aharonian05,
       author = {{Aharonian}, F. and {Akhperjanian}, A.~G. and {Bazer-Bachi}, A.~R. and {Beilicke}, M. and {Benbow}, W. and {Berge}, D. and {Bernl{\"o}hr}, K. and {Boisson}, C. and {Bolz}, O. and {Borrel}, V. and {Braun}, I. and {Breitling}, F. and {Brown}, A.~M. and {Chadwick}, P.~M. and {Chounet}, L. -M. and {Cornils}, R. and {Costamante}, L. and {Degrange}, B. and {Dickinson}, H.~J. and {Djannati-Ata{\"\i}}, A. and {O'C. Drury}, L. and {Dubus}, G. and {Emmanoulopoulos}, D. and {Espigat}, P. and {Feinstein}, F. and {Fontaine}, G. and {Fuchs}, Y. and {Funk}, S. and {Gallant}, Y.~A. and {Giebels}, B. and {Gillessen}, S. and {Glicenstein}, J.~F. and {Goret}, P. and {Hadjichristidis}, C. and {Hauser}, M. and {Heinzelmann}, G. and {Henri}, G. and {Hermann}, G. and {Hinton}, J.~A. and {Hofmann}, W. and {Holleran}, M. and {Horns}, D. and {Jacholkowska}, A. and {de Jager}, O.~C. and {Kh{\'e}lifi}, B. and {Komin}, Nu. and {Konopelko}, A. and {Latham}, I.~J. and {Le Gallou}, R. and {Lemi{\`e}re}, A. and {Lemoine-Goumard}, M. and {Leroy}, N. and {Lohse}, T. and {Martin}, J.~M. and {Martineau-Huynh}, O. and {Marcowith}, A. and {Masterson}, C. and {McComb}, T.~J.~L. and {de Naurois}, M. and {Nolan}, S.~J. and {Noutsos}, A. and {Orford}, K.~J. and {Osborne}, J.~L. and {Ouchrif}, M. and {Panter}, M. and {Pelletier}, G. and {Pita}, S. and {P{\"u}hlhofer}, G. and {Punch}, M. and {Raubenheimer}, B.~C. and {Raue}, M. and {Raux}, J. and {Rayner}, S.~M. and {Reimer}, A. and {Reimer}, O. and {Ripken}, J. and {Rob}, L. and {Rolland}, L. and {Rowell}, G. and {Sahakian}, V. and {Saug{\'e}}, L. and {Schlenker}, S. and {Schlickeiser}, R. and {Schuster}, C. and {Schwanke}, U. and {Siewert}, M. and {Sol}, H. and {Spangler}, D. and {Steenkamp}, R. and {Stegmann}, C. and {Tavernet}, J. -P. and {Terrier}, R. and {Th{\'e}oret}, C.~G. and {Tluczykont}, M. and {Vasileiadis}, G. and {Venter}, C. and {Vincent}, P. and {V{\"o}lk}, H.~J. and {Wagner}, S.~J.},
        title = "{Multi-wavelength observations of PKS 2155-304 with HESS}",
      journal = {\aap},
         year = 2005,
        month = nov,
       volume = {442},
       number = {3},
        pages = {895-907},
          doi = {10.1051/0004-6361:20053353},
archivePrefix = {arXiv},
       eprint = {astro-ph/0506593},
 primaryClass = {astro-ph},
       adsurl = {https://ui.adsabs.harvard.edu/abs/2005A&A...442..895A}
}

@ARTICLE{Aharonian07,
       author = {{Aharonian}, F. and {Akhperjanian}, A.~G. and {Bazer-Bachi}, A.~R. and {Behera}, B. and {Beilicke}, M. and {Benbow}, W. and {Berge}, D. and {Bernl{\"o}hr}, K. and {Boisson}, C. and {Bolz}, O. and {Borrel}, V. and {Boutelier}, T. and {Braun}, I. and {Brion}, E. and {Brown}, A.~M. and {B{\"u}hler}, R. and {B{\"u}sching}, I. and {Bulik}, T. and {Carrigan}, S. and {Chadwick}, P.~M. and {Clapson}, A.~C. and {Chounet}, L. -M. and {Coignet}, G. and {Cornils}, R. and {Costamante}, L. and {Degrange}, B. and {Dickinson}, H.~J. and {Djannati-Ata{\"\i}}, A. and {Domainko}, W. and {Drury}, L. O'C. and {Dubus}, G. and {Dyks}, J. and {Egberts}, K. and {Emmanoulopoulos}, D. and {Espigat}, P. and {Farnier}, C. and {Feinstein}, F. and {Fiasson}, A. and {F{\"o}rster}, A. and {Fontaine}, G. and {Funk}, Seb. and {Funk}, S. and {F{\"u}{\ss}ling}, M. and {Gallant}, Y.~A. and {Giebels}, B. and {Glicenstein}, J.~F. and {Gl{\"u}ck}, B. and {Goret}, P. and {Hadjichristidis}, C. and {Hauser}, D. and {Hauser}, M. and {Heinzelmann}, G. and {Henri}, G. and {Hermann}, G. and {Hinton}, J.~A. and {Hoffmann}, A. and {Hofmann}, W. and {Holleran}, M. and {Hoppe}, S. and {Horns}, D. and {Jacholkowska}, A. and {de Jager}, O.~C. and {Kendziorra}, E. and {Kerschhaggl}, M. and {Kh{\'e}lifi}, B. and {Komin}, Nu. and {Kosack}, K. and {Lamanna}, G. and {Latham}, I.~J. and {Le Gallou}, R. and {Lemi{\`e}re}, A. and {Lemoine-Goumard}, M. and {Lenain}, J. -P. and {Lohse}, T. and {Martin}, J.~M. and {Martineau-Huynh}, O. and {Marcowith}, A. and {Masterson}, C. and {Maurin}, G. and {McComb}, T.~J.~L. and {Moderski}, R. and {Moulin}, E. and {de Naurois}, M. and {Nedbal}, D. and {Nolan}, S.~J. and {Olive}, J. -P. and {Orford}, K.~J. and {Osborne}, J.~L. and {Ostrowski}, M. and {Panter}, M. and {Pedaletti}, G. and {Pelletier}, G. and {Petrucci}, P. -O. and {Pita}, S. and {P{\"u}hlhofer}, G. and {Punch}, M. and {Ranchon}, S. and {Raubenheimer}, B.~C. and {Raue}, M. and {Rayner}, S.~M. and {Renaud}, M. and {Ripken}, J. and {Rob}, L. and {Rolland}, L. and {Rosier-Lees}, S. and {Rowell}, G. and {Rudak}, B. and {Ruppel}, J. and {Sahakian}, V. and {Santangelo}, A. and {Saug{\'e}}, L. and {Schlenker}, S. and {Schlickeiser}, R. and {Schr{\"o}der}, R. and {Schwanke}, U. and {Schwarzburg}, S. and {Schwemmer}, S. and {Shalchi}, A. and {Sol}, H. and {Spangler}, D. and {Stawarz}, {\L}. and {Steenkamp}, R. and {Stegmann}, C. and {Superina}, G. and {Tam}, P.~H. and {Tavernet}, J. -P. and {Terrier}, R. and {van Eldik}, C. and {Vasileiadis}, G. and {Venter}, C. and {Vialle}, J.~P. and {Vincent}, P. and {Vivier}, M. and {V{\"o}lk}, H.~J. and {Volpe}, F. and {Wagner}, S.~J. and {Ward}, M. and {Zdziarski}, A.~A.},
        title = "{An Exceptional Very High Energy Gamma-Ray Flare of PKS 2155-304}",
      journal = {\apjl},
         year = 2007,
        month = aug,
       volume = {664},
       number = {2},
        pages = {L71-L74},
          doi = {10.1086/520635},
archivePrefix = {arXiv},
       eprint = {0706.0797},
 primaryClass = {astro-ph},
       adsurl = {https://ui.adsabs.harvard.edu/abs/2007ApJ...664L..71A}
}

@ARTICLE{Astropy_Collaboration13,
       author = {{Astropy Collaboration} and {Robitaille}, Thomas P. and {Tollerud}, Erik J. and {Greenfield}, Perry and {Droettboom}, Michael and {Bray}, Erik and {Aldcroft}, Tom and {Davis}, Matt and {Ginsburg}, Adam and {Price-Whelan}, Adrian M. and {Kerzendorf}, Wolfgang E. and {Conley}, Alexander and {Crighton}, Neil and {Barbary}, Kyle and {Muna}, Demitri and {Ferguson}, Henry and {Grollier}, Fr{\'e}d{\'e}ric and {Parikh}, Madhura M. and {Nair}, Prasanth H. and {Unther}, Hans M. and {Deil}, Christoph and {Woillez}, Julien and {Conseil}, Simon and {Kramer}, Roban and {Turner}, James E.~H. and {Singer}, Leo and {Fox}, Ryan and {Weaver}, Benjamin A. and {Zabalza}, Victor and {Edwards}, Zachary I. and {Azalee Bostroem}, K. and {Burke}, D.~J. and {Casey}, Andrew R. and {Crawford}, Steven M. and {Dencheva}, Nadia and {Ely}, Justin and {Jenness}, Tim and {Labrie}, Kathleen and {Lim}, Pey Lian and {Pierfederici}, Francesco and {Pontzen}, Andrew and {Ptak}, Andy and {Refsdal}, Brian and {Servillat}, Mathieu and {Streicher}, Ole},
        title = "{Astropy: A community Python package for astronomy}",
      journal = {\aap},
         year = 2013,
        month = oct,
       volume = {558},
          eid = {A33},
        pages = {A33},
          doi = {10.1051/0004-6361/201322068},
archivePrefix = {arXiv},
       eprint = {1307.6212},
 primaryClass = {astro-ph.IM},
       adsurl = {https://ui.adsabs.harvard.edu/abs/2013A&A...558A..33A}
}

@ARTICLE{Balucinska05,
       author = {{Ba{\l}uci{\'n}ska-Church}, M. and {Ostrowski}, M. and {Stawarz}, {\l}. and {Church}, M.~J.},
        title = "{Discovery of hard X-ray features around the hotspots of Cygnus A}",
      journal = {\mnras},
         year = 2005,
        month = feb,
       volume = {357},
       number = {1},
        pages = {L6-L10},
          doi = {10.1111/j.1745-3933.2005.08597.x},
archivePrefix = {arXiv},
       eprint = {astro-ph/0411502},
 primaryClass = {astro-ph},
       adsurl = {https://ui.adsabs.harvard.edu/abs/2005MNRAS.357L...6B}
}

@ARTICLE{Bhatta18,
       author = {{Bhatta}, Gopal and {Mohorian}, Maksym and {Bilinsky}, Illya},
        title = "{Hard X-ray properties of NuSTAR blazars}",
      journal = {\aap},
         year = 2018,
        month = nov,
       volume = {619},
          eid = {A93},
        pages = {A93},
          doi = {10.1051/0004-6361/201833628},
archivePrefix = {arXiv},
       eprint = {1710.09910},
 primaryClass = {astro-ph.HE},
       adsurl = {https://ui.adsabs.harvard.edu/abs/2018A&A...619A..93B}}

@ARTICLE{Bhatta21,
       author = {{Bhatta}, Gopal},
        title = "{Characterizing Long-term Optical Variability Properties of {\ensuremath{\gamma}}-Ray-bright Blazars}",
      journal = {\apj},
         year = 2021,
        month = dec,
       volume = {923},
       number = {1},
          eid = {7},
        pages = {7},
          doi = {10.3847/1538-4357/ac2819},
archivePrefix = {arXiv},
       eprint = {2109.08110},
 primaryClass = {astro-ph.HE},
       adsurl = {https://ui.adsabs.harvard.edu/abs/2021ApJ...923....7B}
}

@ARTICLE{Bhatta25,
       author = {{Bhatta}, Gopal and {Chaudhary}, Suvas Chandra and {Dhital}, Niraj and {Adhikari}, Tek P. and {Mohorian}, Maksym and {Dinesh}, Adithiya and {P{\'a}nis}, Radim and {Neupane}, Raghav and {Maharjan}, Yogesh Singh},
        title = "{Probing X-Ray Timing and Spectral Variability in the Blazar PKS 2155{\textendash}304 over a Decade of XMM-Newton Observations}",
      journal = {\apj},
         year = 2025,
        month = mar,
       volume = {981},
       number = {2},
          eid = {118},
        pages = {118},
          doi = {10.3847/1538-4357/adb0c9},
archivePrefix = {arXiv},
       eprint = {2410.01278},
 primaryClass = {astro-ph.HE},
       adsurl = {https://ui.adsabs.harvard.edu/abs/2025ApJ...981..118B}
}

@ARTICLE{Bhattacharyya20,
       author = {{Bhattacharyya}, Joy and {Ghosh}, Ritesh and {Chatterjee}, Ritaban and {Das}, Nabanita},
        title = "{Blazar Variability: A Study of Nonstationarity and the Flux-Rms Relation}",
      journal = {\apj},
         year = 2020,
        month = jul,
       volume = {897},
       number = {1},
          eid = {25},
        pages = {25},
          doi = {10.3847/1538-4357/ab91a8},
archivePrefix = {arXiv},
       eprint = {2005.05230},
 primaryClass = {astro-ph.HE},
       adsurl = {https://ui.adsabs.harvard.edu/abs/2020ApJ...897...25B}
}

@ARTICLE{Bottcher07,
       author = {{B{\"o}ttcher}, Markus},
        title = "{Modeling the emission processes in blazars}",
      journal = {\apss},
         year = 2007,
        month = jun,
       volume = {309},
       number = {1-4},
        pages = {95-104},
          doi = {10.1007/s10509-007-9404-0},
archivePrefix = {arXiv},
       eprint = {astro-ph/0608713},
 primaryClass = {astro-ph},
       adsurl = {https://ui.adsabs.harvard.edu/abs/2007Ap&SS.309...95B}
}

@ARTICLE{Chatterjee09,
       author = {{Chatterjee}, Ritaban and {Marscher}, Alan P. and {Jorstad}, Svetlana G. and {Olmstead}, Alice R. and {McHardy}, Ian M. and {Aller}, Margo F. and {Aller}, Hugh D. and {L{\"a}hteenm{\"a}ki}, Anne and {Tornikoski}, Merja and {Hovatta}, Talvikki and {Marshall}, Kevin and {Miller}, H. Richard and {Ryle}, Wesley T. and {Chicka}, Benjamin and {Benker}, A.~J. and {Bottorff}, Mark C. and {Brokofsky}, David and {Campbell}, Jeffrey S. and {Chonis}, Taylor S. and {Gaskell}, C. Martin and {Gaynullina}, Evelina R. and {Grankin}, Konstantin N. and {Hedrick}, Cecelia H. and {Ibrahimov}, Mansur A. and {Klimek}, Elizabeth S. and {Kruse}, Amanda K. and {Masatoshi}, Shoji and {Miller}, Thomas R. and {Pan}, Hong-Jian and {Petersen}, Eric A. and {Peterson}, Bradley W. and {Shen}, Zhiqiang and {Strel'nikov}, Dmitriy V. and {Tao}, Jun and {Watkins}, Aaron E. and {Wheeler}, Kathleen},
        title = "{Disk-Jet Connection in the Radio Galaxy 3C 120}",
      journal = {\apj},
         year = 2009,
        month = oct,
       volume = {704},
       number = {2},
        pages = {1689-1703},
          doi = {10.1088/0004-637X/704/2/1689},
archivePrefix = {arXiv},
       eprint = {0909.2051},
 primaryClass = {astro-ph.HE},
       adsurl = {https://ui.adsabs.harvard.edu/abs/2009ApJ...704.1689C}
}

@ARTICLE{Davis01,
       author = {{Davis}, John E.},
        title = "{Event Pileup in Charge-coupled Devices}",
      journal = {\apj},
         year = 2001,
        month = nov,
       volume = {562},
       number = {1},
        pages = {575-582},
          doi = {10.1086/323488},
       adsurl = {https://ui.adsabs.harvard.edu/abs/2001ApJ...562..575D}
}

@INPROCEEDINGS{Drake06,
       author = {{Drake}, Jeremy J. and {Ratzlaff}, Peter and {Kashyap}, Vinay and {Edgar}, Richard and {Izem}, Rima and {Jerius}, Diab and {Siemiginowska}, Aneta and {Vikhlinin}, Alexey},
        title = "{Monte Carlo processes for including Chandra instrument response uncertainties in parameter estimation studies}",
    booktitle = {Observatory Operations: Strategies, Processes, and Systems},
         year = 2006,
       editor = {{Silva}, David R. and {Doxsey}, Rodger E.},
       series = {Society of Photo-Optical Instrumentation Engineers (SPIE) Conference Series},
       volume = {6270},
        month = jun,
          eid = {62701I},
        pages = {62701I},
          doi = {10.1117/12.672226},
       adsurl = {https://ui.adsabs.harvard.edu/abs/2006SPIE.6270E..1ID}
}

@ARTICLE{Edelson01,
       author = {{Edelson}, Rick and {Griffiths}, Gareth and {Markowitz}, Alex and {Sembay}, Steve and {Turner}, Martin J.~L. and {Warwick}, Robert},
        title = "{High Temporal Resolution XMM-Newton Monitoring of PKS 2155-304}",
      journal = {\apj},
         year = 2001,
        month = jun,
       volume = {554},
       number = {1},
        pages = {274-280},
          doi = {10.1086/321332},
archivePrefix = {arXiv},
       eprint = {astro-ph/0102458},
 primaryClass = {astro-ph},
       adsurl = {https://ui.adsabs.harvard.edu/abs/2001ApJ...554..274E}
}

@ARTICLE{Falomo93,
       author = {{Falomo}, R. and {Pesce}, Joseph E. and {Treves}, A.},
        title = "{The Environment of the BL Lacertae Object PKS 2155-304}",
      journal = {\apjl},
         year = 1993,
        month = jul,
       volume = {411},
        pages = {L63},
          doi = {10.1086/186913},
       adsurl = {https://ui.adsabs.harvard.edu/abs/1993ApJ...411L..63F}
}

@INPROCEEDINGS{Freeman01,
       author = {{Freeman}, Peter and {Doe}, Stephen and {Siemiginowska}, Aneta},
        title = "{Sherpa: a mission-independent data analysis application}",
    booktitle = {Astronomical Data Analysis},
         year = 2001,
       editor = {{Starck}, Jean-Luc and {Murtagh}, Fionn D.},
       series = {Society of Photo-Optical Instrumentation Engineers (SPIE) Conference Series},
       volume = {4477},
        month = nov,
        pages = {76-87},
          doi = {10.1117/12.447161},
archivePrefix = {arXiv},
       eprint = {astro-ph/0108426},
 primaryClass = {astro-ph},
       adsurl = {https://ui.adsabs.harvard.edu/abs/2001SPIE.4477...76F}
}

@INPROCEEDINGS{Fruscione06,
       author = {{Fruscione}, Antonella and {McDowell}, Jonathan C. and {Allen}, Glenn E. and {Brickhouse}, Nancy S. and {Burke}, Douglas J. and {Davis}, John E. and {Durham}, Nick and {Elvis}, Martin and {Galle}, Elizabeth C. and {Harris}, Daniel E. and {Huenemoerder}, David P. and {Houck}, John C. and {Ishibashi}, Bish and {Karovska}, Margarita and {Nicastro}, Fabrizio and {Noble}, Michael S. and {Nowak}, Michael A. and {Primini}, Frank A. and {Siemiginowska}, Aneta and {Smith}, Randall K. and {Wise}, Michael},
        title = "{CIAO: Chandra's data analysis system}",
    booktitle = {Society of Photo-Optical Instrumentation Engineers (SPIE) Conference Series},
         year = 2006,
       editor = {{Silva}, David R. and {Doxsey}, Rodger E.},
       series = {Society of Photo-Optical Instrumentation Engineers (SPIE) Conference Series},
       volume = {6270},
        month = jun,
          eid = {62701V},
        pages = {62701V},
          doi = {10.1117/12.671760},
       adsurl = {https://ui.adsabs.harvard.edu/abs/2006SPIE.6270E..1VF}
}

@INPROCEEDINGS{Garmire03,
       author = {{Garmire}, Gordon P. and {Bautz}, Mark W. and {Ford}, Peter G. and {Nousek}, John A. and {Ricker}, George R., Jr.},
        title = "{Advanced CCD imaging spectrometer (ACIS) instrument on the Chandra X-ray Observatory}",
    booktitle = {X-Ray and Gamma-Ray Telescopes and Instruments for Astronomy.},
         year = 2003,
       editor = {{Truemper}, Joachim E. and {Tananbaum}, Harvey D.},
       series = {Society of Photo-Optical Instrumentation Engineers (SPIE) Conference Series},
       volume = {4851},
        month = mar,
        pages = {28-44},
          doi = {10.1117/12.461599},
       adsurl = {https://ui.adsabs.harvard.edu/abs/2003SPIE.4851...28G}
}

@ARTICLE{Ghisellini09,
       author = {{Ghisellini}, G. and {Tavecchio}, F. and {Ghirlanda}, G.},
        title = "{Jet and accretion power in the most powerful Fermi blazars}",
      journal = {\mnras},
         year = 2009,
        month = nov,
       volume = {399},
       number = {4},
        pages = {2041-2054},
          doi = {10.1111/j.1365-2966.2009.15397.x},
archivePrefix = {arXiv},
       eprint = {0906.2195},
 primaryClass = {astro-ph.HE},
       adsurl = {https://ui.adsabs.harvard.edu/abs/2009MNRAS.399.2041G}
}

@ARTICLE{Giannios09,
       author = {{Giannios}, Dimitrios and {Uzdensky}, Dmitri A. and {Begelman}, Mitchell C.},
        title = "{Fast TeV variability in blazars: jets in a jet}",
      journal = {\mnras},
         year = 2009,
        month = may,
       volume = {395},
       number = {1},
        pages = {L29-L33},
          doi = {10.1111/j.1745-3933.2009.00635.x},
archivePrefix = {arXiv},
       eprint = {0901.1877},
 primaryClass = {astro-ph.HE},
       adsurl = {https://ui.adsabs.harvard.edu/abs/2009MNRAS.395L..29G}
}

@ARTICLE{Haggard19,
       author = {{Haggard}, Daryl and {Nynka}, Melania and {Mon}, Brayden and {de la Cruz Hernandez}, Noelia and {Nowak}, Michael and {Heinke}, Craig and {Neilsen}, Joseph and {Dexter}, Jason and {Fragile}, P. Chris and {Baganoff}, Fred and {Bower}, Geoffrey C. and {Corrales}, Lia R. and {Coti Zelati}, Francesco and {Degenaar}, Nathalie and {Markoff}, Sera and {Morris}, Mark R. and {Ponti}, Gabriele and {Rea}, Nanda and {Wilms}, J{\"o}ern and {Yusef-Zadeh}, Farhad},
        title = "{Chandra Spectral and Timing Analysis of Sgr A*'s Brightest X-Ray Flares}",
      journal = {\apj},
         year = 2019,
        month = dec,
       volume = {886},
       number = {2},
          eid = {96},
        pages = {96},
          doi = {10.3847/1538-4357/ab4a7f},
archivePrefix = {arXiv},
       eprint = {1908.01781},
 primaryClass = {astro-ph.HE},
       adsurl = {https://ui.adsabs.harvard.edu/abs/2019ApJ...886...96H}
}

@ARTICLE{Harrison13,
       author = {{Harrison}, Fiona A. and {Craig}, William W. and {Christensen}, Finn E. and {Hailey}, Charles J. and {Zhang}, William W. and {Boggs}, Steven E. and {Stern}, Daniel and {Cook}, W. Rick and {Forster}, Karl and {Giommi}, Paolo and {Grefenstette}, Brian W. and {Kim}, Yunjin and {Kitaguchi}, Takao and {Koglin}, Jason E. and {Madsen}, Kristin K. and {Mao}, Peter H. and {Miyasaka}, Hiromasa and {Mori}, Kaya and {Perri}, Matteo and {Pivovaroff}, Michael J. and {Puccetti}, Simonetta and {Rana}, Vikram R. and {Westergaard}, Niels J. and {Willis}, Jason and {Zoglauer}, Andreas and {An}, Hongjun and {Bachetti}, Matteo and {Barri{\`e}re}, Nicolas M. and {Bellm}, Eric C. and {Bhalerao}, Varun and {Brejnholt}, Nicolai F. and {Fuerst}, Felix and {Liebe}, Carl C. and {Markwardt}, Craig B. and {Nynka}, Melania and {Vogel}, Julia K. and {Walton}, Dominic J. and {Wik}, Daniel R. and {Alexander}, David M. and {Cominsky}, Lynn R. and {Hornschemeier}, Ann E. and {Hornstrup}, Allan and {Kaspi}, Victoria M. and {Madejski}, Greg M. and {Matt}, Giorgio and {Molendi}, Silvano and {Smith}, David M. and {Tomsick}, John A. and {Ajello}, Marco and {Ballantyne}, David R. and {Balokovi{\'c}}, Mislav and {Barret}, Didier and {Bauer}, Franz E. and {Blandford}, Roger D. and {Brandt}, W. Niel and {Brenneman}, Laura W. and {Chiang}, James and {Chakrabarty}, Deepto and {Chenevez}, Jerome and {Comastri}, Andrea and {Dufour}, Francois and {Elvis}, Martin and {Fabian}, Andrew C. and {Farrah}, Duncan and {Fryer}, Chris L. and {Gotthelf}, Eric V. and {Grindlay}, Jonathan E. and {Helfand}, David J. and {Krivonos}, Roman and {Meier}, David L. and {Miller}, Jon M. and {Natalucci}, Lorenzo and {Ogle}, Patrick and {Ofek}, Eran O. and {Ptak}, Andrew and {Reynolds}, Stephen P. and {Rigby}, Jane R. and {Tagliaferri}, Gianpiero and {Thorsett}, Stephen E. and {Treister}, Ezequiel and {Urry}, C. Megan},
        title = "{The Nuclear Spectroscopic Telescope Array (NuSTAR) High-energy X-Ray Mission}",
      journal = {\apj},
         year = 2013,
        month = jun,
       volume = {770},
       number = {2},
          eid = {103},
        pages = {103},
          doi = {10.1088/0004-637X/770/2/103},
archivePrefix = {arXiv},
       eprint = {1301.7307},
 primaryClass = {astro-ph.IM},
       adsurl = {https://ui.adsabs.harvard.edu/abs/2013ApJ...770..103H}
}

@ARTICLE{Harutyunyan26,
       author = {{Harutyunyan}, G. and {Sahakyan}, N. and {B{\'e}gu{\'e}}, D. and {Khachatryan}, M.},
        title = "{A comprehensive view of PKS 2155{\ensuremath{-}}304 from 2008 to 2023 through a multi-epoch modelling of its spectral energy distributions}",
      journal = {\mnras},
         year = 2026,
        month = feb,
       volume = {546},
       number = {1},
          eid = {staf2282},
        pages = {staf2282},
          doi = {10.1093/mnras/staf2282},
archivePrefix = {arXiv},
       eprint = {2509.11827},
 primaryClass = {astro-ph.HE},
       adsurl = {https://ui.adsabs.harvard.edu/abs/2026MNRAS.546f2282H}
}

@ARTICLE{Hu24,
       author = {{Hu}, Xin-Ke and {Yu}, Yu-Wei and {Zhang}, Jin and {Wu}, Tan-Zheng and {Lian}, Ji-Shun and {Wang}, Xiang-Gao and {Zhang}, Hai-Ming and {Liang}, En-Wei},
        title = "{X-Ray Polarization Variability of High Spectral Peak BL Lacertaes: Cases of 1ES 1959+650 and PKS 2155─304}",
      journal = {\apjl},
         year = 2024,
        month = mar,
       volume = {963},
       number = {2},
          eid = {L41},
        pages = {L41},
          doi = {10.3847/2041-8213/ad2a4f},
archivePrefix = {arXiv},
       eprint = {2312.01755},
 primaryClass = {astro-ph.HE},
       adsurl = {https://ui.adsabs.harvard.edu/abs/2024ApJ...963L..41H}
}

@ARTICLE{Hu25,
       author = {{Hu}, W. and {Kang}, J.~L. and {Wang}, J.~X. and {Xiao}, G.~C. and {Ren}, G.~W.},
        title = "{The origin of X-ray intra-day variability in HBL PKS 2155─304}",
      journal = {\aap},
         year = 2025,
        month = may,
       volume = {697},
          eid = {A71},
        pages = {A71},
          doi = {10.1051/0004-6361/202453490},
archivePrefix = {arXiv},
       eprint = {2504.10773},
 primaryClass = {astro-ph.HE},
       adsurl = {https://ui.adsabs.harvard.edu/abs/2025A&A...697A..71H}
}

@ARTICLE{Hunter07,
       author = {{Hunter}, John D.},
        title = "{Matplotlib: A 2D Graphics Environment}",
      journal = {Computing in Science and Engineering},
         year = 2007,
        month = may,
       volume = {9},
       number = {3},
        pages = {90-95},
          doi = {10.1109/MCSE.2007.55},
       adsurl = {https://ui.adsabs.harvard.edu/abs/2007CSE.....9...90H}
}

@INPROCEEDINGS{Joye03,
       author = {{Joye}, W.~A. and {Mandel}, E.},
        title = "{New Features of SAOImage DS9}",
    booktitle = {Astronomical Data Analysis Software and Systems XII},
         year = 2003,
       editor = {{Payne}, H.~E. and {Jedrzejewski}, R.~I. and {Hook}, R.~N.},
       series = {Astronomical Society of the Pacific Conference Series},
       volume = {295},
        month = jan,
        pages = {489},
       adsurl = {https://ui.adsabs.harvard.edu/abs/2003ASPC..295..489J}
}

@ARTICLE{Kapanadze14,
       author = {{Kapanadze}, B. and {Romano}, P. and {Vercellone}, S. and {Kapanadze}, S.},
        title = "{The X-ray behaviour of the high-energy peaked BL Lacertae source PKS 2155-304 in the 0.3-10 keV band}",
      journal = {\mnras},
         year = 2014,
        month = oct,
       volume = {444},
       number = {2},
        pages = {1077-1094},
          doi = {10.1093/mnras/stu1504},
       adsurl = {https://ui.adsabs.harvard.edu/abs/2014MNRAS.444.1077K}
}

@ARTICLE{Kouch24,
       author = {{Kouch}, Pouya M. and {Liodakis}, Ioannis and {Middei}, Riccardo and {Kim}, Dawoon E. and {Tavecchio}, Fabrizio and {Marscher}, Alan P. and {Marshall}, Herman L. and {Ehlert}, Steven R. and {Di Gesu}, Laura and {Jorstad}, Svetlana G. and {Agudo}, Iv{\'a}n and {Madejski}, Grzegorz M. and {Romani}, Roger W. and {Errando}, Manel and {Lindfors}, Elina and {Nilsson}, Kari and {Toppari}, Ella and {Potter}, Stephen B. and {Imazawa}, Ryo and {Sasada}, Mahito and {Fukazawa}, Yasushi and {Kawabata}, Koji S. and {Uemura}, Makoto and {Mizuno}, Tsunefumi and {Nakaoka}, Tatsuya and {Akitaya}, Hiroshi and {McCall}, Callum and {Jermak}, Helen E. and {Steele}, Iain A. and {Myserlis}, Ioannis and {Gurwell}, Mark and {Keating}, Garrett K. and {Rao}, Ramprasad and {Kang}, Sincheol and {Lee}, Sang-Sung and {Kim}, Sang-Hyun and {Cheong}, Whee Yeon and {Jeong}, Hyeon-Woo and {Angelakis}, Emmanouil and {Kraus}, Alexander and {Aceituno}, Francisco Jos{\'e} and {Bonnoli}, Giacomo and {Casanova}, V{\'\i}ctor and {Escudero}, Juan and {Ag{\'\i}s-Gonz{\'a}lez}, Beatriz and {Husillos}, C{\'e}sar and {Morcuende}, Daniel and {Otero-Santos}, Jorge and {Sota}, Alfredo and {Bachev}, Rumen and {Antonelli}, Lucio Angelo and {Bachetti}, Matteo and {Baldini}, Luca and {Baumgartner}, Wayne H. and {Bellazzini}, Ronaldo and {Bianchi}, Stefano and {Bongiorno}, Stephen D. and {Bonino}, Raffaella and {Brez}, Alessandro and {Bucciantini}, Niccol{\`o} and {Capitanio}, Fiamma and {Castellano}, Simone and {Cavazzuti}, Elisabetta and {Chen}, Chien-Ting and {Ciprini}, Stefano and {Costa}, Enrico and {De Rosa}, Alessandra and {Del Monte}, Ettore and {Di Lalla}, Niccol{\`o} and {Di Marco}, Alessandro and {Donnarumma}, Immacolata and {Doroshenko}, Victor and {Dov{\v{c}}iak}, Michal and {Enoto}, Teruaki and {Evangelista}, Yuri and {Fabiani}, Sergio and {Ferrazzoli}, Riccardo and {Garcia}, Javier A. and {Gunji}, Shuichi and {Hayashida}, Kiyoshi and {Heyl}, Jeremy and {Iwakiri}, Wataru and {Kaaret}, Philip and {Karas}, Vladimir and {Kislat}, Fabian and {Kitaguchi}, Takao and {Kolodziejczak}, Jeffery J. and {Krawczynski}, Henric and {La Monaca}, Fabio and {Latronico}, Luca and {Maldera}, Simone and {Manfreda}, Alberto and {Marin}, Fr{\'e}d{\'e}ric and {Marinucci}, Andrea and {Massaro}, Francesco and {Matt}, Giorgio and {Mitsuishi}, Ikuyuki and {Muleri}, Fabio and {Negro}, Michela and {Ng}, Chi-Yung and {O'Dell}, Stephen L. and {Omodei}, Nicola and {Oppedisano}, Chiara and {Papitto}, Alessandro and {Pavlov}, George G. and {Peirson}, Abel Lawrence and {Perri}, Matteo and {Pesce-Rollins}, Melissa and {Petrucci}, Pierre-Olivier and {Pilia}, Maura and {Possenti}, Andrea and {Poutanen}, Juri and {Puccetti}, Simonetta and {Ramsey}, Brian D. and {Rankin}, John and {Ratheesh}, Ajay and {Roberts}, Oliver J. and {Sgr{\`o}}, Carmelo and {Slane}, Patrick and {Soffitta}, Paolo and {Spandre}, Gloria and {Swartz}, Douglas A. and {Tamagawa}, Toru and {Taverna}, Roberto and {Tawara}, Yuzuru and {Tennant}, Allyn F. and {Thomas}, Nicholas E. and {Tombesi}, Francesco and {Trois}, Alessio and {Tsygankov}, Sergey S. and {Turolla}, Roberto and {Vink}, Jacco and {Weisskopf}, Martin C. and {Wu}, Kinwah and {Xie}, Fei and {Zane}, Silvia},
        title = "{IXPE observation of PKS 2155─304 reveals the most highly polarized blazar}",
      journal = {\aap},
         year = 2024,
        month = sep,
       volume = {689},
          eid = {A119},
        pages = {A119},
          doi = {10.1051/0004-6361/202449166},
archivePrefix = {arXiv},
       eprint = {2406.01693},
 primaryClass = {astro-ph.HE},
       adsurl = {https://ui.adsabs.harvard.edu/abs/2024A&A...689A.119K}
}

@ARTICLE{Lee11,
       author = {{Lee}, Hyunsook and {Kashyap}, Vinay L. and {van Dyk}, David A. and {Connors}, Alanna and {Drake}, Jeremy J. and {Izem}, Rima and {Meng}, Xiao-Li and {Min}, Shandong and {Park}, Taeyoung and {Ratzlaff}, Pete and {Siemiginowska}, Aneta and {Zezas}, Andreas},
        title = "{Accounting for Calibration Uncertainties in X-ray Analysis: Effective Areas in Spectral Fitting}",
      journal = {\apj},
         year = 2011,
        month = apr,
       volume = {731},
       number = {2},
          eid = {126},
        pages = {126},
          doi = {10.1088/0004-637X/731/2/126},
archivePrefix = {arXiv},
       eprint = {1102.4610},
 primaryClass = {astro-ph.IM},
       adsurl = {https://ui.adsabs.harvard.edu/abs/2011ApJ...731..126L}
}

@ARTICLE{Madejski16,
       author = {{Madejski}, G.~M. and {Nalewajko}, K. and {Madsen}, K.~K. and {Chiang}, J. and {Balokovi{\'c}}, M. and {Paneque}, D. and {Furniss}, A.~K. and {Hayashida}, M. and {Urry}, C.~M. and {Sikora}, M. and {Ajello}, M. and {Blandford}, R.~D. and {Harrison}, F.~A. and {Sanchez}, D. and {Giebels}, B. and {Stern}, D. and {Alexander}, D.~M. and {Barret}, D. and {Boggs}, S.~E. and {Christensen}, F.~E. and {Craig}, W.~W. and {Forster}, K. and {Giommi}, P. and {Grefenstette}, B. and {Hailey}, C. and {Hornstrup}, A. and {Kitaguchi}, T. and {Koglin}, J.~E. and {Mao}, P.~H. and {Miyasaka}, H. and {Mori}, K. and {Perri}, M. and {Pivovaroff}, M.~J. and {Puccetti}, S. and {Rana}, V. and {Westergaard}, N.~J. and {Zhang}, W.~W. and {Zoglauer}, A.},
        title = "{First NuSTAR Observations of the BL Lac-type Blazar PKS 2155-304: Constraints on the Jet Content and Distribution of Radiating Particles}",
      journal = {\apj},
         year = 2016,
        month = nov,
       volume = {831},
       number = {2},
          eid = {142},
        pages = {142},
          doi = {10.3847/0004-637X/831/2/142},
archivePrefix = {arXiv},
       eprint = {1609.02203},
 primaryClass = {astro-ph.HE},
       adsurl = {https://ui.adsabs.harvard.edu/abs/2016ApJ...831..142M}
}

@ARTICLE{Marscher14,
       author = {{Marscher}, Alan P.},
        title = "{Turbulent, Extreme Multi-zone Model for Simulating Flux and Polarization Variability in Blazars}",
      journal = {\apj},
         year = 2014,
        month = jan,
       volume = {780},
       number = {1},
          eid = {87},
        pages = {87},
          doi = {10.1088/0004-637X/780/1/87},
archivePrefix = {arXiv},
       eprint = {1311.7665},
 primaryClass = {astro-ph.HE},
       adsurl = {https://ui.adsabs.harvard.edu/abs/2014ApJ...780...87M}
}

@ARTICLE{Massaro04,
       author = {{Massaro}, E. and {Perri}, M. and {Giommi}, P. and {Nesci}, R.},
        title = "{Log-parabolic spectra and particle acceleration in the BL Lac object Mkn 421: Spectral analysis of the complete BeppoSAX wide band X-ray data set}",
      journal = {\aap},
         year = 2004,
        month = jan,
       volume = {413},
        pages = {489-503},
          doi = {10.1051/0004-6361:20031558},
archivePrefix = {arXiv},
       eprint = {astro-ph/0312260},
 primaryClass = {astro-ph},
       adsurl = {https://ui.adsabs.harvard.edu/abs/2004A&A...413..489M}
}

@inproceedings{McKinney10,
  author    = {McKinney, W.},
  title     = {Data Structures for Statistical Computing in {Python}},
  booktitle = {Proceedings of the 9th Python in Science Conference},
  year      = {2010},
  pages     = {51--56},
  address   = {Austin, TX}
}

@ARTICLE{Nandra15,
       author = {{Nandra}, K. and {Laird}, E.~S. and {Aird}, J.~A. and {Salvato}, M. and {Georgakakis}, A. and {Barro}, G. and {Perez-Gonzalez}, P.~G. and {Barmby}, P. and {Chary}, R. -R. and {Coil}, A. and {Cooper}, M.~C. and {Davis}, M. and {Dickinson}, M. and {Faber}, S.~M. and {Fazio}, G.~G. and {Guhathakurta}, P. and {Gwyn}, S. and {Hsu}, L. -T. and {Huang}, J. -S. and {Ivison}, R.~J. and {Koo}, D.~C. and {Newman}, J.~A. and {Rangel}, C. and {Yamada}, T. and {Willmer}, C.},
        title = "{AEGIS-X: Deep Chandra Imaging of the Central Groth Strip}",
      journal = {\apjs},
         year = 2015,
        month = sep,
       volume = {220},
       number = {1},
          eid = {10},
        pages = {10},
          doi = {10.1088/0067-0049/220/1/10},
archivePrefix = {arXiv},
       eprint = {1503.09078},
 primaryClass = {astro-ph.HE},
       adsurl = {https://ui.adsabs.harvard.edu/abs/2015ApJS..220...10N}
}

@ARTICLE{Price-Whelan18,
       author = {{Astropy Collaboration} and {Price-Whelan}, A.~M. and {Sip{\H{o}}cz}, B.~M. and {G{\"u}nther}, H.~M. and {Lim}, P.~L. and {Crawford}, S.~M. and {Conseil}, S. and {Shupe}, D.~L. and {Craig}, M.~W. and {Dencheva}, N. and {Ginsburg}, A. and {VanderPlas}, J.~T. and {Bradley}, L.~D. and {P{\'e}rez-Su{\'a}rez}, D. and {de Val-Borro}, M. and {Aldcroft}, T.~L. and {Cruz}, K.~L. and {Robitaille}, T.~P. and {Tollerud}, E.~J. and {Ardelean}, C. and {Babej}, T. and {Bach}, Y.~P. and {Bachetti}, M. and {Bakanov}, A.~V. and {Bamford}, S.~P. and {Barentsen}, G. and {Barmby}, P. and {Baumbach}, A. and {Berry}, K.~L. and {Biscani}, F. and {Boquien}, M. and {Bostroem}, K.~A. and {Bouma}, L.~G. and {Brammer}, G.~B. and {Bray}, E.~M. and {Breytenbach}, H. and {Buddelmeijer}, H. and {Burke}, D.~J. and {Calderone}, G. and {Cano Rodr{\'\i}guez}, J.~L. and {Cara}, M. and {Cardoso}, J.~V.~M. and {Cheedella}, S. and {Copin}, Y. and {Corrales}, L. and {Crichton}, D. and {D'Avella}, D. and {Deil}, C. and {Depagne}, {\'E}. and {Dietrich}, J.~P. and {Donath}, A. and {Droettboom}, M. and {Earl}, N. and {Erben}, T. and {Fabbro}, S. and {Ferreira}, L.~A. and {Finethy}, T. and {Fox}, R.~T. and {Garrison}, L.~H. and {Gibbons}, S.~L.~J. and {Goldstein}, D.~A. and {Gommers}, R. and {Greco}, J.~P. and {Greenfield}, P. and {Groener}, A.~M. and {Grollier}, F. and {Hagen}, A. and {Hirst}, P. and {Homeier}, D. and {Horton}, A.~J. and {Hosseinzadeh}, G. and {Hu}, L. and {Hunkeler}, J.~S. and {Ivezi{\'c}}, {\v{Z}}. and {Jain}, A. and {Jenness}, T. and {Kanarek}, G. and {Kendrew}, S. and {Kern}, N.~S. and {Kerzendorf}, W.~E. and {Khvalko}, A. and {King}, J. and {Kirkby}, D. and {Kulkarni}, A.~M. and {Kumar}, A. and {Lee}, A. and {Lenz}, D. and {Littlefair}, S.~P. and {Ma}, Z. and {Macleod}, D.~M. and {Mastropietro}, M. and {McCully}, C. and {Montagnac}, S. and {Morris}, B.~M. and {Mueller}, M. and {Mumford}, S.~J. and {Muna}, D. and {Murphy}, N.~A. and {Nelson}, S. and {Nguyen}, G.~H. and {Ninan}, J.~P. and {N{\"o}the}, M. and {Ogaz}, S. and {Oh}, S. and {Parejko}, J.~K. and {Parley}, N. and {Pascual}, S. and {Patil}, R. and {Patil}, A.~A. and {Plunkett}, A.~L. and {Prochaska}, J.~X. and {Rastogi}, T. and {Reddy Janga}, V. and {Sabater}, J. and {Sakurikar}, P. and {Seifert}, M. and {Sherbert}, L.~E. and {Sherwood-Taylor}, H. and {Shih}, A.~Y. and {Sick}, J. and {Silbiger}, M.~T. and {Singanamalla}, S. and {Singer}, L.~P. and {Sladen}, P.~H. and {Sooley}, K.~A. and {Sornarajah}, S. and {Streicher}, O. and {Teuben}, P. and {Thomas}, S.~W. and {Tremblay}, G.~R. and {Turner}, J.~E.~H. and {Terr{\'o}n}, V. and {van Kerkwijk}, M.~H. and {de la Vega}, A. and {Watkins}, L.~L. and {Weaver}, B.~A. and {Whitmore}, J.~B. and {Woillez}, J. and {Zabalza}, V. and {Astropy Contributors}},
        title = "{The Astropy Project: Building an Open-science Project and Status of the v2.0 Core Package}",
      journal = {\aj},
         year = 2018,
        month = sep,
       volume = {156},
       number = {3},
          eid = {123},
        pages = {123},
          doi = {10.3847/1538-3881/aabc4f},
archivePrefix = {arXiv},
       eprint = {1801.02634},
 primaryClass = {astro-ph.IM},
       adsurl = {https://ui.adsabs.harvard.edu/abs/2018AJ....156..123A}
}

@ARTICLE{Siemiginowska07,
       author = {{Siemiginowska}, Aneta and {Stawarz}, {\L}ukasz and {Cheung}, C.~C. and {Harris}, D.~E. and {Sikora}, Marek and {Aldcroft}, Thomas L. and {Bechtold}, Jill},
        title = "{The 300 kpc Long X-Ray Jet in PKS 1127-145, z = 1.18 Quasar: Constraining X-Ray Emission Models}",
      journal = {\apj},
         year = 2007,
        month = mar,
       volume = {657},
       number = {1},
        pages = {145-158},
          doi = {10.1086/510898},
archivePrefix = {arXiv},
       eprint = {astro-ph/0611406},
 primaryClass = {astro-ph},
       adsurl = {https://ui.adsabs.harvard.edu/abs/2007ApJ...657..145S}
}

@ARTICLE{Thimmappa22,
       author = {{Thimmappa}, R. and {Stawarz}, {\L}. and {Neilsen}, J. and {Ostrowski}, M. and {Reville}, B.},
        title = "{X-Ray Spectral Analysis of the Jet Termination Shock in Pictor A on Subarcsecond Scales with Chandra}",
      journal = {\apj},
         year = 2022,
        month = dec,
       volume = {941},
       number = {2},
          eid = {204},
        pages = {204},
          doi = {10.3847/1538-4357/aca472},
archivePrefix = {arXiv},
       eprint = {2208.10648},
 primaryClass = {astro-ph.HE},
       adsurl = {https://ui.adsabs.harvard.edu/abs/2022ApJ...941..204T}
}

@ARTICLE{Tramacere07,
       author = {{Tramacere}, A. and {Massaro}, F. and {Cavaliere}, A.},
        title = "{Signatures of synchrotron emission and of electron acceleration in the X-ray spectra of Mrk 421}",
      journal = {\aap},
         year = 2007,
        month = may,
       volume = {466},
       number = {2},
        pages = {521-529},
          doi = {10.1051/0004-6361:20066723},
archivePrefix = {arXiv},
       eprint = {astro-ph/0702151},
 primaryClass = {astro-ph},
       adsurl = {https://ui.adsabs.harvard.edu/abs/2007A&A...466..521T}
}

@ARTICLE{Tramacere09,
       author = {{Tramacere}, A. and {Giommi}, P. and {Perri}, M. and {Verrecchia}, F. and {Tosti}, G.},
        title = "{Swift observations of the very intense flaring activity of Mrk 421 during 2006. I. Phenomenological picture of electron acceleration and predictions for MeV/GeV emission}",
      journal = {\aap},
         year = 2009,
        month = jul,
       volume = {501},
       number = {3},
        pages = {879-898},
          doi = {10.1051/0004-6361/200810865},
archivePrefix = {arXiv},
       eprint = {0901.4124},
 primaryClass = {astro-ph.HE},
       adsurl = {https://ui.adsabs.harvard.edu/abs/2009A&A...501..879T}
}

@ARTICLE{Urry95,
       author = {{Urry}, C. Megan and {Padovani}, Paolo},
        title = "{Unified Schemes for Radio-Loud Active Galactic Nuclei}",
      journal = {\pasp},
         year = 1995,
        month = sep,
       volume = {107},
        pages = {803},
          doi = {10.1086/133630},
archivePrefix = {arXiv},
       eprint = {astro-ph/9506063},
 primaryClass = {astro-ph},
       adsurl = {https://ui.adsabs.harvard.edu/abs/1995PASP..107..803U}
}

@ARTICLE{Uttley01,
       author = {{Uttley}, Philip and {McHardy}, Ian M.},
        title = "{The flux-dependent amplitude of broadband noise variability in X-ray binaries and active galaxies}",
      journal = {\mnras},
         year = 2001,
        month = may,
       volume = {323},
       number = {2},
        pages = {L26-L30},
          doi = {10.1046/j.1365-8711.2001.04496.x},
archivePrefix = {arXiv},
       eprint = {astro-ph/0103367},
 primaryClass = {astro-ph},
       adsurl = {https://ui.adsabs.harvard.edu/abs/2001MNRAS.323L..26U}
}

@ARTICLE{Uttley05,
       author = {{Uttley}, P. and {McHardy}, I.~M. and {Vaughan}, S.},
        title = "{Non-linear X-ray variability in X-ray binaries and active galaxies}",
      journal = {\mnras},
         year = 2005,
        month = may,
       volume = {359},
       number = {1},
        pages = {345-362},
          doi = {10.1111/j.1365-2966.2005.08886.x},
archivePrefix = {arXiv},
       eprint = {astro-ph/0502112},
 primaryClass = {astro-ph},
       adsurl = {https://ui.adsabs.harvard.edu/abs/2005MNRAS.359..345U}
}

@ARTICLE{van_der_Walt11,
       author = {{van der Walt}, St{\'e}fan and {Colbert}, S. Chris and {Varoquaux}, Ga{\"e}l},
        title = "{The NumPy Array: A Structure for Efficient Numerical Computation}",
      journal = {Computing in Science and Engineering},
         year = 2011,
        month = mar,
       volume = {13},
       number = {2},
        pages = {22-30},
          doi = {10.1109/MCSE.2011.37},
archivePrefix = {arXiv},
       eprint = {1102.1523},
 primaryClass = {cs.MS},
       adsurl = {https://ui.adsabs.harvard.edu/abs/2011CSE....13b..22V}
}

@ARTICLE{Vaughan03,
       author = {{Vaughan}, S. and {Edelson}, R. and {Warwick}, R.~S. and {Uttley}, P.},
        title = "{On characterizing the variability properties of X-ray light curves from active galaxies}",
      journal = {\mnras},
         year = 2003,
        month = nov,
       volume = {345},
       number = {4},
        pages = {1271-1284},
          doi = {10.1046/j.1365-2966.2003.07042.x},
archivePrefix = {arXiv},
       eprint = {astro-ph/0307420},
 primaryClass = {astro-ph},
       adsurl = {https://ui.adsabs.harvard.edu/abs/2003MNRAS.345.1271V}
}

@ARTICLE{Virtanen20,
       author = {{Virtanen}, Pauli and {Gommers}, Ralf and {Oliphant}, Travis E. and {Haberland}, Matt and {Reddy}, Tyler and {Cournapeau}, David and {Burovski}, Evgeni and {Peterson}, Pearu and {Weckesser}, Warren and {Bright}, Jonathan and {van der Walt}, St{\'e}fan J. and {Brett}, Matthew and {Wilson}, Joshua and {Millman}, K. Jarrod and {Mayorov}, Nikolay and {Nelson}, Andrew R.~J. and {Jones}, Eric and {Kern}, Robert and {Larson}, Eric and {Carey}, C.~J. and {Polat}, {\.I}lhan and {Feng}, Yu and {Moore}, Eric W. and {VanderPlas}, Jake and {Laxalde}, Denis and {Perktold}, Josef and {Cimrman}, Robert and {Henriksen}, Ian and {Quintero}, E.~A. and {Harris}, Charles R. and {Archibald}, Anne M. and {Ribeiro}, Ant{\^o}nio H. and {Pedregosa}, Fabian and {van Mulbregt}, Paul and {SciPy 1. 0 Contributors}},
        title = "{SciPy 1.0: fundamental algorithms for scientific computing in Python}",
      journal = {Nature Methods},
         year = 2020,
        month = feb,
       volume = {17},
        pages = {261-272},
          doi = {10.1038/s41592-019-0686-2},
archivePrefix = {arXiv},
       eprint = {1907.10121},
 primaryClass = {cs.MS},
       adsurl = {https://ui.adsabs.harvard.edu/abs/2020NatMe..17..261V}
}

@INPROCEEDINGS{Weisskopf00,
       author = {{Weisskopf}, Martin C. and {Tananbaum}, Harvey D. and {Van Speybroeck}, Leon P. and {O'Dell}, Stephen L.},
        title = "{Chandra X-ray Observatory (CXO): overview}",
    booktitle = {X-Ray Optics, Instruments, and Missions III},
         year = 2000,
       editor = {{Truemper}, Joachim E. and {Aschenbach}, Bernd},
       series = {Society of Photo-Optical Instrumentation Engineers (SPIE) Conference Series},
       volume = {4012},
        month = jul,
        pages = {2-16},
          doi = {10.1117/12.391545},
archivePrefix = {arXiv},
       eprint = {astro-ph/0004127},
 primaryClass = {astro-ph},
       adsurl = {https://ui.adsabs.harvard.edu/abs/2000SPIE.4012....2W}
}

@ARTICLE{Wierzcholska26,
       author = {{Wierzcholska}, A. and {Zacharias}, M.},
        title = "{20 years of monitoring: PKS 2155-304 and PKS 1510-089 in the eyes of Swift and Fermi. I. The case of PKS 2155-304}",
      journal = {arXiv e-prints},
         year = 2026,
        month = apr,
          eid = {arXiv:2604.00876},
        pages = {arXiv:2604.00876},
          doi = {10.48550/arXiv.2604.00876},
archivePrefix = {arXiv},
       eprint = {2604.00876},
 primaryClass = {astro-ph.HE},
       adsurl = {https://ui.adsabs.harvard.edu/abs/2026arXiv260400876W}
}

@ARTICLE{Xu14,
       author = {{Xu}, Jin and {van Dyk}, David A. and {Kashyap}, Vinay L. and {Siemiginowska}, Aneta and {Connors}, Alanna and {Drake}, Jeremy and {Meng}, Xiao-Li and {Ratzlaff}, Pete and {Yu}, Yaming},
        title = "{A Fully Bayesian Method for Jointly Fitting Instrumental Calibration and X-Ray Spectral Models}",
      journal = {\apj},
         year = 2014,
        month = oct,
       volume = {794},
       number = {2},
          eid = {97},
        pages = {97},
          doi = {10.1088/0004-637X/794/2/97},
       adsurl = {https://ui.adsabs.harvard.edu/abs/2014ApJ...794...97X}
}

@ARTICLE{Zhang99,
       author = {{Zhang}, Y.~H. and {Celotti}, A. and {Treves}, A. and {Chiappetti}, L. and {Ghisellini}, G. and {Maraschi}, L. and {Pian}, E. and {Tagliaferri}, G. and {Tavecchio}, F. and {Urry}, C.~M.},
        title = "{Rapid X-Ray Variability of the BL Lacertae Object PKS 2155-304}",
      journal = {\apj},
         year = 1999,
        month = dec,
       volume = {527},
       number = {2},
        pages = {719-732},
          doi = {10.1086/308116},
archivePrefix = {arXiv},
       eprint = {astro-ph/9907325},
 primaryClass = {astro-ph},
       adsurl = {https://ui.adsabs.harvard.edu/abs/1999ApJ...527..719Z}
}

@ARTICLE{Zhang21,
       author = {{Zhang}, Zhongli and {Gupta}, Alok C. and {Gaur}, Haritma and {Wiita}, Paul J. and {An}, Tao and {Lu}, Yang and {Fan}, Shida and {Xu}, Haiguang},
        title = "{X-Ray Intraday Variability of the TeV Blazar PKS 2155-304 with Suzaku during 2005-2014}",
      journal = {\apj},
         year = 2021,
        month = mar,
       volume = {909},
       number = {2},
          eid = {103},
        pages = {103},
          doi = {10.3847/1538-4357/abdd38},
archivePrefix = {arXiv},
       eprint = {2101.05977},
 primaryClass = {astro-ph.HE},
       adsurl = {https://ui.adsabs.harvard.edu/abs/2021ApJ...909..103Z}
}

@ARTICLE{Jagan21,
       author = {{Jagan}, Sitha K. and {Sahayanathan}, S. and {Rieger}, Frank M. and {Ravikumar}, C.~D.},
        title = "{Convex X-ray spectra of PKS 2155-304 and constraints on the minimum electron energy}",
      journal = {\mnras},
         year = 2021,
        month = sep,
       volume = {506},
       number = {3},
        pages = {3996-4006},
          doi = {10.1093/mnras/stab1993},
archivePrefix = {arXiv},
       eprint = {2107.04534},
 primaryClass = {astro-ph.HE},
       adsurl = {https://ui.adsabs.harvard.edu/abs/2021MNRAS.506.3996J}
}

@ARTICLE{Wierzcholska15,
       author = {{Wierzcholska}, Alicja and {Ostrowski}, Micha{\l} and {Stawarz}, {\L}ukasz and {Wagner}, Stefan and {Hauser}, Marcus},
        title = "{Longterm optical monitoring of bright BL Lacertae objects with ATOM: Spectral variability and multiwavelength correlations}",
      journal = {\aap},
         year = 2015,
        month = jan,
       volume = {573},
          eid = {A69},
        pages = {A69},
          doi = {10.1051/0004-6361/201423967},
archivePrefix = {arXiv},
       eprint = {1409.3022},
 primaryClass = {astro-ph.HE},
       adsurl = {https://ui.adsabs.harvard.edu/abs/2015A&A...573A..69W}
}

@ARTICLE{HESS2014,
       author = {{H.~E.~S.~S. Collaboration} and {Abramowski}, A. and {Aharonian}, F. and {Ait Benkhali}, F. and {Akhperjanian}, A.~G. and {Ang{\"u}ner}, E.~O. and {Backes}, M. and {Balenderan}, S. and {Balzer}, A. and {Barnacka}, A. and {Becherini}, Y. and {Becker Tjus}, J. and {Berge}, D. and {Bernhard}, S. and {Bernl{\"o}hr}, K. and {Birsin}, E. and {Biteau}, J. and {B{\"o}ttcher}, M. and {Boisson}, C. and {Bolmont}, J. and {Bordas}, P. and {Bregeon}, J. and {Brun}, F. and {Brun}, P. and {Bryan}, M. and {Bulik}, T. and {Carrigan}, S. and {Casanova}, S. and {Chadwick}, P.~M. and {Chakraborty}, N. and {Chalme-Calvet}, R. and {Chaves}, R.~C.~G. and {Chr{\'e}tien}, M. and {Colafrancesco}, S. and {Cologna}, G. and {Conrad}, J. and {Couturier}, C. and {Cui}, Y. and {Dalton}, M. and {Davids}, I.~D. and {Degrange}, B. and {Deil}, C. and {deWilt}, P. and {Djannati-Ata{\"\i}}, A. and {Domainko}, W. and {Donath}, A. and {O'C. Drury}, L. and {Dubus}, G. and {Dutson}, K. and {Dyks}, J. and {Dyrda}, M. and {Edwards}, T. and {Egberts}, K. and {Eger}, P. and {Espigat}, P. and {Farnier}, C. and {Fegan}, S. and {Feinstein}, F. and {Fernandes}, M.~V. and {Fernandez}, D. and {Fiasson}, A. and {Fontaine}, G. and {F{\"o}rster}, A. and {F{\"u}{\ss}ling}, M. and {Gabici}, S. and {Gajdus}, M. and {Gallant}, Y.~A. and {Garrigoux}, T. and {Giavitto}, G. and {Giebels}, B. and {Glicenstein}, J.~F. and {Gottschall}, D. and {Grondin}, M.-H. and {Grudzi{\'n}ska}, M. and {Hadsch}, D. and {H{\"a}ffner}, S. and {Hahn}, J. and {Harris}, J. and {Heinzelmann}, G. and {Henri}, G. and {Hermann}, G. and {Hervet}, O. and {Hillert}, A. and {Hinton}, J.~A. and {Hofmann}, W. and {Hofverberg}, P. and {Holler}, M. and {Horns}, D. and {Ivascenko}, A. and {Jacholkowska}, A. and {Jahn}, C. and {Jamrozy}, M. and {Janiak}, M. and {Jankowsky}, F. and {Jung}, I. and {Kastendieck}, M.~A. and {Katarzy{\'n}ski}, K. and {Katz}, U. and {Kaufmann}, S. and {Kh{\'e}lifi}, B. and {Kieffer}, M. and {Klepser}, S. and {Klochkov}, D. and {Klu{\'z}niak}, W. and {Kolitzus}, D. and {Komin}, Nu. and {Kosack}, K. and {Krakau}, S. and {Krayzel}, F. and {Kr{\"u}ger}, P.~P. and {Laffon}, H. and {Lamanna}, G. and {Lefaucheur}, J. and {Lefranc}, V. and {Lemi{\`e}re}, A. and {Lemoine-Goumard}, M. and {Lenain}, J.-P. and {Lohse}, T. and {Lopatin}, A. and {Lu}, C.-C. and {Marandon}, V. and {Marcowith}, A. and {Marx}, R. and {Maurin}, G. and {Maxted}, N. and {Mayer}, M. and {McComb}, T.~J.~L. and {M{\'e}hault}, J. and {Meintjes}, P.~J. and {Menzler}, U. and {Meyer}, M. and {Mitchell}, A.~M.~W. and {Moderski}, R. and {Mohamed}, M. and {Mor{\r{a}}}, K. and {Moulin}, E. and {Murach}, T. and {de Naurois}, M. and {Niemiec}, J. and {Nolan}, S.~J. and {Oakes}, L. and {Odaka}, H. and {Ohm}, S. and {Opitz}, B. and {Ostrowski}, M. and {Oya}, I. and {Panter}, M. and {Parsons}, R.~D. and {Paz Arribas}, M. and {Pekeur}, N.~W. and {Pelletier}, G. and {Perez}, J. and {Petrucci}, P.-O. and {Peyaud}, B. and {Pita}, S. and {Poon}, H. and {P{\"u}hlhofer}, G. and {Punch}, M. and {Quirrenbach}, A. and {Raab}, S. and {Reichardt}, I. and {Reimer}, A. and {Reimer}, O. and {Renaud}, M. and {de los Reyes}, R. and {Rieger}, F. and {Rob}, L. and {Romoli}, C. and {Rosier-Lees}, S. and {Rowell}, G. and {Rudak}, B. and {Rulten}, C.~B. and {Sahakian}, V. and {Salek}, D. and {Sanchez}, D.~A. and {Santangelo}, A. and {Schlickeiser}, R. and {Sch{\"u}ssler}, F. and {Schulz}, A. and {Schwanke}, U. and {Schwarzburg}, S. and {Schwemmer}, S. and {Sol}, H. and {Spanier}, F. and {Spengler}, G. and {Spies}, F. and {Stawarz}, {\L}. and {Steenkamp}, R. and {Stegmann}, C. and {Stinzing}, F. and {Stycz}, K. and {Sushch}, I. and {Tavernet}, J.-P. and {Tavernier}, T. and {Taylor}, A.~M. and {Terrier}, R. and {Tluczykont}, M. and {Trichard}, C. and {Valerius}, K. and {van Eldik}, C.},
        title = "{Long-term monitoring of PKS 2155-304 with ATOM and H.E.S.S.: investigation of optical/{\ensuremath{\gamma}}-ray correlations in different spectral states}",
      journal = {\aap},
         year = 2014,
        month = nov,
       volume = {571},
          eid = {A39},
        pages = {A39},
          doi = {10.1051/0004-6361/201424142},
archivePrefix = {arXiv},
       eprint = {1409.0253},
 primaryClass = {astro-ph.HE},
       adsurl = {https://ui.adsabs.harvard.edu/abs/2014A&A...571A..39H}
}

@BOOK{Rybicki79,
       author = {{Rybicki}, George B. and {Lightman}, Alan P.},
        title = "{Radiative processes in astrophysics}",
         year = 1979,
       adsurl = {https://ui.adsabs.harvard.edu/abs/1979rpa..book.....R}
}

@ARTICLE{Tavecchio98,
       author = {{Tavecchio}, Fabrizio and {Maraschi}, Laura and {Ghisellini}, Gabriele},
        title = "{Constraints on the Physical Parameters of TeV Blazars}",
      journal = {\apj},
         year = 1998,
        month = dec,
       volume = {509},
       number = {2},
        pages = {608-619},
          doi = {10.1086/306526},
archivePrefix = {arXiv},
       eprint = {astro-ph/9809051},
 primaryClass = {astro-ph},
       adsurl = {https://ui.adsabs.harvard.edu/abs/1998ApJ...509..608T}
}

@ARTICLE{Pandey17,
       author = {{Pandey}, Ashwani and {Gupta}, Alok C. and {Wiita}, Paul J.},
        title = "{X-Ray Intraday Variability of Five TeV Blazars with NuSTAR}",
      journal = {\apj},
         year = 2017,
        month = jun,
       volume = {841},
       number = {2},
          eid = {123},
        pages = {123},
          doi = {10.3847/1538-4357/aa705e},
archivePrefix = {arXiv},
       eprint = {1705.02719},
 primaryClass = {astro-ph.HE},
       adsurl = {https://ui.adsabs.harvard.edu/abs/2017ApJ...841..123P}
}

@ARTICLE{Mannheim93,
       author = {{Mannheim}, K.},
        title = "{The proton blazar.}",
      journal = {\aap},
         year = 1993,
        month = mar,
       volume = {269},
        pages = {67-76},
          doi = {10.48550/arXiv.astro-ph/9302006},
archivePrefix = {arXiv},
       eprint = {astro-ph/9302006},
 primaryClass = {astro-ph},
       adsurl = {https://ui.adsabs.harvard.edu/abs/1993A&A...269...67M}
}

@ARTICLE{Mucke01,
       author = {{M{\"u}cke}, A. and {Protheroe}, R.~J.},
        title = "{A proton synchrotron blazar model for flaring in Markarian 501}",
      journal = {Astroparticle Physics},
         year = 2001,
        month = mar,
       volume = {15},
       number = {1},
        pages = {121-136},
          doi = {10.1016/S0927-6505(00)00141-9},
archivePrefix = {arXiv},
       eprint = {astro-ph/0004052},
 primaryClass = {astro-ph},
       adsurl = {https://ui.adsabs.harvard.edu/abs/2001APh....15..121M}
}

@ARTICLE{Bottcher13,
       author = {{B{\"o}ttcher}, M. and {Reimer}, A. and {Sweeney}, K. and {Prakash}, A.},
        title = "{Leptonic and Hadronic Modeling of Fermi-detected Blazars}",
      journal = {\apj},
         year = 2013,
        month = may,
       volume = {768},
       number = {1},
          eid = {54},
        pages = {54},
          doi = {10.1088/0004-637X/768/1/54},
archivePrefix = {arXiv},
       eprint = {1304.0605},
 primaryClass = {astro-ph.HE},
       adsurl = {https://ui.adsabs.harvard.edu/abs/2013ApJ...768...54B}
}

@ARTICLE{IceCube18,
       author = {{IceCube Collaboration} and {Aartsen}, M.~G. and {Ackermann}, M. and {Adams}, J. and {Aguilar}, J.~A. and {Ahlers}, M. and {Ahrens}, M. and {Al Samarai}, I. and {Altmann}, D. and {Andeen}, K. and {Anderson}, T. and {Ansseau}, I. and {Anton}, G. and {Arg{\"u}elles}, C. and {Auffenberg}, J. and {Axani}, S. and {Bagherpour}, H. and {Bai}, X. and {Barron}, J.~P. and {Barwick}, S.~W. and {Baum}, V. and {Bay}, R. and {Beatty}, J.~J. and {Becker Tjus}, J. and {Becker}, K.-H. and {BenZvi}, S. and {Berley}, D. and {Bernardini}, E. and {Besson}, D.~Z. and {Binder}, G. and {Bindig}, D. and {Blaufuss}, E. and {Blot}, S. and {Bohm}, C. and {B{\"o}rner}, M. and {Bos}, F. and {B{\"o}ser}, S. and {Botner}, O. and {Bourbeau}, E. and {Bourbeau}, J. and {Bradascio}, F. and {Braun}, J. and {Brenzke}, M. and {Bretz}, H.-P. and {Bron}, S. and {Brostean-Kaiser}, J. and {Burgman}, A. and {Busse}, R.~S. and {Carver}, T. and {Cheung}, E. and {Chirkin}, D. and {Christov}, A. and {Clark}, K. and {Classen}, L. and {Coenders}, S. and {Collin}, G.~H. and {Conrad}, J.~M. and {Coppin}, P. and {Correa}, P. and {Cowen}, D.~F. and {Cross}, R. and {Dave}, P. and {Day}, M. and {de Andr{\'e}}, J.~P.~A.~M. and {De Clercq}, C. and {DeLaunay}, J.~J. and {Dembinski}, H. and {De Ridder}, S. and {Desiati}, P. and {de Vries}, K.~D. and {de Wasseige}, G. and {de With}, M. and {DeYoung}, T. and {D{\'\i}az-V{\'e}lez}, J.~C. and {di Lorenzo}, V. and {Dujmovic}, H. and {Dumm}, J.~P. and {Dunkman}, M. and {Dvorak}, E. and {Eberhardt}, B. and {Ehrhardt}, T. and {Eichmann}, B. and {Eller}, P. and {Evenson}, P.~A. and {Fahey}, S. and {Fazely}, A.~R. and {Felde}, J. and {Filimonov}, K. and {Finley}, C. and {Flis}, S. and {Franckowiak}, A. and {Friedman}, E. and {Fritz}, A. and {Gaisser}, T.~K. and {Gallagher}, J. and {Gerhardt}, L. and {Ghorbani}, K. and {Glauch}, T. and {Gl{\"u}senkamp}, T. and {Goldschmidt}, A. and {Gonzalez}, J.~G. and {Grant}, D. and {Griffith}, Z. and {Haack}, C. and {Hallgren}, A. and {Halzen}, F. and {Hanson}, K. and {Hebecker}, D. and {Heereman}, D. and {Helbing}, K. and {Hellauer}, R. and {Hickford}, S. and {Hignight}, J. and {Hill}, G.~C. and {Hoffman}, K.~D. and {Hoffmann}, R. and {Hoinka}, T. and {Hokanson-Fasig}, B. and {Hoshina}, K. and {Huang}, F. and {Huber}, M. and {Hultqvist}, K. and {H{\"u}nnefeld}, M. and {Hussain}, R. and {In}, S. and {Iovine}, N. and {Ishihara}, A. and {Jacobi}, E. and {Japaridze}, G.~S. and {Jeong}, M. and {Jero}, K. and {Jones}, B.~J.~P. and {Kalaczynski}, P. and {Kang}, W. and {Kappes}, A. and {Kappesser}, D. and {Karg}, T. and {Karle}, A. and {Katz}, U. and {Kauer}, M. and {Keivani}, A. and {Kelley}, J.~L. and {Kheirandish}, A. and {Kim}, J. and {Kim}, M. and {Kintscher}, T. and {Kiryluk}, J. and {Kittler}, T. and {Klein}, S.~R. and {Koirala}, R. and {Kolanoski}, H. and {K{\"o}pke}, L. and {Kopper}, C. and {Kopper}, S. and {Koschinsky}, J.~P. and {Koskinen}, D.~J. and {Kowalski}, M. and {Krings}, K. and {Kroll}, M. and {Kr{\"u}ckl}, G. and {Kunwar}, S. and {Kurahashi}, N. and {Kuwabara}, T. and {Kyriacou}, A. and {Labare}, M. and {Lanfranchi}, J.~L. and {Larson}, M.~J. and {Lauber}, F. and {Leonard}, K. and {Lesiak-Bzdak}, M. and {Leuermann}, M. and {Liu}, Q.~R. and {Lozano Mariscal}, C.~J. and {Lu}, L. and {L{\"u}nemann}, J. and {Luszczak}, W. and {Madsen}, J. and {Maggi}, G. and {Mahn}, K.~B.~M. and {Mancina}, S. and {Maruyama}, R. and {Mase}, K. and {Maunu}, R. and {Meagher}, K. and {Medici}, M. and {Meier}, M. and {Menne}, T. and {Merino}, G. and {Meures}, T. and {Miarecki}, S. and {Micallef}, J. and {Moment{\'e}}, G. and {Montaruli}, T. and {Moore}, R.~W. and {Morse}, R. and {Moulai}, M. and {Nahnhauer}, R. and {Nakarmi}, P. and {Naumann}, U. and {Neer}, G.},
        title = "{Multimessenger observations of a flaring blazar coincident with high-energy neutrino IceCube-170922A}",
      journal = {Science},
         year = 2018,
        month = jul,
       volume = {361},
       number = {6398},
          eid = {eaat1378},
        pages = {eaat1378},
          doi = {10.1126/science.aat1378},
archivePrefix = {arXiv},
       eprint = {1807.08816},
 primaryClass = {astro-ph.HE},
       adsurl = {https://ui.adsabs.harvard.edu/abs/2018Sci...361.1378I}
}

@ARTICLE{HI4PI16,
       author = {{HI4PI Collaboration} and {Ben Bekhti}, N. and {Fl{\"o}er}, L. and {Keller}, R. and {Kerp}, J. and {Lenz}, D. and {Winkel}, B. and {Bailin}, J. and {Calabretta}, M.~R. and {Dedes}, L. and {Ford}, H.~A. and {Gibson}, B.~K. and {Haud}, U. and {Janowiecki}, S. and {Kalberla}, P.~M.~W. and {Lockman}, F.~J. and {McClure-Griffiths}, N.~M. and {Murphy}, T. and {Nakanishi}, H. and {Pisano}, D.~J. and {Staveley-Smith}, L.},
        title = "{HI4PI: A full-sky H I survey based on EBHIS and GASS}",
      journal = {\aap},
         year = 2016,
        month = oct,
       volume = {594},
          eid = {A116},
        pages = {A116},
          doi = {10.1051/0004-6361/201629178},
archivePrefix = {arXiv},
       eprint = {1610.06175},
 primaryClass = {astro-ph.GA},
       adsurl = {https://ui.adsabs.harvard.edu/abs/2016A&A...594A.116H}
}
\appendix

\begin{figure*}[!htbp]
	\centering 
        \includegraphics[width=0.52\columnwidth]{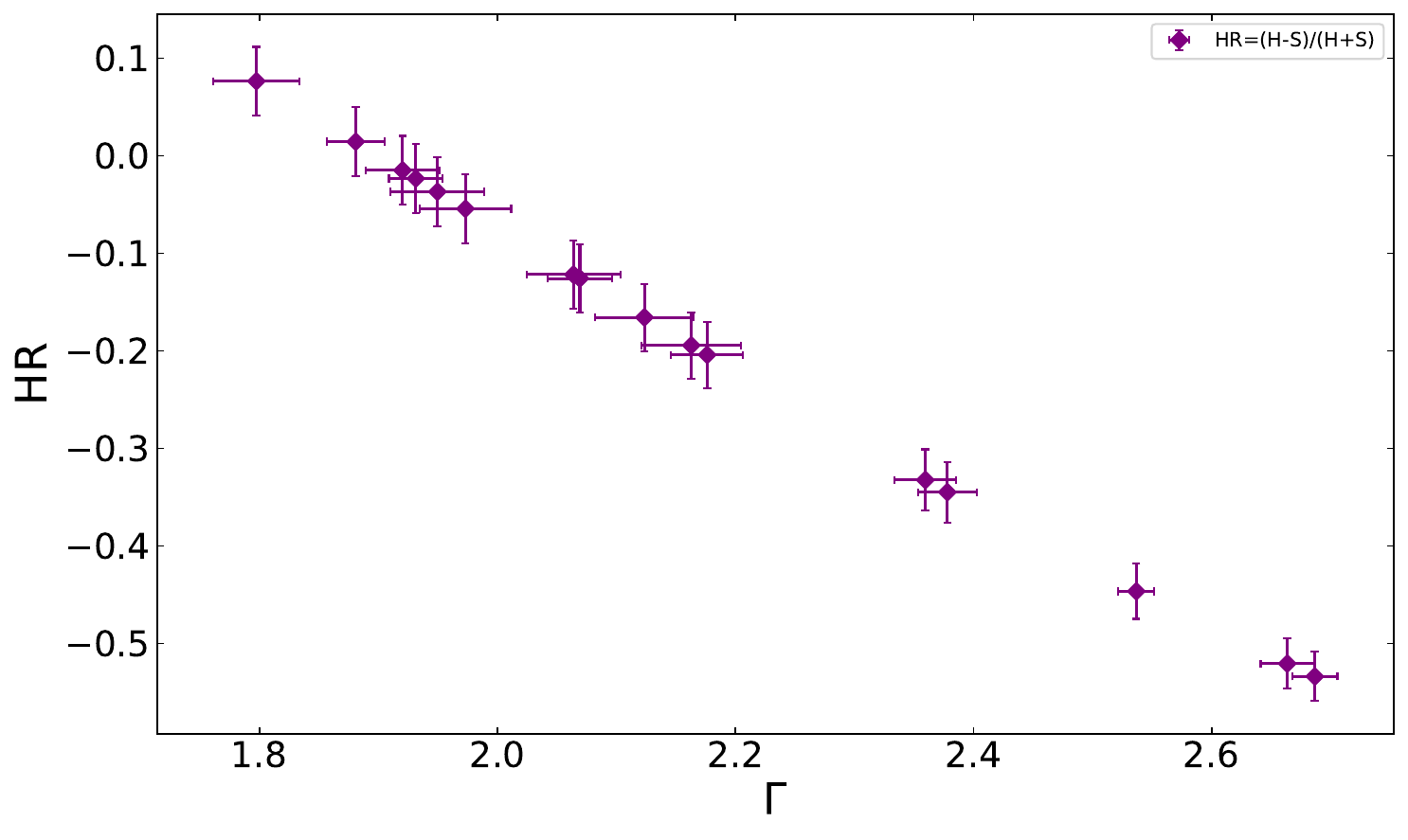}
        \includegraphics[width=0.38\columnwidth]{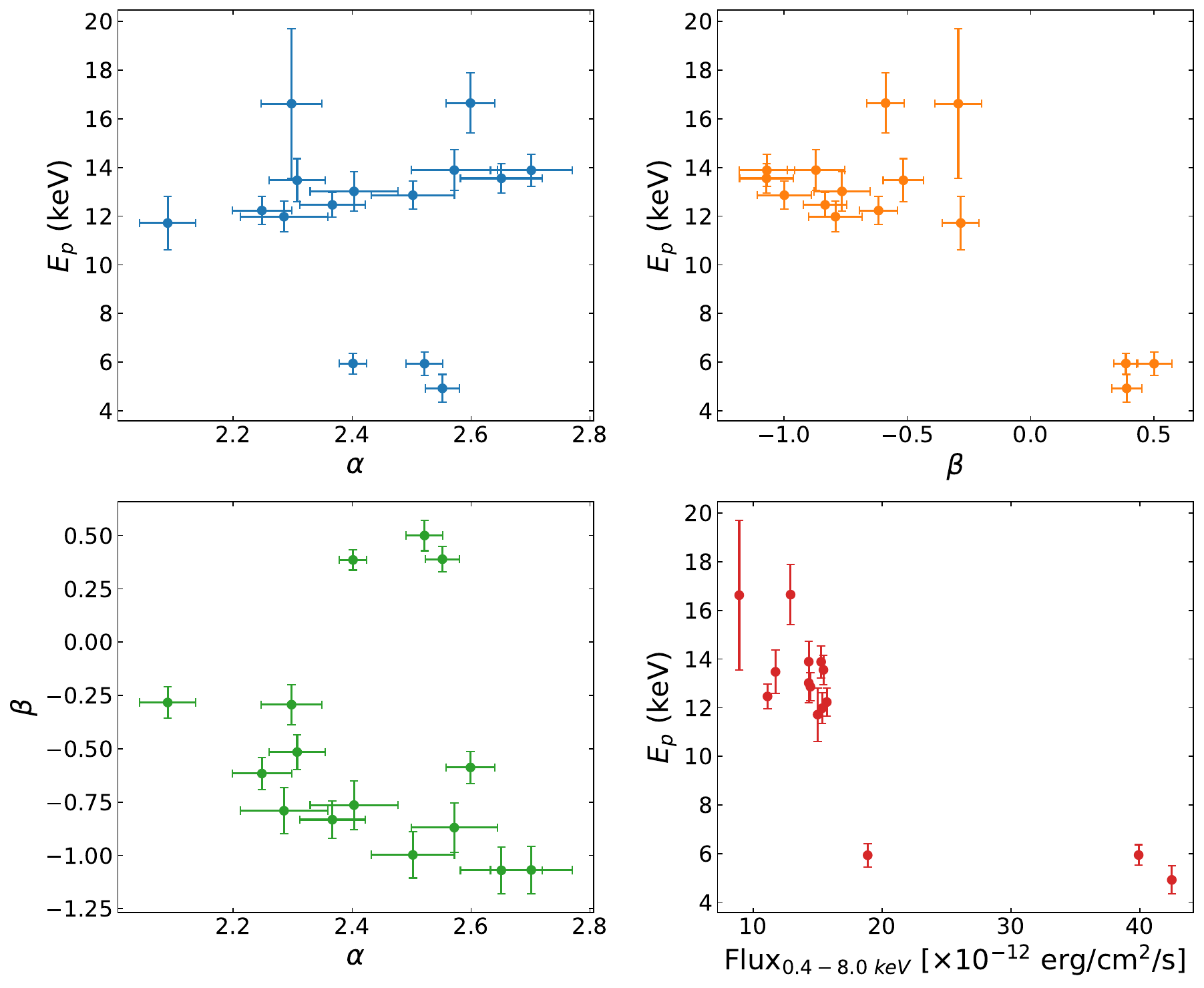}
	\caption{\textit{left}: HR vs $\Gamma$ of PKS~2155$-$304 from {\it Chandra} using PL model. \textit{right}: The plots of $\alpha$ vs $\beta$ vs flux and E$_p$ of PKS~2155$-$304 from Chandra using LP model.}
  \label{fig:PL_Gamma_vs_HR_Chandra}
\end{figure*}

\begin{figure*}[!htbp]
	\centering 
    \includegraphics[width=0.45\columnwidth]{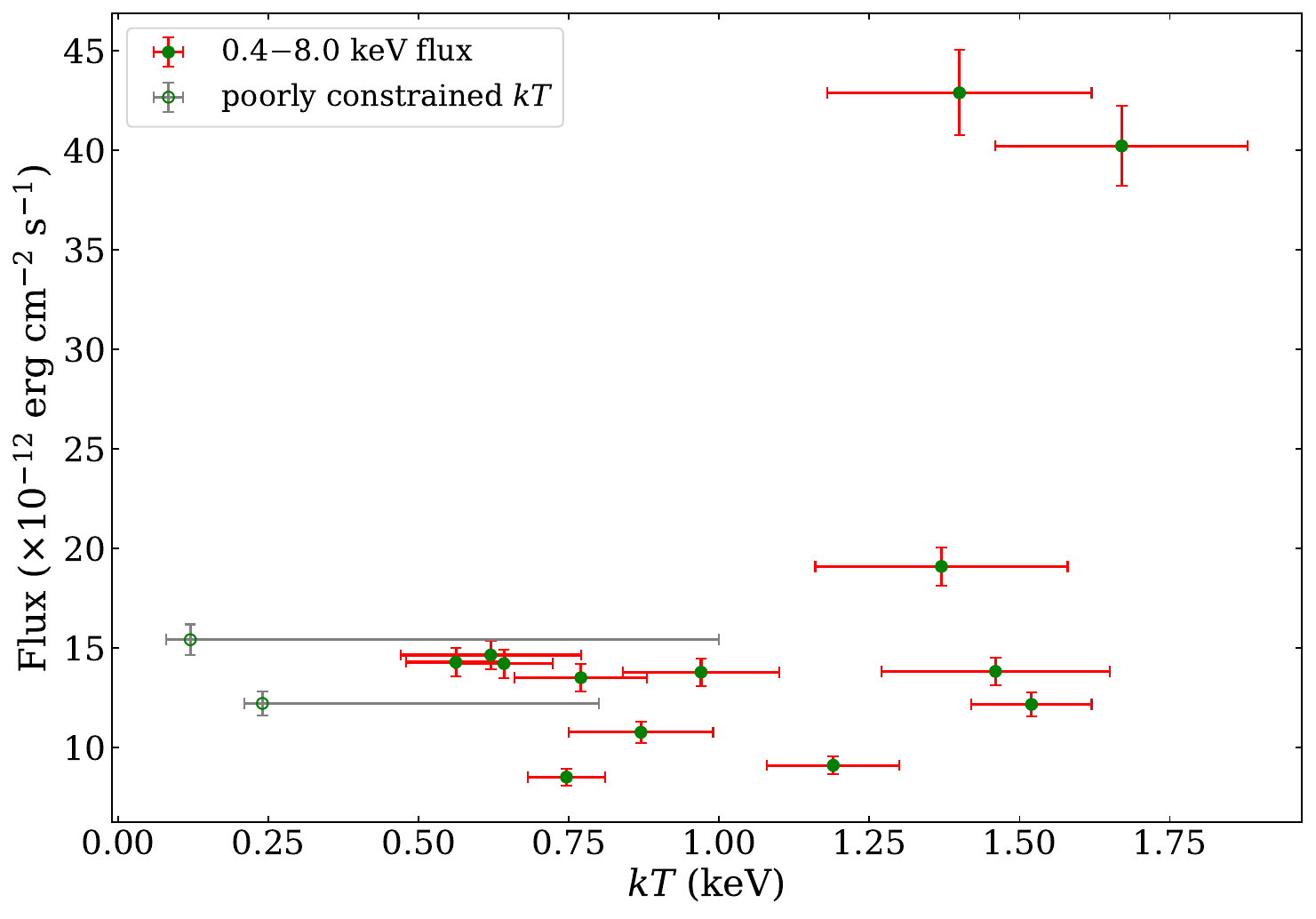}
    \includegraphics[width=0.45\columnwidth]{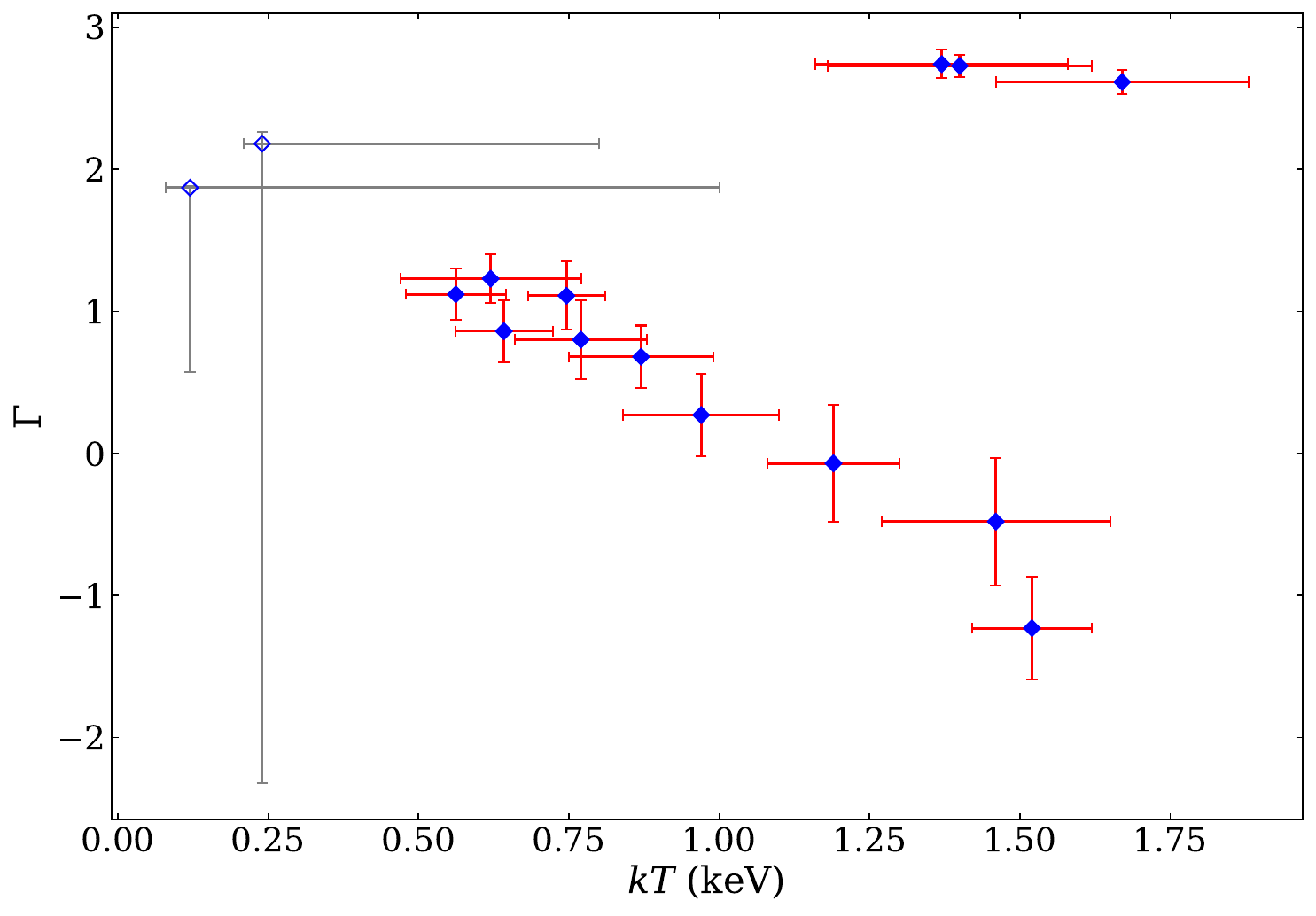}
    \caption{{\it Left}: 0.4$-$8.0 keV flux vs $kT$ temperature in keV; {\it right}: photon index $\Gamma$ vs $kT$ temperature in keV of PKS~2155$-$304 from {\it Chandra} using BR+PL model. \textbf{Open symbols mark ObsIDs 3167 and 5173, for which $kT$ is poorly constrained (asymmetric $1\sigma$ intervals); ObsID 3706, for which $kT$ is unconstrained, is not shown.} }

   \label{fig:Flux_vs_kT_Gamma_vs_kT}
\end{figure*}

\end{document}